\documentclass[aps,pre,amsmath,amsfonts,amssymb,floatfix,superscriptaddress,
nofootinbib]{revtex4}
\usepackage{mathrsfs} \usepackage{graphicx,color} \usepackage{bbm} 
\usepackage{multirow}
\usepackage{algorithm,algpseudocode}

\usepackage{rotate}
\usepackage{epsfig}
\usepackage{psfrag}
\usepackage{mathtools}

\let\a=\alpha \let\b=\beta \let\g=\gamma \let\d=\delta
\let\e=\zeta  \let\h=\eta 
\let\l=\lambda    
 \let\t=\tau \let\f=\varphi 
   
 \let\Th=\Theta\let\L=\Lambda  

\let\r=\rho \let\th=\theta \let\io=\infty

\def\PP{{\cal P}}\def\EE{{\cal E}} 
 
 \def\II{{\cal I}}
\def\LL{{\cal L}}  
\def\DD{{\cal D}}

\def\ola{{\overline \lambda}}

\def\dla{{\overline {\delta \lambda}}_1}
\def\dlaT{{\overline {\delta \lambda}}_2}
\def\dP{{\delta P}}

\def\to{\rightarrow}

\newcommand{\beq}{\begin{equation}} \newcommand{\eeq}{\end{equation}}
 \newcommand{\wt}{\widetilde}

\newcommand{\sgn}{\operatorname{sign}}
\newcommand\be{\begin{equation}}
\newcommand\bea{\begin{eqnarray} \nonumber }
\newcommand\ee{\end{equation}}
\newcommand\eea{\end{eqnarray}}

\begin{document}

\title{Tipping points in macroeconomic Agent-Based models}

\author{Stanislao Gualdi}
\affiliation{
Universit\'e Pierre et Marie Curie - Paris 6, Laboratoire de Physique 
Th\'eorique de la Mati\`ere 
Condens\'ee, 4, Place Jussieu, 
Tour 12, 75252 Paris Cedex 05, France and 
IRAMIS, CEA-Saclay, 91191 Gif sur Yvette Cedex, France
}
\email{stanislao.gualdi@gmail.com}

\author{Marco Tarzia}
\affiliation{
Universit\'e Pierre et Marie Curie - Paris 6, Laboratoire de Physique 
Th\'eorique de la Mati\`ere 
Condens\'ee, 4, Place Jussieu, 
Tour 12, 75252 Paris Cedex 05, France }

\author{Francesco Zamponi}
\affiliation{Laboratoire de Physique Th\'eorique,
\'Ecole Normale Sup\'erieure, UMR 8549 CNRS, 24 Rue Lhomond, 75231 Paris Cedex 
05, France}

\author{Jean-Philippe Bouchaud}
\affiliation{CFM, 23 rue de l'Universit\'e, 75007 Paris, France, and Ecole 
Polytechnique, 91120 Palaiseau, France.}

\begin{abstract} 

\medskip

The aim of this work is to explore the possible types of {\it phenomena} that
simple macroeconomic Agent-Based models (ABM) can reproduce. 
{
We propose a methodology, inspired by statistical physics, 
that characterizes a model through its ``phase diagram'' in the 
space of parameters.
Our first motivation is to understand the large macro-economic fluctuations 
observed
in the ``Mark I'' ABM devised by D. Delli Gatti and collaborators. 
In this regard, our major finding is the generic existence of a {\it phase
transition} between a ``good economy'' where unemployment is low, and a 
``bad economy'' where unemployment is high. 
}
We {then} introduce a simpler framework that allows us to show that this transition is
{\it robust} against many modifications of the model, and is generically 
induced by an asymmetry between the rate of hiring and the rate of firing of the firms. 
The unemployment level remains small until a tipping point, beyond
which the economy suddenly collapses. If the parameters are such that the system
is close to this transition, any small fluctuation is amplified as the system
jumps between the two equilibria. 
We have explored several natural
extensions of the model. One is to introduce a bankruptcy threshold, {limiting 
the firms maximum level of debt-to-sales ratio}. This leads to a rich phase diagram with, in particular, a region
where {\it acute endogenous crises} occur, during which
the unemployment rate shoots up before the economy can recover.
We also introduce simple wage policies. This 
leads to inflation (in the ``good'' phase) or deflation (in the ``bad'' phase), 
but leaves 
the overall phase diagram of the model essentially unchanged. 
We have also explored the effect of simple monetary policies 
that attempt to contain rising unemployment and defang crises. We end the paper
with general comments on the usefulness of ABMs to model macroeconomic
phenomena, in particular in view of the time needed to reach a steady state 
that raises the issue of {\it ergodicity} in these models.
\end{abstract} 

\maketitle

{\it It is human nature to think wisely and to act absurdly} -- Anatole France 

\clearpage
\tableofcontents

\clearpage

\section{Introduction}

\subsection{From micro-rules to macro-behaviour}

Inferring the behaviour of large assemblies from the behaviour of its elementary
constituents is arguably one of 
the most important problems in a host of different disciplines: physics, 
material
sciences, biology, computer sciences, sociology and,
of course, economics and finance. It is also a notoriously hard problem. 
Statistical
physics has developed in the last 150 years essentially to understand 
this micro-macro link. Clearly, when interactions are absent or small enough,
the system as a whole merely reflects the properties of individual
entities. This is the canvas of traditional macro-economic approaches. Economic
agents are assumed to be identical, non-interacting, rational agents, an
idealization known as the ``Representative Agent'' (RA). In this framework,
micro and macro trivially coincide. However, we know (in particular from
physics) that discreteness, heterogeneities and/or interactions can lead to
totally unexpected phenomena. Think for example of super-conductivity or
super-fluidity\footnote{
See e.g. Ref.~\cite{Balibar} for an history of the discovery of super-fluidity 
and a list of references.
}: 
before their experimental discovery, it was simply beyond human
imagination that individual electrons or atoms could ``conspire'' to create a
collective state that can flow without friction. Micro and macro behaviour not
only do not coincide in general, but genuinely {\it surprising} behaviour can
emerge through aggregation. From the point of view of economic theory, this is
interesting,
because financial and economic history is strewn with bubbles, crashes, crises
and upheavals of all sorts. These are very hard to fathom within a
Representative Agent framework~\cite{Kirman}, within which crises would require 
large
aggregate shocks, when in fact small local shocks can trigger large systemic
effects when heterogeneities, interactions and network effects are taken into
account~\cite{Brock,Complexity,JPB,Julius}. The need to include these effects has 
spurred a large activity in
``Agent-Based models'' (ABMs) \cite{ABM-collective, ABM-economics, ABM-review}. 
These models need numerical simulations, but are extremely versatile because 
any possible behavioural rules, interactions,
heterogeneities can be taken into account.

In fact, these models are so versatile that they suffer from the ``wilderness of
high dimensional spaces'' (paraphrasing Sims \cite{Sims}). The number of 
parameters and explicit or implicit choices of behavioural rules is so large 
($\sim 10$ or more, even in the simplest models, see below) that 
the results of the model may appear unreliable and arbitrary, and the 
calibration of the
parameters is an hopeless (or highly unstable) task. 
Mainstream RA ``Dynamic Stochastic General Equilibrium'' models (DSGE), on the other hand, 
are simple enough to lead to closed
form analytical results, with simple narratives and 
well-trodden calibration avenues~\cite{DSGE}.  
{
In spite of their unrealistic character, these models appear to perform 
satisfactorily in `normal' times, when fluctuations are small. However, they become deeply flawed 
in times
of economic instability \cite{Buiter}, suggesting different assumptions are needed to 
understand what is observed in
reality.} But even after the 2008 crisis, these traditional models are still favoured by most 
economists, both in academia and in institutional and professional circles.
ABMs are seen at best as a promising research direction and at worst as an unwarranted ``black box'' (see 
\cite{Roventini} for an enlightening discussion on the debate between traditional DSGE models and 
ABMs, and \cite{Fagiolo1,Fagiolo2,Kirman2,Caballero} for further insights).

\subsection{A methodological manifesto}

\begin{figure}
\centering
\includegraphics[scale=0.7]{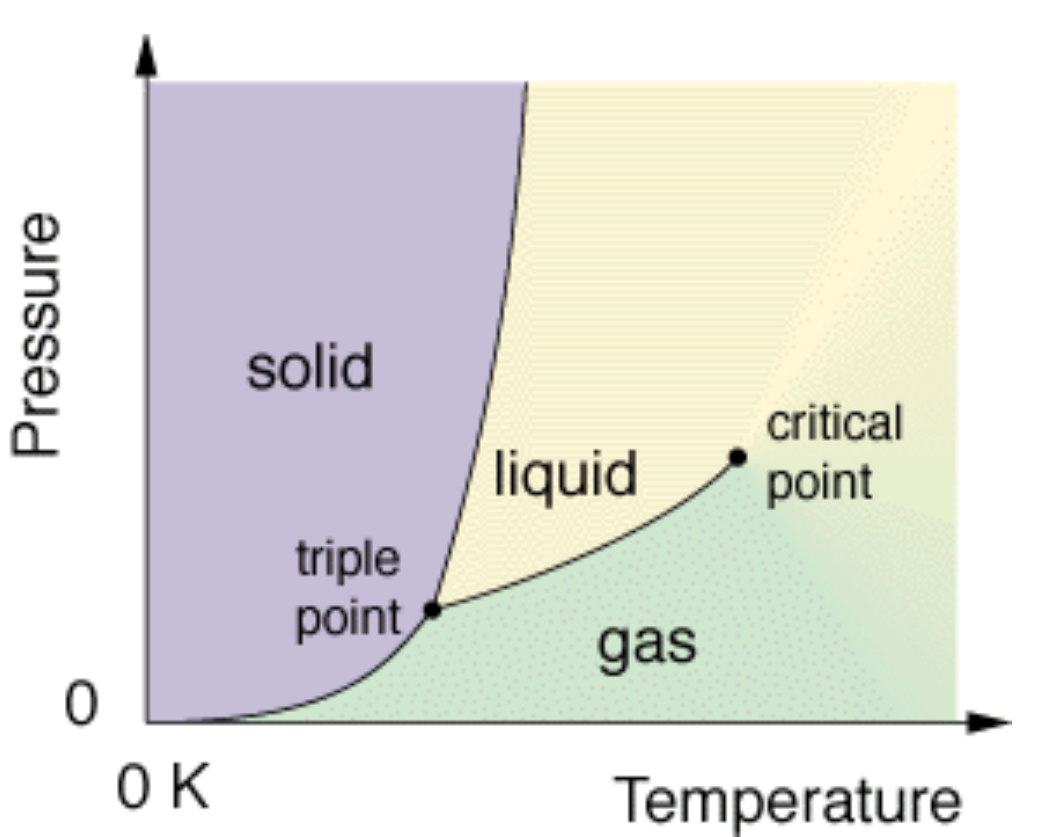}
\caption{
A typical ``phase diagram'', here the solid-liquid-gas phases in the
temperature-pressure plane. Far from phase transitions, within a given phase, 
the behaviour of the
system is qualitatively similar for all values of the parameters. 
Close to phase boundaries, on the other hand, small changes of parameters can
lead to dramatic macroscopic changes.
}
\label{fig:typ_pd}
\end{figure}

At this stage, it seems to us that some clarifications are indeed needed,
concerning both the objectives and methodology of Agent-Based models. 
ABMs do indeed suffer from the wilderness of high dimensional spaces, and some
guidance is necessary to put these models on a firm footing. 
In this respect, statistical physics offers a key concept: the {\it phase 
diagram} in
parameter space~\cite{Goldenfeld}. A classic example, shown in Fig. 1, is the 
phase diagram of usual substances as a function of two parameters, here
temperature and pressure. The generic picture is that the number
of distinct phases is usually small (e.g. three in the example of
Fig.~\ref{fig:typ_pd}: solid, liquid, gas). Well within each phase, the
properties are qualitatively similar 
and small changes of parameters have a small effect. Macroscopic (aggregate)
properties do not fluctuate any more for very large systems and are robust
against changes of microscopic details. This is the ``nice'' scenario, where the
dynamics of the system can be described in terms of a small number of
macroscopic (aggregate) variables, with some effective parameters that
encode the microscopic details. But other scenarios are of course possible; for
example, if one sits close to the boundary between two phases, fluctuations can
remain large even for large systems and small changes of parameters can 
radically
change the macroscopic behaviour of the system. There may be mechanisms 
naturally
driving the system close to criticality (like in Self Organized 
Criticality~\cite{Bak}),
or, alternatively, situations in which whole phases are critical, like for 
``spin-glasses''~\cite{SG}.

In any case, the very first step in exploring the properties of an Agent-Based
model should be, we believe, to identify the different possible phases in 
parameter space 
and the location of the phase boundaries. In order to do this, numerical
simulations turn out to be very helpful \cite{Buchanan,Foley}: aggregate 
behaviour usually quickly sets in, even
for small sizes. Some parameters usually turn out to be crucial, while others
are found to play little role. This is useful to establish a qualitative {\it
phenomenology} of the model -- what kind of behaviour can the model reproduce,
which basic mechanisms are important, which effects are potentially missing?
This first, qualitative step allows one to unveil the ``skeleton'' of the ABM.
Simplified models that retain most of the phenomenology can then be constructed
and perhaps solved analytically, enhancing the understanding of the important
mechanisms, and providing some narrative to make sense of the observed effects.
In our opinion, calibration of an ABM using real data can only start to make
sense 
after this initial qualitative investigation is in full command -- which is in
itself not trivial when the number of parameters is large. The phase diagram of
the model allows one to restrict the region of parameters that lead to
``reasonable'' outcomes (see for example the discussion in 
\cite{Irene,financeABM}).

\subsection{Outline, results {and limitations} of the paper}

The aim of this paper is to put these ideas into practice in the context of a
well-studied macroeconomic Agent-Based model (called ``Mark I'' below), devised
by Delli Gatti and collaborators \cite{MarkIref,MarkIbook}. This model is at 
the core of the European
project ``CRISIS'', which {partly} justifies our attempt to shed some theoretical light
on this framework.\footnote{see www.crisis-economics.eu} In the first part of 
the paper, we briefly recall the main
ingredients of the model and show that as one increases the baseline interest
rate, there is a phase transition between a ``good'' state of the economy, where
unemployment is low and a ``bad'' state of the 
economy where production and demand collapse. In the second part of the paper,
we study the phase diagram of a highly simplified version of Mark I, dubbed 
`Mark 0' that aims at capturing the main
drivers of this phase transition. {The model is a ``hybrid'' macro/ABM model where firms are 
treated individually but households are only described in aggregate}. Mark 0 does not include any exogeneous 
shock; crises can only be of endogeneous origin.
We show that the most important parameter in this regard is
the asymmetry between the firms' propensity to hire (when business is good) or
to fire (when business is bad). In Mark I, this asymmetry is induced by the
reaction of firms to the level of interest rates, but other plausible mechanisms
would lead to the same effect. The simplest version of the model is amenable to
an analytic treatment and exhibits a ``tipping point'' (i.e. a discontinuous
transition) between high employment and high unemployment.
{
When a bankruptcy condition is introduced {(in the form of a maximum level of 
debt-to-sales ratio)}, the model reveals an extremely rich
phenomenology: the ``good'' phase
of the economy is further split into three distinct phases: one of full employment, 
a second one with a substantial level of residual unemployment, and a third, 
highly interesting region, where endogeneous crises appear. 
We find that both the amount of credit available to firms (which in 
our model sets the bankrupcty threshold), and the way default 
costs are absorbed by the system, are the most important aspects 
in shaping the qualitative behavior of the economy.
}
Finally, we allow wages to adapt (whereas they are kept fixed in Mark I) and 
allow inflation or deflation to set in. Still, the overall shape of the phase diagram 
is not modified. We investigate further enhancements of the models, in particular 
simple policy experiments. Open questions
and future directions, in particular concerning macroeconomic ABMs in general,
are discussed in our final section.

{Before embarking to the core of our results, we want to clearly state 
what our ambition and objectives are, and what they are not. We do claim that the methodology
proposed here is interesting and general, and could help improving the relevance of macroeconomic ABMs.
We do believe that large aggregate volatility and crises (in particular those appearing in 
Mark I) can be understood through the lens of instabilities and phase transitions, as exemplified by our 
highly stylized Mark 0 model. We are also convinced 
that simple ``skeleton'' ABMs (for which most of the phenomenology can be fully dissected) 
must be developed and compared with traditional DSGE models before embarking into full-fledged models of the
economy.
On the other hand, we do {\it not} claim that our basic Mark 0 framework is necessarily the best starting point. Mark 0
was primarily set up as a simplification of Mark I. Still, we find that Mark 0 leads to a surprisingly 
rich and to some extent realistic set of possible behaviours, including business cycles and crises, inflation, 
policy experiments, etc. However, we do {\it not} wish to claim that Mark 0 is able to reproduce all the known 
empirical stylized facts and could well be in contradiction with some of them\footnote{
The idea of building {\it mathematical models of reality} that reproduce some phenomena, but might even be in contradiction with others,
is at the heart of the development of physics. Most probably, its use was introduced in the Hellenistic period~\cite{Russo}.
A striking example~\cite{Russo} is Archimedes' {\it On Floating Bodies}. In the first of the two books, in fact, Archimedes
provides a mathematical proof of the sphericity of Earth (assumed to be liquid and at rest). However, in the second book, he
assumes the surface of Earth to be flat for the purpose of describing other phenomena, for which the sphericity of Earth is irrelevant.
%This kind of reasoning, which is typical of Hellenistic scientists and of modern physicists, is not always easily understood outside these communities.
%For example, several intellectuals in the Roman age criticized Hellenistic scientists, blaming them for ``contradicting themselves''~\cite{Russo}.
}. In view of the simplicity of the model, this has to be expected. 
But we believe that Mark 0 can serve as a useful building block in the quest of a more comprehensive ABM, as more and more effects are 
progressively incorporated, in a controlled manner, to the model. Even if Mark 0 turns out to be little more 
than a methodological exercice, we hope that the ABM community (and perhaps beyond) will find it inspiring.
}

{To conclude this (long) introduction, let us insist that all the claims made in this paper only refer to the studied {\it models}, 
but do not necessarily apply to economic reality. If fact, our central point is that a model has to be understood inside-out 
before even trying to match any empirical fact.}

\section{A phase transition in ``Mark I''}

\subsection{Description of the model in a nutshell}

The Mark I family of agent-based models was proposed by Delli Gatti and 
collaborators as a family of simple stylized
macroeconomic models \cite{MarkIref,MarkIbook}. Note that
several other macroeconomic Agent-Based models have been put forth in the recent
years, see~\cite{Eurace,Eurace2,SantAnna,Dosi,Lagom,Gordon}. Mark I is {particularly}
interesting because large fluctuations in
unemployment and output seem to persist in the stationary state (a feature in 
fact shared by many ABMs cited above). 

The Mark I economy~\cite{MarkIref} 
is made up of a set of firms, households, firms owners and a bank. 
Firms produce a certain quantity of a single (and not storable) good, 
proportional to the
number of their employees, that is sold at a time-dependent and firm-dependent 
prices. Firms pay identical time-independent wages. 
When the cash owned by a firm is not enough to pay the wages, it asks banks for 
a loan. The
bank provides loans to the firms at an interest rate
that depends on the financial fragility of the firm. Households provide
workforce in exchange of a salary and want to spend a fixed fraction of their 
savings (or wealth). 
The owners of the firms do not work but receive dividends if the firms make 
profits.
Firms are adaptive, in the sense that they continuously update their production
(i.e. they hire/fire workers) and their prices, in an attempt to
match their production with the demand of goods issued by the households. They
also choose how much extra loan they want to take on, as a function of the
offered interest rate. This last feature, combined with the price and 
production update rules, will turn out to be crucial in the
dynamics of the model. 

The above description defines the basic structure of the Mark I family, but it 
is of course totally insufficient to code the model, since
many additional choices have to be made, leading to several different possible 
implementations of the model. 
Here we will use as a baseline model one of the simplest implementation of Mark 
I, whose
description can be found in~\cite{Mark1_Milan}; the total number of parameters 
in this version is $10$
(but some parameters are actually implicitly fixed from the beginning). 
We have recoded this basic version and also a slightly different version
that we call ``Mark I+'',  which differs on minor details 
(some that we will specify below) but also on one major aspect: our version of
the model {\it strictly conserves the amount of money in circulation}, i.e. the 
money
in the bank + total firm assets + total households wealth (here savings), in order to avoid 
-- at this stage -- any
effect due to uncontrolled money creation. A detailed
pseudo-code of Mark I+ is provided in Appendix~\ref{app:Mark1+}. 

\subsection{State variables}

In short (see Appendix~\ref{app:Mark1+} for a complete description), the 
dynamic evolution of the model
is defined by the following state variables. 
The state of each firm $i = 1\cdots N_{\rm F}$ is specified by its price 
$p_i(t)$, the salary it offers $W_i(t)$, its production $Y_i(t)$,
its target production $Y^T_i(t)$, its demand $D_i(t)$, its liquidity 
$\LL_i(t)$, its total debt $\DD_i^T(t)$.
Moreover, each firm is owned by a household and has a list of employees that is 
dynamically updated.
The state of each household $a = 1 \cdots N_{\rm H}$ is specified by its 
wealth (in the form of savings) $S_a(t)$ and by the firm for which it works (if any).

\subsection{Update rules for prices and production}

Among all the micro-rules that any Agent-Based model has to postulate, some seem
to be more crucial than others. An important item in Mark I is the behavioural
rule for firms adaptation to their economic environment. Instead of the
standard, infinite horizon, profit optimizing firm framework (that is both
unrealistic and intractable), Mark I postulates a heuristic rule for production
$Y_i(t)$ and price $p_i(t)$ update, which reads as follows:
\beq     \label{update}
\begin{split}
     Y_i(t) = D_i(t) \mbox{\ \& } p_i(t)>\bar{p}(t)  &\Rightarrow
Y^T_i(t+1)=Y_i(t)[1+\gamma_y \xi_i(t)]  \\
     Y_i(t) = D_i(t) \mbox{\ \& } p_i(t)<\bar{p}(t)  &\Rightarrow
p_i(t+1)=p_i(t)[1+\gamma_p \xi_i(t)]  \\
     Y_i(t) > D_i(t) \mbox{\ \& } p_i(t)<\bar{p}(t)  &\Rightarrow 
Y^T_i(t+1)=Y_i(t)[1-\gamma_y \xi_i(t)]   \\
     Y_i(t) > D_i(t) \mbox{\ \& } p_i(t)>\bar{p}(t)  &\Rightarrow 
p_i(t+1)=p_i(t)[1-\gamma_p \xi_i(t)],  
\end{split}
\eeq
where $D_i(t)$ is the total demand for the goods produced by firm $i$ at time
$t$, and
\beq
\bar{p}(t) = \frac{\sum_i p_i(t) D_i(t) }{\sum_i D_i(t)}
\eeq
is the average price of sold goods at time $t$, $\xi_i(t)$ a
$U[0,1]$ random variable, independent across firms and across times, and
$\gamma_y, \gamma_p$ two  parameters in $[0,1]$. 
The quantity $Y^T_i(t)$ is the {\it target} production at time $t$, not
necessarily the realized one, as described below. 
These heuristic rules can be interpreted as a plausible {\it t\^atonnement}
process of the firms, that attempt to guess their correct production level and
price based on the information on the last time step. In spirit, each unit time
step might correspond to a quarter, so the order of magnitude of the $\gamma$
parameters should be a few percent. Note that in the version of Mark I that we 
consider, wages are fixed to a constant
value $W_i(t) \equiv 1$, for all times and all firms.

{
As we shall see later, the adaptive price/production adjustments described
in Eq. \eqref{update} leads to two stable attractors (full employment and full
unemployment). Which of the two is be reached by the dynamics depends mainly on the
level of asymmetry between an upward and downward production adjustments.
In Eq. \eqref{update}, the production adjustment depends
on a single parameter $\gamma_y$ and in this case the system evolves towards a full
employment state in the absence of any other constraint. However, as it will become clear in the following section,
financial constraints on loans may lead to an effective ``weakening'' of the upward adjustment, 
possibly driving the dynamics towards the full unemployment state.
As long as such asymmetries between upward and downward production adjustments exist 
(together with some noise in the price dynamics), the scenario described above and in the
following sections is very general and in fact do not depend on details of the update process.
}

\subsection{Debt and loans}

The model further assumes linear productivity, hence the target production 
corresponds
to a target workforce $Y^T_i(t)/\alpha$, where $\alpha$ is a constant
coefficient that can always be set to unity (gains in productivity are not
considered at this stage). The financial need of the firm is
$\max{[0,Y^T_i(t)W_i(t)-\mathcal{L}_i(t)]}$, where $\mathcal{L}_i(t)$ is the
cash available. The total current debt of the firm is $\mathcal{D}^T_i(t)$. The
financial fragility of the firm $\ell_i(t)$ is defined in Mark I as the ratio of
debt over cash. The offered rate by the banks for the 
loan is given by:
\be \label{rate}
\rho_i(t)=\rho_0 G\left(\ell_i(t)\right) \times (1 + \xi_i^\prime(t)),
\ee
where $\rho_0$ is the baseline (central bank) interest rate, $G$ is an
increasing function (taken to be $G(\ell) = 1 + \tanh(\ell)$ in the reference 
Mark I 
and $G(\ell)=1+\ln(1+\ell)$ in Mark I+), and
$\xi^\prime$ another noise term
{drawn from a uniform distribution $U[0,1]$}.
Depending on the rate offered, firms decide to take the full loan
they need or only a fraction $F(\rho)$ of it, where $F \leq 1$ is a decreasing
function of $\rho$, called ``credit contraction''. For example, in the 
reference Mark I,
$F(\rho \leq 5 \%)=1$ and $F(\rho > 5 \%)=0.8$. We have played with the choice
of the two functions $F,G$ and the phase transition reported below is in fact
robust whenever these functions are reasonable. In Mark I+, 
we chose a continuous function, such as to avoid built in discontinuities:\footnote{ {Note however that the arbitrary thresholds ($5 \%$ and $10 \%$) in 
Eq. (\ref{creditcontraction}) are of little importance and only affect the precise location of the phase transition.}}
\be
\label{creditcontraction}
F(x) = 
\begin{cases} 
1 \quad\mbox{if  } x < 5 \% \\ 
1-\frac{x-5 \%}{5\%} \quad \mbox{if  } 5 \% <  x < 10 \%\\
0 \quad \mbox{if  } x > 10 \%.
 \end{cases} 
\ee
The important feature here is that when $F<1$, the firm does not have enough
money to hire the target workforce $Y^T_i(t)$ and is therefore obliged to
hire less, or even to start firing in order to match its financial 
constraints. This financial constraint therefore induces an {\it asymmetry} in the hiring/firing
process: when firms are indebted, hiring will be slowed down by the cost of
further loans. As we will see later, this asymmetry is responsible for an abrupt
change in the steady state of the economy.

\subsection{Spending budget and bankruptcy}

Firms pay salaries to workers and households determine their budget as a
fraction $c$ (constant in time and across households) of their total wealth
(including the latest salary). Each household then selects $M$ firms at random
and sorts them according to their price; it then buys all it can buy from each
firm sequentially, from the lower price to the highest price\footnote{
{
In this 
sense, the good market is not efficient since the household
demand is not necessarily satisfied when $M$ is small. 
The job market instead, though not efficient 
due to the presence of unvoluntarely unemployment,  
is characterized by perfect information
since all the workers can contact all the firms until all the open positions 
are filled.
}}.
The budget left-over is added to the savings. Each firm sells a quantity $D_i(t)
\leq Y_i(t)$, compute its profits (that includes interests paid on debt), and
updates its cash and debt accordingly. Moreover, each firm pays back to the 
bank a fraction $\t$ of its
total debt $\DD^T_i(t)$.
It also pays dividends to the firm owners
if profits are positive. Firms with negative liquidity $\mathcal{L}_i(t)<0$ go
bankrupt. In Mark I+, the cost of the bankruptcy (i.e. $-\mathcal{L}_i(t)$) is
spread over healthy firms and on households. Once a firm is bankrupt it is
re-initialized in the next time step with the owner's money, to a firm with a
price and production equal to their corresponding average values at that moment
in time, and zero debt (see Appendix~\ref{app:Mark1+} for more precise 
statements).

\subsection{Numerical results: a phase transition}

When exploring the phase space of Mark I, it soon becomes clear that the 
baseline 
interest rate $\rho_0$ plays a major role. In order not to mix different
effects, we remove altogether the noise term $\xi$ in Eq. (\ref{rate}) that 
affects the actual
rate offered to the firms. We find that as long as $\rho_0$ is smaller than a
certain threshold $\rho_c$, firms are on average below the credit contraction 
threshold and always manage to have enough loans to pay wages. In this case the
economy is stable and after few ($\sim 100$) time steps reaches a stationary 
state where the
unemployment rate is low. If on the other hand the baseline interest rate
$\rho_0$ exceeds a critical value $\rho_c$, firms cannot afford to take as much
loans as they would need to hire (or keep) the desired amount of workers. 
Surprisingly, this induces a sudden, catastrophic breakdown of the economy.
Production collapses to very small values and unemployment sky-rockets. This
transition between two states of the economy takes place in both the reference
Mark I and in the modified Mark I+; as we shall show in the next section, this
transition is actually generic and occurs in simplified models as well. Note in particular
that $\rho_c$ is  {\it different} from the value at which $F(x)$ starts decreasing.

The data we show in Fig.~\ref{fig:M1_PT} corresponds to Mark I+ with parameters
$\gamma_p=\gamma_y=0.1$ and $M=3$ (see Appendix~\ref{app:Mark1+} for the 
general parameter
setting of the model). While the qualitative behaviour of the model is robust, 
the
details of the transition may change with other parameter settings. For example,
smaller values of $\gamma_p, \gamma_y$ lead to lower critical thresholds
$\rho_c$ (as well as smaller values of $M$) and to longer equilibration times
($T_{eq}$ scales approximately as $1/\gamma_{y,p}$ for $\rho_0<\rho_c$). 
Increasing the
size of the economy only affects the magnitude of the fluctuations within one
phase leaving the essential features of the transition unchanged.
Interestingly, although it is not clear from Fig.~\ref{fig:M1_PT}, the model 
exhibits
oscillatory patterns of the employment rate. The presence of these oscillations
can be seen in the frequency domain of the employment rate time series (not
shown here), which is essentially characterized by a white noise power spectrum
with a well defined peak at intermediate frequencies. All these effects will
become clearer within the reduced model described in the next section.

\begin{figure}
\centering
\includegraphics[scale=0.25]{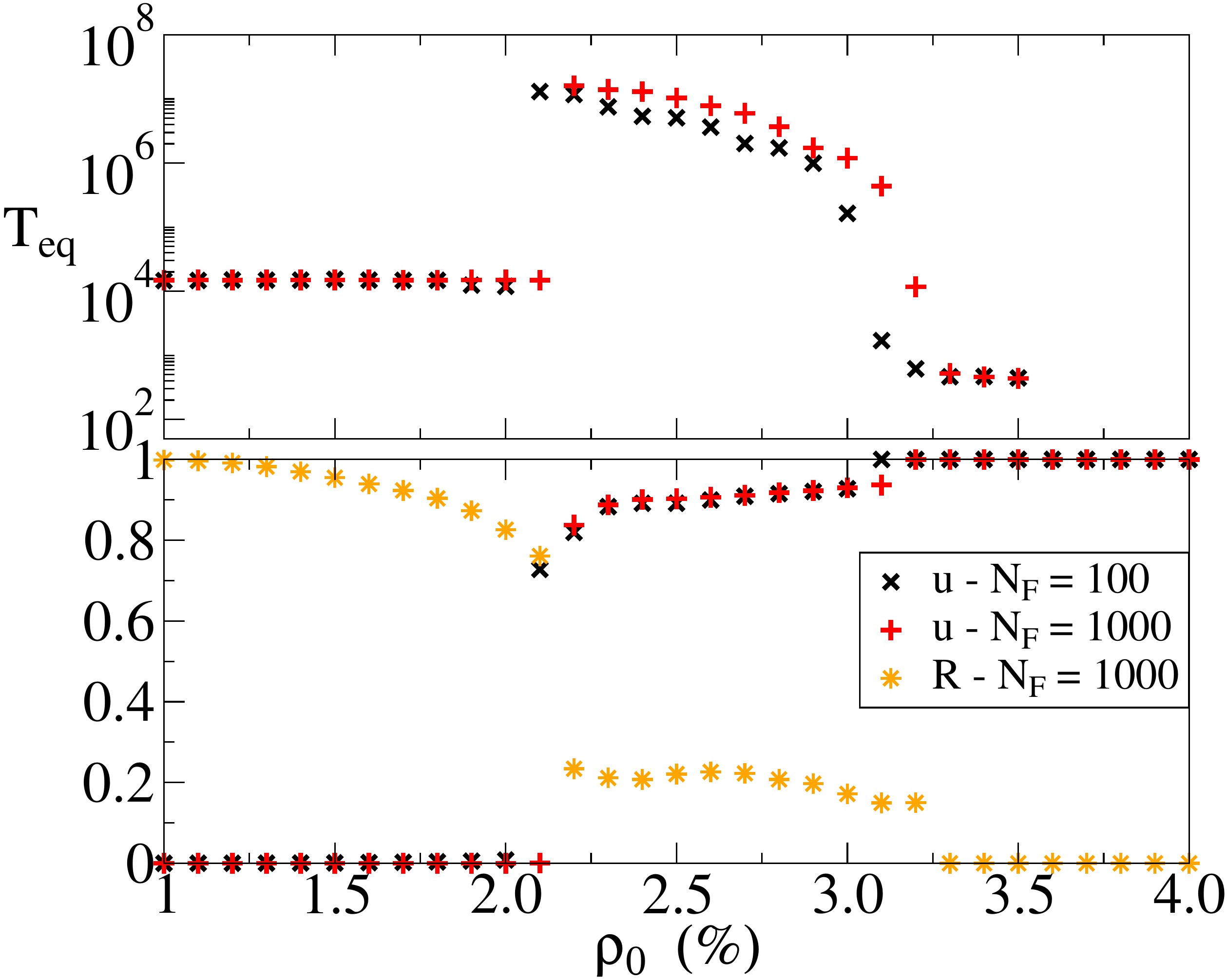}
\includegraphics[scale=0.25]{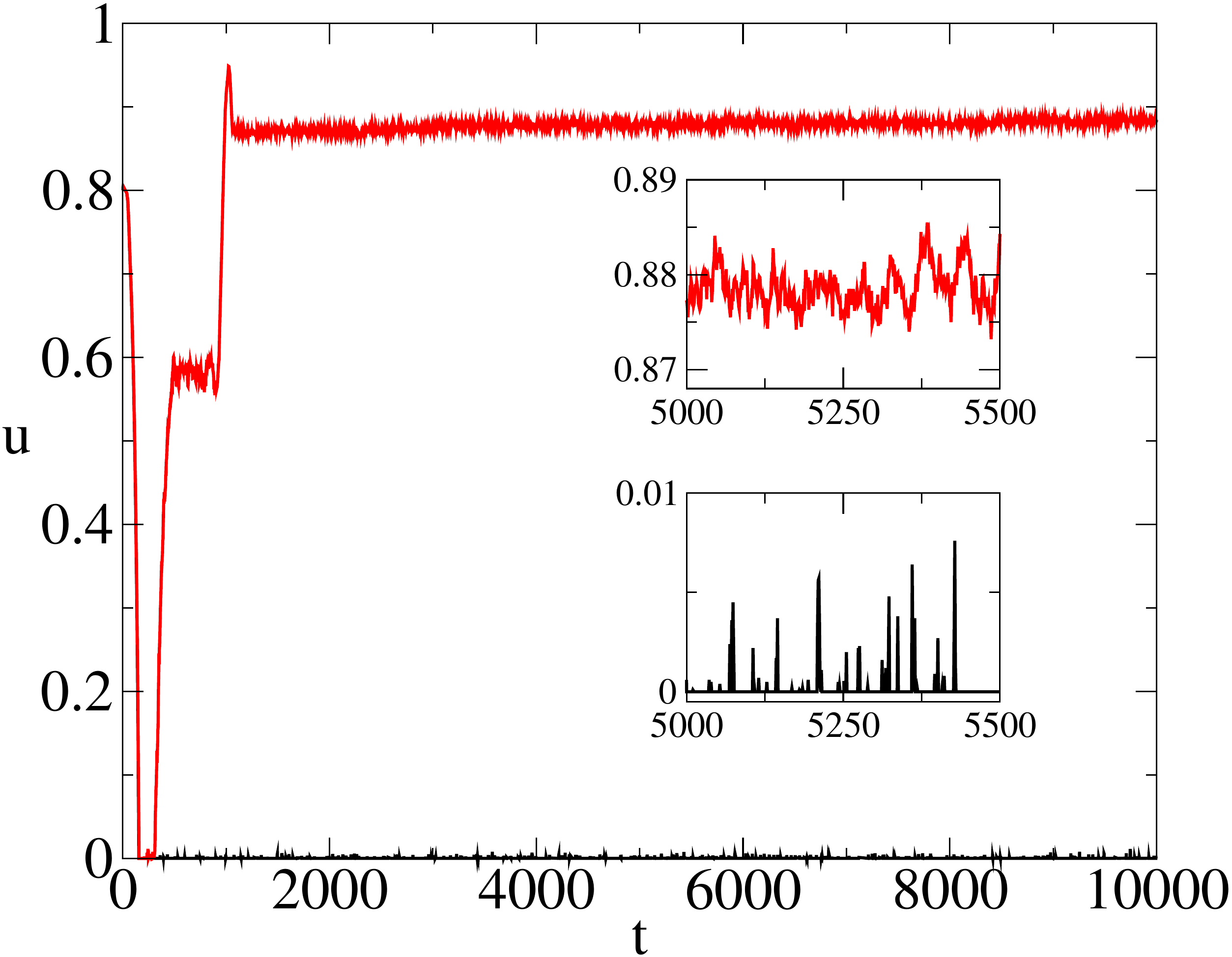}
\caption{\emph{Left:} Average unemployment rate $u$ as a function of the
interest rate $\rho_0$ for two system sizes (with $N_{\rm H} = 10 N_{\rm F}$ in 
both cases). 
The average is over $100\,000$ time steps discarding the first $50\,000$ time 
steps.
The phase transition at $\rho_{c}=2.1\%$ is of first order with $u$ jumping 
discontinuously from small values to $1$ (intermediate values obtained for
$\rho_0 \in [2.1\%,3.1\%]$ are only due to a much longer equilibration time 
near the critical point). 
Note that in the bottom graph we show averages for $T_{eq}=50\,000$ regardless 
of $\rho_0$ while an estimate
of the time needed to reach the steady state as a function of $\rho_0$ is 
plotted in logarithmic scale in the top graph.
{
In the bottom graph we also show the average value of the asymmetry measure $R$ (see main text)
for $N_{\rm F}=1000$. As one can see $R$ is decreasing with the interest rate up to
a point where the asymmetry is strong enough to drive the economy into the high unemployment
phase. 
}
\emph{Right:} Two trajectories of the unemployment rate with $N_{\rm F}=1000$ at
$\rho_0=1.9\%$ and $\rho_0=2.5\%$.
}
\label{fig:M1_PT}
\end{figure}

{
Anticipating the results of the next section, we have characterized, as 
a function of $\rho_0$, the asymmetry ratio $R$ that characterizes how firms react to the need to hire or to
fire. More precisely, we compute the ratio between the {\it target} number of job creations (resp. destructions) -- as a reaction to excess demand (or supply) in the
previous time step -- to the {\it realized} number of hires (fires) after financial constraints
are met. At each time step we can then extract the average target-to-realized ratios for firms in excess supply
and demand respectively. The ratio $R$ (averaged over time) between these two numbers is a proxy of the asymmetry 
with which firms react to the need to hire or fire. 
It is clear that increasing $\rho_0$ hobbles the capacity of
firms to hire when they need to, and therefore decreases the ratio $R$, as seen in Fig.~\ref{fig:M1_PT},
which suggests that the employment collapse is related to this ratio. The results of the next section
will fully confirm this scenario.
}

To sum up, our most salient finding is that Mark I (or Mark I+) has essentially
two stationary states, with a first order (discontinuous) transition line 
separating the two. If
the parameters are such that the system lies close to this critical line, then
any small modulation of these parameters -- such as the 
noise
term that appears in Eq. (\ref{rate}) -- will be {\it amplified} by the 
proximity
of the
transition, and lead to interesting boom/bust oscillations, of the kind
originally observed in Mark I~\cite{Mark1_Milan}, and perhaps of economic 
relevance. 
The question, 
of course, is how generic this
scenario is. We will now show, by studying much simplified versions of Mark
I, that this transition is generic, and can indeed be induced by the asymmetry
between hiring and firing. We will then progressively enrich our watered-down
model (call ``Mark 0'' below) and see how the qualitative picture that we
propose is affected by additional features.

\section{Hybrid ABM's: the minimal ``Mark 0'' model}

Moving away from the RA framework, Agent-Based Modeling bites the bullet and
attempts to represent in details  ~{\it all} the individual components of the
economy (as, for example, in~\cite{Eurace,Eurace2}). This might however be 
counter-productive, 
at least in the research stage we are still in: keeping too many details is not 
only computer-time
consuming, it may also hobble the understanding of the effects that ABMs attempt
to capture. It may well be that some sectors of the economy can be adequately
represented in terms of aggregate variables, while discreteness, heterogeneities
and interactions are crucial
in other sectors. In our attempt to simplify Mark I, we posit that the whole 
household
sector can be represented by aggregate variables: total wealth (again entirely in the form of savings) 
$S(t)$, total
income wage $W(t)$ and total consumption budget $C_B(t)$ {(which, as we will emphasize below, 
is in general larger than the {\it actual} consumption $C(t)$)}. We also remove the banking
sector and treat the loans in the simplest possible way -- see below. While the
interest rate is zero in the simplest version, the incentive to hire/fire 
provided by the interest rate, that was at play in Mark I, will be encoded in a
phenomenological way in the update rule for production. The firms, on the other
hand, are kept as individual entities (but the above simplifications will allow
us to simulate very large economies, with $N_{\rm F}=100\,000$ firms or more). 

\subsection{Set-up of the model}
\label{sec:Mark0_desc}

The minimal version of the Mark 0 model is defined as follows.
The salient features are:
\begin{itemize}
\item There are $N_{\rm F}$ firms in total and $\mu N_{\rm F}$ households, $\mu 
> 1$.\footnote{
Actually, households are treated as a unique aggregate variable, therefore $\mu$
is not a relevant parameter: one can see that its value is irrelevant and one 
can always
set $\mu=1$ for simplicity. Yet it is useful to think that the aggregate 
variables represents
in an effective way a certain number of individual households, hence we keep 
the parameter
$\mu$ explicit in the following.
} Each firm $i = 1\cdots N_{\rm F}$ pays a salary $W_i(t)$
{
and produces output $Y_i(t)$ by means of a one-to-one technology that uses only 
labor as an input.
}
Productivity is chosen to be {\it constant in time} and fixed to $\a=1$. We therefore
neglect any productivity shocks in our model; interestingly, crises (when they occur) will
be of endogeneous origin. With $\a=1$, $Y_i(t)$ is simply
equal to the number of employees of firm~$i$.  Hence, the employment rate
$\varepsilon(t)$ and unemployment rate $u(t)$
are:
\beq
\begin{split}
\varepsilon(t) &= \frac{1}{\mu N_{\rm F}} \sum_i Y_i(t) \ , \\
u(t) &= 1 - \varepsilon(t) \ .
\end{split}
\eeq
\item Households are described by their total accumulated savings
$S(t)$ (which at this stage are always non-negative) and by their total wage income 
$\sum_i W_i(t) Y_i(t)$.
At each time step, they set a total consumption budget\footnote{Of course, one 
could choose different
$c$'s for the fraction of savings and the fraction of wages devoted to 
spending, or any other non-linear spending schedule.}
\beq
\label{cons_budget}
C_B(t) = c \, [ S(t)+ \sum_i W_i(t) Y_i(t) ]
\eeq
which is distributed
among firms using an intensity of choice model~\cite{Anderson}. The demand of 
goods for firm $i$ is therefore:
\be\label{intensity} 
D_i(t) = \frac{C_B(t)}{p_i(t)} \frac{e^{- \beta p_i(t) / \overline p(t)}}{Z(t)} \ ,\qquad Z(t)=\sum_i 
e^{-\beta p_i(t) / \overline p(t)}
\ee
where $\beta$ is the price sensitivity parameter {determining an 
exponential dependence of households demand
in the price offered by the firm; $\beta = 0$ corresponds to
complete price insensitivity and $\beta \to \infty$ means that households select
only the firm with the lowest price.}\footnote{
{In this sense, as long as $\beta>0$ firms 
compete on prices. An averaged scatter plot of firms profits versus the price 
offered 
(not shown here) indeed displays a well-shaped concave profit function.}
} The normalization is such that $C_B(t) = \sum_i p_i(t) D_i(t)$, as it should be.
\item Firms are described by their price $p_i(t)$, their salary $W_i(t)$, and 
their production $Y_i(t)$.
\begin{itemize}
\item
For simplicity, we fix the salary $W_i(t) \equiv 1$ -- an extension that 
includes wage dynamics is discussed below.
\item
For the price, we keep the Mark I price update rule
(\ref{update}), with the average production-weighted price:
\beq\label{pbardef}
\overline p(t) = \frac{\sum_i p_i(t) Y_i(t)}{\sum_i Y_i(t)} \ .
\eeq
{Note that this price update rule only makes sense if firms 
anticipate that households are price sensitive, i.e. if $\beta > 0$, which we will 
assume in the following. Still, 
the dynamics of the model as defined remains perfectly well-behaved when $\beta=0$, even
if in this limit, the rational behaviour of firms would be to increase their 
price indefinitely and produce very little.}
\item
For production, we assume that firms are more careful with the way they deal
with their workforce than posited in Mark I. Independently of their price 
level, firms try to adjust
their production to the observed demand. 
When firms want to hire, they open positions on the job market; we assume that 
the total number of unemployed workers, 
which is $\mu N_{\rm F} u(t)$, is distributed among firms according to an 
intensity of choice of model which depends on both the wage 
offered by the firm\footnote{Since at this stage wages are equal 
among firms the distribution is uniform. Below we allow firms to update their 
wage. A higher wage will then translate in the availability 
of a larger share of unemployed workers in the hiring process.} 
and on the same parameter $\beta$ as it is for Eq.~\eqref{intensity}; 
therefore the maximum number of available workers to each firm is:
\be\label{wintensity} 
\mu \tilde{u}_i(t) = \frac{e^{\beta W_i(t) / \overline w(t)}}{\sum_i e^{\beta 
W_i(t) / \overline w(t)}} \mu N_{\rm F} u(t) \ .
\ee
where
\beq\label{wbardef}
\overline w(t) = \frac{\sum_i W_i(t) Y_i(t)}{\sum_i Y_i(t)} \ .
\eeq
\end{itemize}
In summary, we have
\beq\label{update0}
\begin{split}
    \text{If } Y_i(t) < D_i(t)  &\hskip10pt \Rightarrow \hskip10pt 
    \begin{cases}
&     Y_i(t+1)=Y_i(t)+ \min\{ \eta_+ ( D_i(t)-Y_i(t)), \mu \tilde{u}_i(t) \} \\
&  \text{If } p_i(t) < \overline{p}(t) \hskip10pt \Rightarrow \hskip10pt   
p_i(t+1) = p_i(t) (1 + \g_p \xi_i(t) ) \\
&  \text{If } p_i(t) \geq \overline{p}(t) \hskip10pt \Rightarrow \hskip10pt   
p_i(t+1) = p_i(t) \\
\end{cases}
\\
  \text{If }   Y_i(t) > D_i(t)  &\hskip10pt  \Rightarrow \hskip10pt
  \begin{cases}
&    Y_i(t+1)=\max\{ Y_i(t) - \eta_- [Y_i(t)-D_i(t)] , 0\} \\
&  \text{If } p_i(t) > \overline{p}(t) \hskip10pt \Rightarrow \hskip10pt   
p_i(t+1) = p_i(t) (1 - \g_p \xi_i(t) ) \\
&  \text{If } p_i(t) \leq \overline{p}(t) \hskip10pt \Rightarrow \hskip10pt   
p_i(t+1) = p_i(t)\\
    \end{cases}
\end{split}
\eeq
where $\eta_\pm \in [0,1]$ are what we denote as the hiring/firing propensity 
of the firms. Note that this rule ensures that there is
no overshoot in production; furthermore the $\max$ in the second rule is not 
necessary 
mathematically when $\eta_- \leq 1$, but we kept it for clarity.
{
Each row of Eq.~\eqref{update0} specifies an adjustment mechanism for output 
and the individual
price. According to this mechanism, there is an increase in output if excess demand is positive, 
and a decrease in output if excess demand is negative, i.e. if there is excess supply. The 
propensities to hire/fire $\eta_\pm$ can be seen as the sensitivity of the output change to excess 
demand/supply. These sensitivities are generally less than unity (i.e. the firm is not adjusting 
output one to one with excess demand/supply), because there are hiring (firing) costs of different
kinds (real costs, time-to-hire, administrative constraints, inertia, etc.).  Firing costs generate ``labour hoarding'', 
while hiring costs prevents the economy to adapt quickly to excess demand. 
Hiring and firing costs may not be identical. Therefore the sensitivity to excess demand (or hiring propensity $\eta_+$) may 
be different from the sensitivity to excess supply (or firing propensity 
$\eta_-$).
}
Note that because of the $ \min\{ D_i(t)-Y_i(t), \mu u(t) \}$ term in 
Eq.~\eqref{update0}, the total production of
the model is bounded by $\sum_i Y_i(t) = \mu N_{\rm F} \varepsilon(t) \leq \mu 
N_{\rm F}$, as it should be because $\varepsilon(t)=1$ corresponds
to full employment and in that case $\sum_i Y_i(t) = \mu N_{\rm F}$. However, in
the following we will sometimes (when stated) remove this
bound for a better mathematical tractability. This amounts to replace 
$ \min\{ D_i(t)-Y_i(t), \mu u(t) \} \to D_i(t)-Y_i(t) $ in Eq.~\eqref{update0} 
(it corresponds to choosing $\mu = \io$).
Removing the bound corresponds to a situation where labor resources can freely 
exceed the working population, such that the high employment phase translates 
into an exponential explosion of the economy output, which is of course 
unrealistic.
\item {\it Accounting of firms and households}. Each firm $i=1\cdots N_{\rm F}$ pays a 
total wage bill $W_i(t) Y_i(t)$ and has
{total sales}
$p_i(t) \min\{ Y_i(t), D_i(t) \}$. Moreover, if the profit of the firm $\PP_i(t) = 
p_i(t) \min\{ Y_i(t), D_i(t) \} - W_i(t) Y_i(t)$
is positive, the firm pays a dividend $\d \times \PP_i(t)$ to the 
households.\footnote{We have also considered the case where firms distribute 
a fraction $\delta$ of the profits {\it plus} the reserves. See below.}

{Note that if $D_i > Y_i$, the demand for goods of firm $i$ cannot be immediately satisfied, and we assume that
in this case households involuntarily save and delay their consumption until the next round (but still using Eq. (\ref{cons_budget}) with the 
correctly updated savings).  The {\it actual} consumption $C(t)$ 
(limited by production) is therefore given by:
\be
C(t):= \sum_{i=1}^{N_{\rm F}} p_i(t) \min\{ Y_i(t), D_i(t) \} \leq C_B(t) =  \sum_{i=1}^{N_{\rm F}} p_i(t) D_i(t).
\ee
}
In summary, the accounting equations for total accumulated savings $S(t)$ and
firms' {net deposits $\EE_i(t)$ -- possibly negative -- are the following (here $\th(x\geq 0)=1$ and $\th(x < 0)=0$  
is the Heaviside step function}):
\beq\begin{split}
& \PP_i(t) = p_i(t) \min\{ Y_i(t), D_i(t) \} - W_i(t) Y_i(t) \ , \\
& \EE_i(t+1) = \EE_i(t) + \PP_i(t)  - \d \PP_i(t) \th(\PP_i(t)) \ , \\
& S(t+1) = S(t) - \sum_i \PP_i(t) + \d \sum_i  \PP_i(t) \th(\PP_i(t)) \equiv S(t) + {\cal I}(t) - C(t)\ ,
\end{split}\eeq
{where ${\cal I}(t)$ is the total income of the households (wages plus dividends) and $C(t)$ is the money
actually spent by households, which is in general less than their consumption budget $C_B(t)$. This corresponds to \emph{unvoluntarly} 
households savings whenever production is below demand.}

Note that total money $S(t) + \sum_i \EE_i(t)$ is clearly conserved since 
$\Delta S(t) = -\Delta[\sum_i\EE_i(t)]$ for cash-flow consistency in a closed 
economy.
\item {\it Bank accounting}. 
{As mentioned above, we allow the firms' net deposits to become negative, which we 
interpret as the firm being in need of an immediate extra line of credit. Depending on the financial fragility of 
the firm (defined below), the bank may or may not agree to restructure the debt and provide this extra credit. 
If it does, our accounting procedure can be rephrased in the 
following way. In case of negative net deposits $\EE_i(t)<0$ the bank provides the firm with the extra liquidity needed 
to pay the wages. The equity of the firm is therefore equal to 
its deposits when positive, and is close to zero, but still equal to what 
it needs, when the net deposits is negative. However, when the firm becomes too indebted, the bank will not
provide the liquidity needed to pay the wages, leading to negative equity and bankruptcy. 
From an accounting perspective the matrix balance sheet of 
our model can be summarized as follows. We assume the bank's equity to be constant in time (let it be $0$
for simplicity). The bank holds at the beginning of the simulation a quantity
$M$ of currency (issued by the central bank, not modeled here) and therefore has assets equal to $M$
throughout the simulation.}

Households have deposits $S>0$ while firms have either deposits (if 
$\EE_i>0$) or liabilities (if $\EE_i<0$). Defining $\EE^+=\sum_i\max{(\EE_i,0)}$ and $\EE^-= -\sum_i\min{(\EE_i,0)}$, the 
balance sheet of the banking system is therefore:
\be
M + \EE^-(t) = S(t) + \EE^+(t) \equiv \mathcal{X}(t),
\ee
which means that the total amount of deposits $\mathcal{X}$ at any time is
equal to initial deposits plus deposits created by the banking system when 
issuing loans. In this setting reserves $M$ at the bank are kept unchanged when loans are 
issued (as outlined 
for example in~\cite{boe}) and deposits increase (decrease) only when loans are 
issued (repaid).
Correspondingly, the fundamental time-invariant macroeconomic accounting identity
\be
S(t) + \EE^+(t) - \EE^-(t) = S(t) + \sum_i\EE_i(t) = M
\ee
is obeyed and amounts to our money conserving equation. 

\item {\it Financial fragility and bankruptcy resolution.} We measure the indebtement level of a firm through the ratio of (negative) 
net deposits over payroll (equal, to a good approximation, to total sales):
\be
\Phi_i=-\EE_i/(W_i Y_i)
\ee
which we interpret as a measure of financial fragility. {(Implicitly, this assumes that
the ``real assets'' of the firms -- not modeled here -- are proportional to total payroll)}. If $\Phi_i(t) < \Theta $, 
i.e. when the level of debt is not too high compared to the size
of the company, the firm is allowed to continue
its activity. If on the other hand $\Phi_i(t) \geq \Theta$, the firm
defaults. 
 
When firms exceeds the bankrupcty threshold the default resolution we choose is 
the following. We first define the set ${\cal H}_i$ of financially 
``healthy'' firms that are potential buyers for the defaulted firm $i$. 
The condition for this is that $\EE_j(t) > \max(- \EE_i(t),\Theta Y_j(t)W_j(t))$, meaning that the firm 
$j$ has a strongly positive net deposits and can take on the debt of $i$ without going under water. 

\begin{itemize}
\item With probability $1-f$, $f \in [0,1]$ being a new parameter, a firm $j$ 
is chosen at random in ${\cal H}_i$; $j$ transfers to $i$ the needed money
to pay the debts, hence $\EE_j \rightarrow \EE_j + \EE_i$ (remember that 
$\EE_i$ is negative) and
$\EE_i \rightarrow 0$ after the transaction. Furthermore, we set $p_i = p_j$ 
and $W_i = W_j$, and the firm $i$ keeps
its employees and its current level of production. 
\item With probability $f$, or whenever ${\cal H}_i = \emptyset$, the firm $i$ is 
not bailed out, goes bankrupt and its production is set to zero.
In this case its debt $\EE_i(t) < 0$ is transferred to the households' accumulated savings,
$S(t) \to S(t) + \EE_i(t)$ in order to keep total money fixed\footnote{
One can interpret this by imagining the presence of a bank that collects the 
deposits of households and lend money to firms, 
at zero interest rate. If a firm goes bankrupt, the bank looses its loan, which 
means that its deposits (the households' savings) are reduced.}.
\end{itemize}
Hence, when $f$ is large, most of the bankruptcies load weighs on the 
households, reducing their savings, whereas when $f$ is small, bankruptcies 
tend to fragilise the firm sector. (The Mark I+ model discussed above corresponds to $f=1/2$.)
{It is important to stress that changing the details of the bankruptcy rules while maintaining 
proper money conservation does not modify the main qualitative message of our paper.
The important point here is that the default costs are transferred to households and firms (to 
ensure money
conservation) and have some repercussion on demand (through households accumulated savings)
or on firms fragility (through firms net deposits). This can create default avalanches and crises. 
}
\item {\it Firm revival.} A defaulted firm has a finite probability $\varphi$ per unit time to get
revived; when it does so its price is fixed to $p_i(t) = \overline p(t)$, its
workforce is the available workforce, $Y_i(t) = \mu u(t)$,
and its net deposits is the amount needed to pay the wage bill, $\EE_i(t) = 
W_i(t) Y_i(t)$.
This liquidity is provided by the households, therefore $S(t) \to S(t) - 
W_i(t) Y_i(t)$ when the firm is revived,
again to ensure total money conservation. Note that during this 
bankrupt/revival phase, the households' savings $S(t)$ might become
negative: if this happens, then we set $S(t)=0$ and the debt of households is 
spread over the firms with positive liquidity\footnote{
Again, this can be interpreted by imagining that the firms with positive 
liquidity deposit their cash in the bank. When the bank needs to provide
a loan to a revived firm, or loses money due to a bankrupt, it prefers to take 
this money from households' deposits, but if these are not available,
then it takes the money from firms' deposits.}, proportionally to their current value of $\EE_i$, again in order to ensure 
total money conservation and $S(t) \geq 0$.
\end{itemize}
The above description contains all the details of the definition of the model, 
however for full clarification a pseudo-code
of this minimal Mark 0 model is provided in Appendix~\ref{app:Mark0} (together 
with the extensions discussed in Sec.~\ref{section:extensions}). 
The total number of relevant parameter of Mark 0 is equal to $9$:
$c,\beta,\gamma_p,\eta_\pm,\delta,\Theta,\varphi,f$ plus the number of firms
$N_{\rm F}$. However, most of these parameters end up playing very little role 
in 
determining the {\it qualitative}, long-time aggregate behaviour of the model. 
Only two quantities play an important role, and turn out to be:
\begin{enumerate}
\item the ratio $R = \eta_+/\eta_-$
between $\eta_+$ and $\eta_-$, which is meant to capture any asymmetry in the
hiring/firing process. As noted above (see Fig. \ref{fig:M1_PT}), a rising interest rate endogeneously 
leads to such a hiring/firing asymmetry in Mark I and
Mark I+. But other sources of asymmetry can also be envisaged: for example, 
overreaction of the firms to bad news and
under-reaction to good news, leading to an over-prudent
hiring schedule. Capital inertia can also cause a delay in hiring, whereas
firing can be immediate.
\item the default threshold $\Th$, which controls the ratio between total debt 
and total circulating currency.
In our minimal setting with no banks, it plays the role of a money multiplier. 
Monetary policy within Mark 0 boils down to
setting of the maximum acceptable debt to payroll ratio $\Th$.
\end{enumerate}

The other parameters change the phase diagram of the model quantitatively but 
not qualitatively. In order of 
importance, the most notable ones are $f$ (the redistribution of debt over 
households or firms upon bankruptcies) and $\beta$ (the sensitivity to price) --
see below. 

\subsection{Numerical results \& Phase diagram}
\label{numerical_results0}

\begin{figure}
\centering
\includegraphics[scale=0.35]{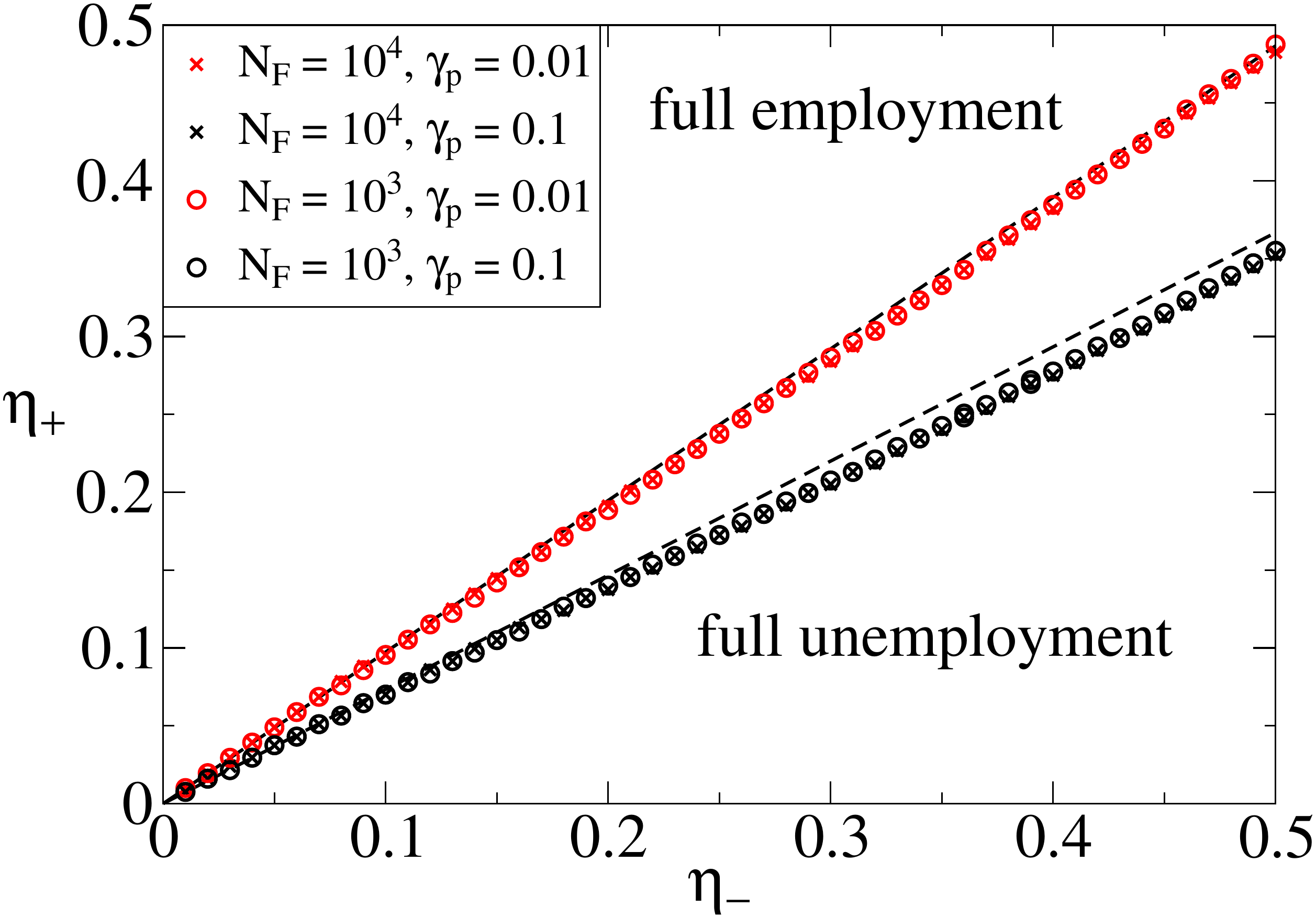}
\includegraphics[scale=0.35]{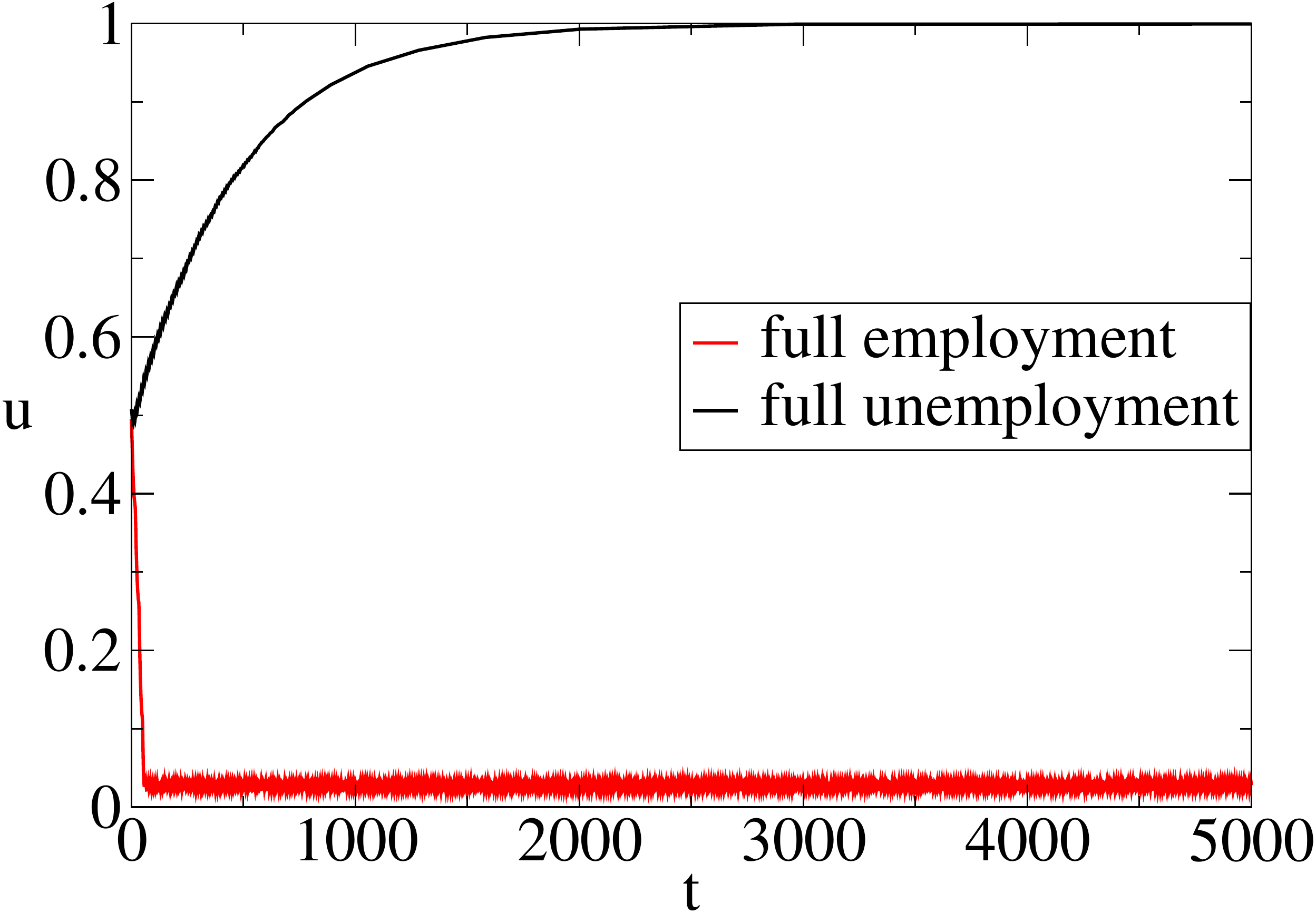}
\caption{
({\it Left})
Phase diagram of the basic Mark 0 model with $\beta=2$ and $\Theta=\infty$. 
There
are two distinct phases separated by a critical line which depends on the
parameters $\gamma_p$ and $\beta$. 
The dashed lines correspond to the analytical result in Eq.~\eqref{critic_line} 
which agrees well
with numerical simulations for small values of $\eta_+,\eta_-$ where the 
perturbative
method of Section III-C and Appendix~\ref{app:D1} is justified. For simplicity 
we show here only results with $\beta=2$
but Eq.~\eqref{critic_line} is in good agreement with numerical results up to 
$\beta\sim4$.
({\it Right})
Two typical trajectories of $u(t)$ in the two phases ($R=5/3$ in the full 
employment line and $R=3/5$ in the full unemployment line).
The other parameters are: $N_{\rm F}=10\,000$, $\gamma_p=0.1$, $\Theta=\infty$, 
$\varphi=0.1$, $\beta=2$, $c=0.5$. $\delta=0.02$ and $\f=1$. 
}
\label{fig:PDG0}
\end{figure}

When running numerical simulations of Mark 0, we find (Fig.~\ref{fig:PDG0}) 
that 
after a transient that can be surprisingly long\footnote{Think
of one time step as a quarter, which seems reasonable for the frequency of 
price and workforce updates. The equilibration time is then 
20 years or so, or even much longer 
as for the convergence to the `bad' state in Mark I, see Fig.~\ref{fig:M1_PT}. 
Albeit studying a very different ABM, 
similarly long time scales can be observed in the plots shown in \cite{Gordon}.
See also the discussion in the conclusion on this point}, the unemployment rate 
settles around a well
defined average value, with some fluctuations (except in some cases where 
endogenous crises appear, see below).

We find the same qualitative phase
diagram for all parameters
$\beta,\gamma_p,\varphi,\delta,f$. 
For a given set of 
parameters, there is a critical value $R_c$ of $R = \eta_+/\eta_-$ separating a
full unemployment phase for $R<R_c$ from a phase where some of the labour force 
is employed for $R>R_c$. Here $R_c \leq 1$ is a value that depends on
all other parameters. In Sec.~\ref{analytical} we explain how the phase transition at $R_c$ and 
other features of the model can be understood by means of approximate analytical
calculations.

The transition is particularly abrupt in the limit $\Theta\to\infty$ (no 
indebtment limit), in which the unemployment
rate jumps all the way from 0 to 1 at $R_c$, see Fig.~\ref{fig:PDG0}.
The figure shows that indeed only the ratio $R = \eta_+/\eta_-$ is relevant, 
the actual values of $\eta_\pm$ only change the time scale over which
the production fluctuates. Moreover, the phase diagram is almost independent of 
$N_{\rm F}$, which confirms that we are effectively in a limit
where the number of firms can be considered to be very large\footnote{Actually, 
in the specific case $\Theta=\infty$, even $N_{\rm F}=1$ provides a 
similar phase diagram, although the critical $R_c$ slightly depends on $N_{\rm 
F}$ when this number is small}.

\begin{figure}
\centering
\includegraphics[scale=0.35]{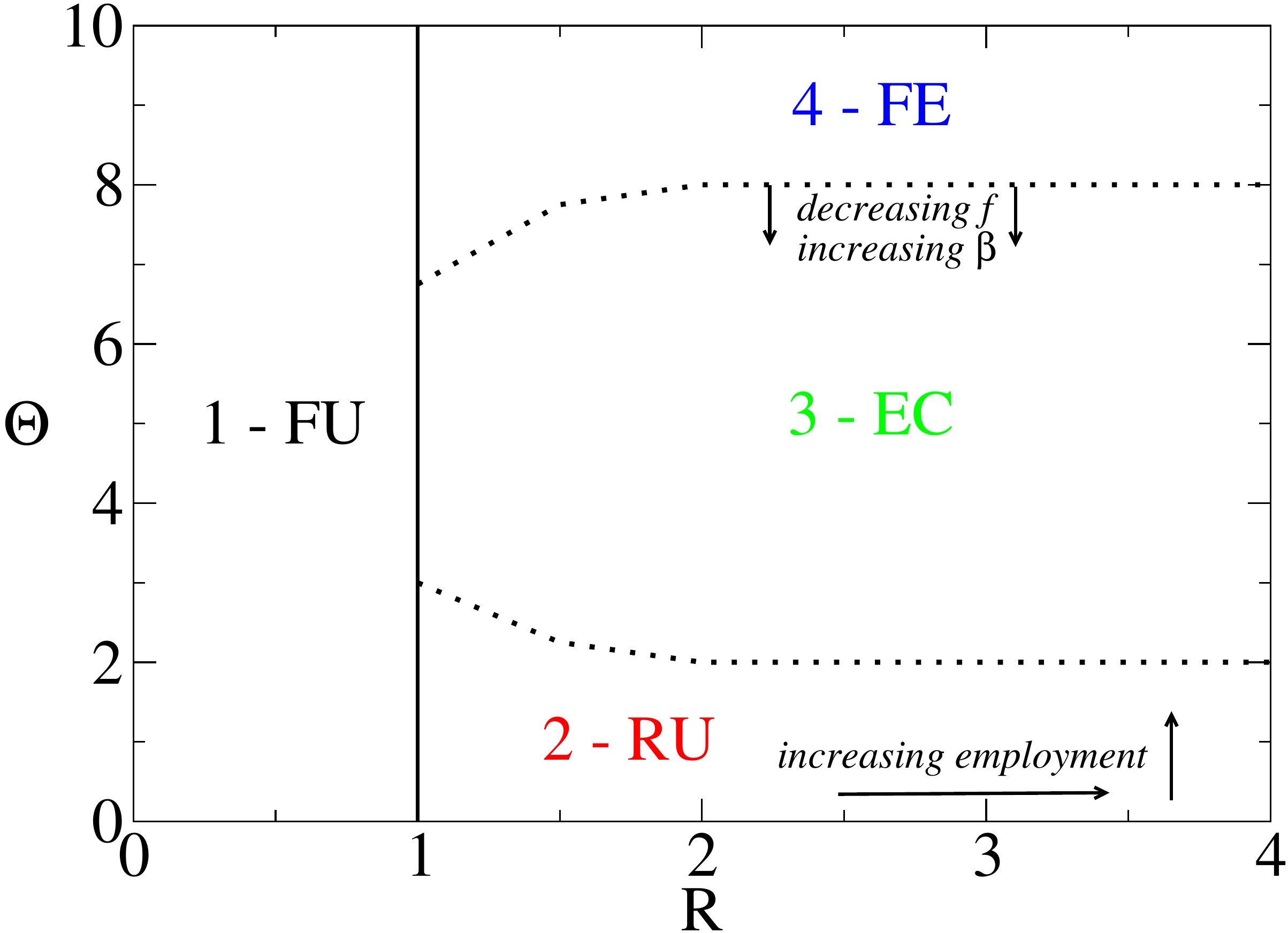}
\includegraphics[scale=0.35]{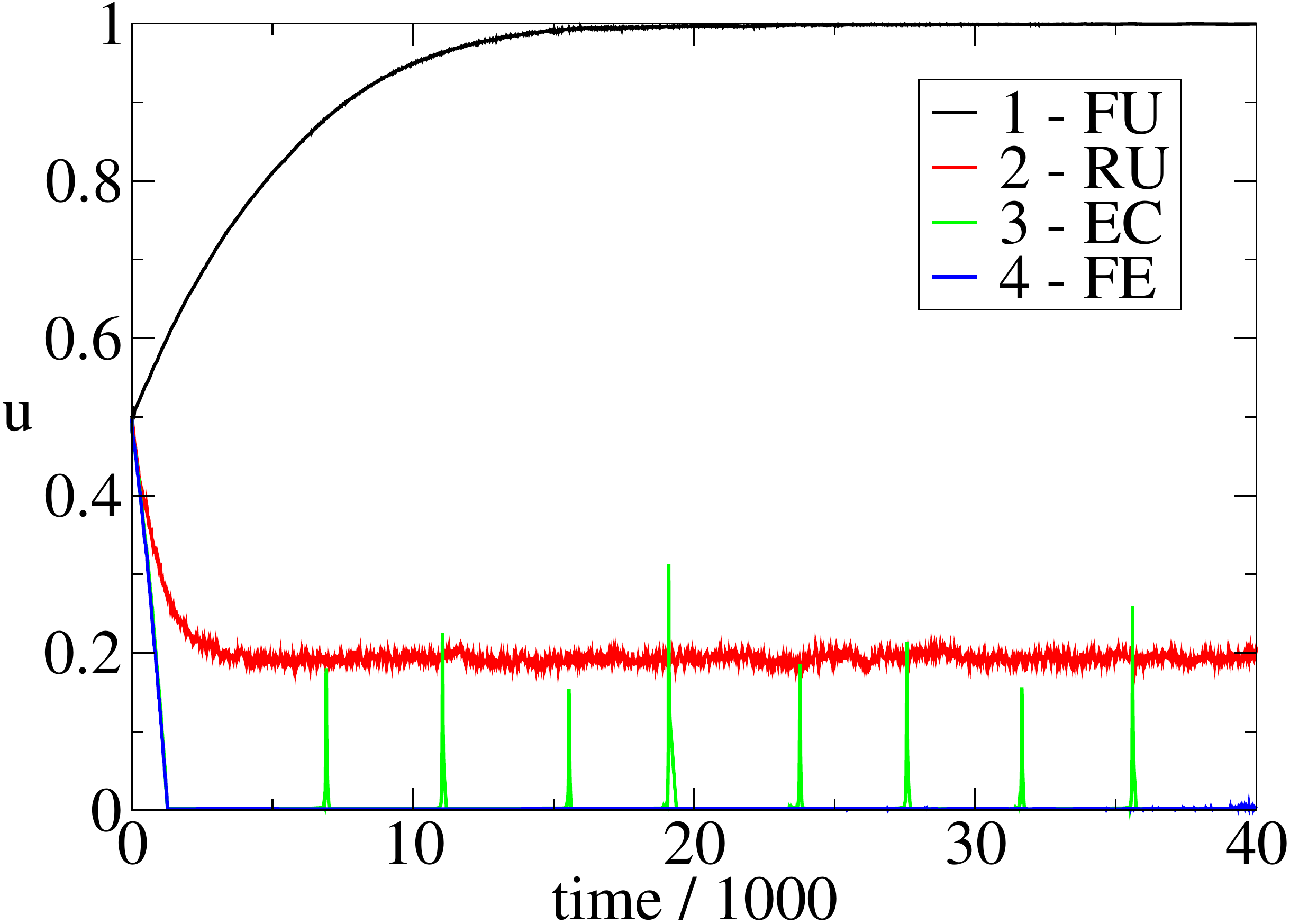}
\includegraphics[scale=0.35]{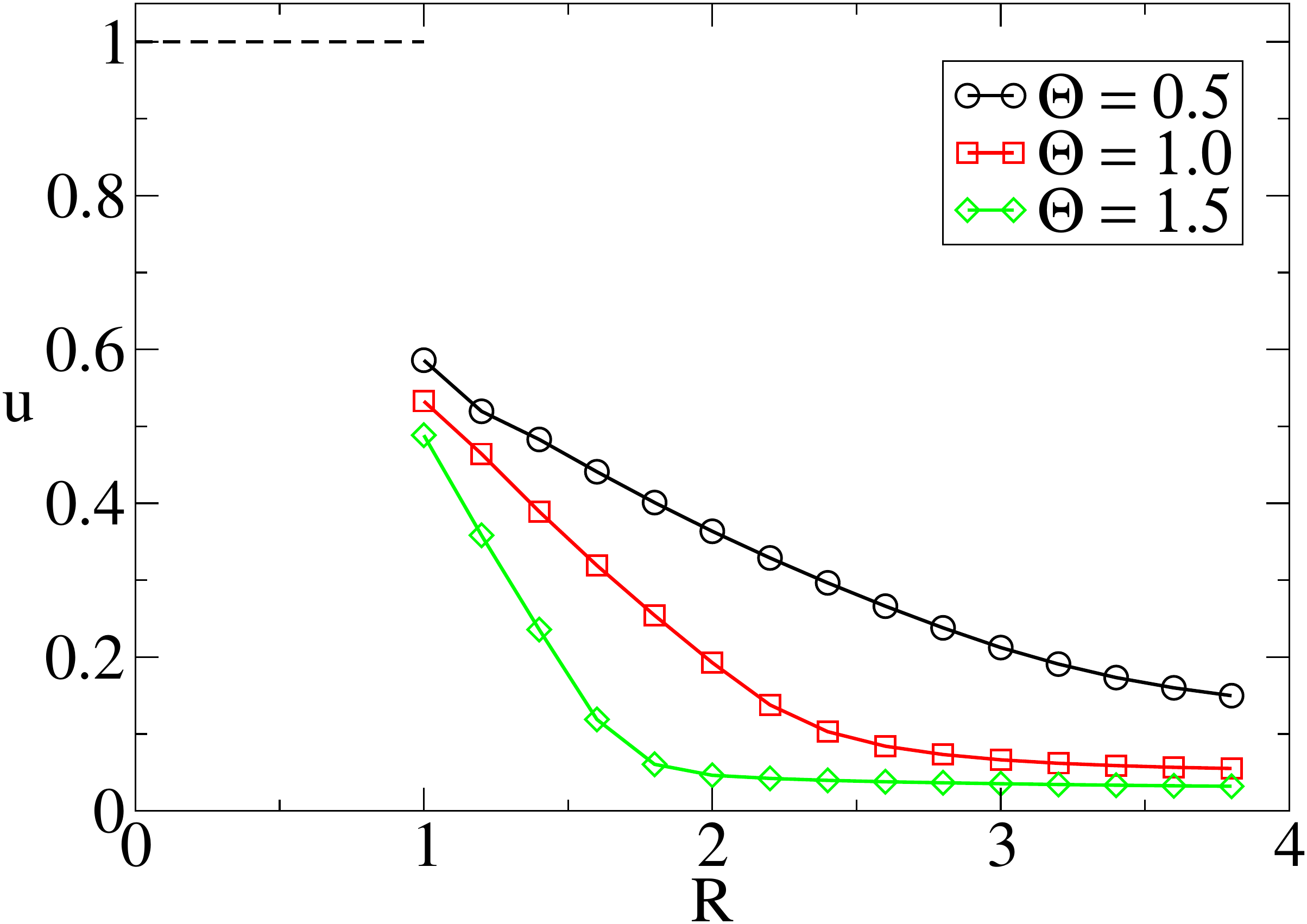}
\includegraphics[scale=0.35]{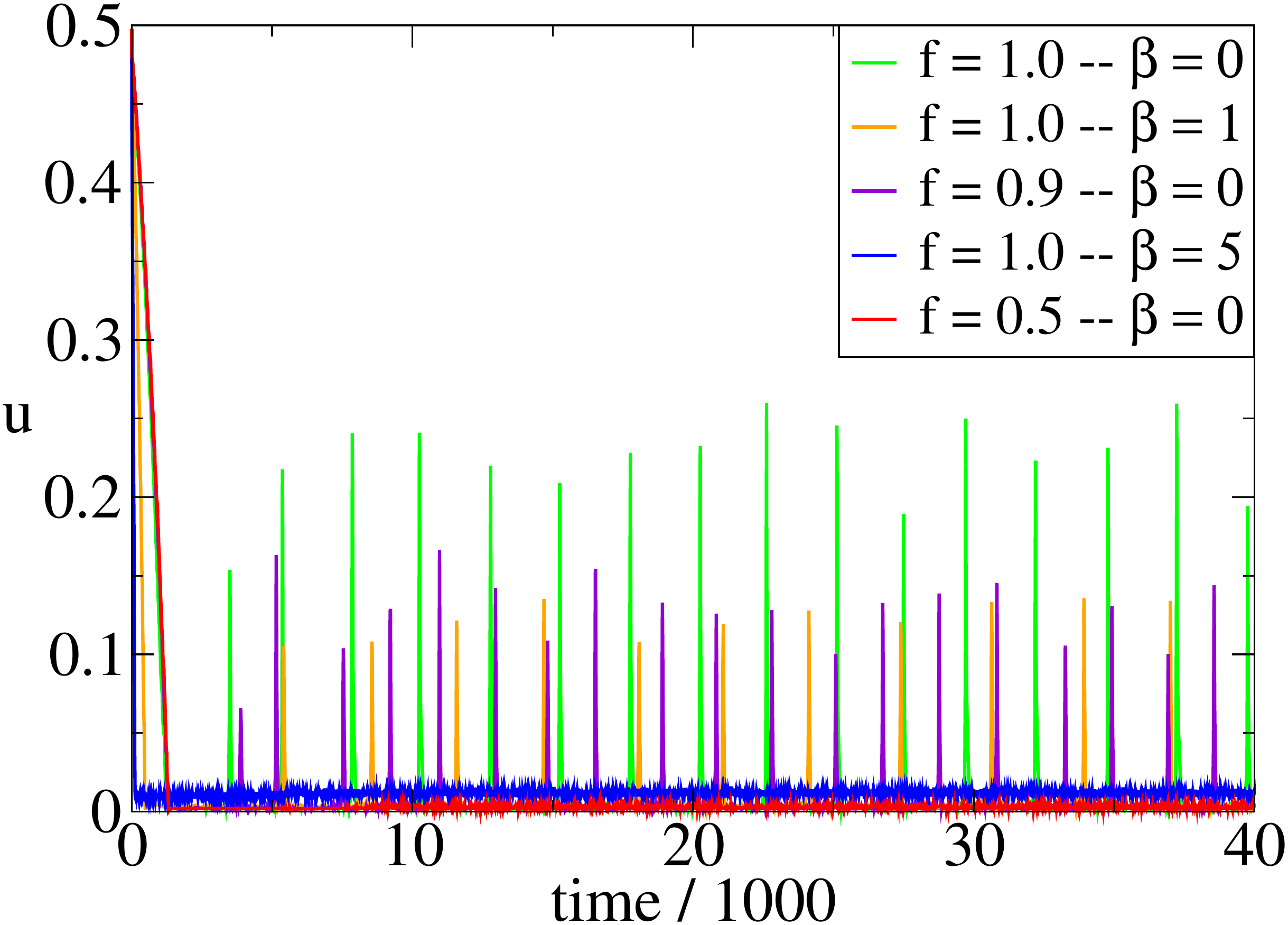}
\caption{
({\it Top Left}) 
Phase diagram of the basic Mark 0 model in the $R-\Theta$ plane, 
here with $N_{\rm F}=5000$, $c=0.5$, $\g_p=0.05$, $\d=0.02$, $\f=0.1$. Here we 
keep the 'extreme' cases $\beta=0$ and $f=1$ since the dependence of phase 
boundaries on $\beta$ and on $f$ (here reported schematically) is shown in 
detail in Fig.~\ref{thetac_b}. 
The other parameters are irrelevant.
There are four distinct phases separated by critical lines. 
({\it Top Right})
Typical time series of $u(t)$ for each of the phases.
({\it Bottom Left})
Stationary value of the unemployment rate as a function of $R$ for different 
values of $\Th$ in phase 2.
({\it Bottom Right})
Typical trajectories of $u(t)$ in the Endogenous Crisis phase (region 3) for 
different values of $f$ and $\b$, the other parameters are kept fixed. Note 
that 
increasing $\b$ or decreasing $f$ lead to small crisis amplitudes.
}
\label{fig:PDGsummary}
\end{figure}

Interestingly, for finite values of $\Theta$, the phase diagram is more complex and shown 
in Fig.~\ref{fig:PDGsummary}.
Besides the full unemployment phase (region 1, 
which always prevails when $R < R_c$, we find three other different regions for 
$R>R_c$, that actually survive many 
extensions of Mark 0 that we have considered (see section 
\ref{section:extensions}):
\begin{itemize}
\item
At very large $\Theta$ (region 4, ``FE''), the full employment phase persists, 
although a small value of the unemployment appears in a narrow region around 
$R \approx R_c$.
The width of this region of small unemployment vanishes as $\Theta$ increases. 
\item
At very low $\Theta$ (region 2, ``RU''), one finds persistent ``Residual 
Unemployment'' in a large region of $R > R_c$. The unemployment rate decreases 
continuously with $R$ and $\Th$ but reaches values as large as $0.5$ close to 
$R=R_c$ (see Fig.~\ref{fig:PDGsummary}, Bottom Left). 
\item
A very interesting {\it endogenous crises} phase appears for intermediate values 
of $\Theta$ (region 3, ``EC''), where the unemployment rate is most of the time 
very close to zero, but 
endogenous crises occur, which manifest themselves as sharp spikes of the 
unemployment that can reach quite large values.
These spikes appear almost periodically, and their frequency and amplitude 
depend on some of the other parameters of the model, in particular $f$ and 
$\b$, see Fig.~\ref{fig:PDGsummary}, 
Bottom Right.
\end{itemize}
The phase diagram in the plane $R-\Theta$ is presented in 
Fig.~\ref{fig:PDGsummary}, 
together with typical time series of the unemployment rate, for each of the 
four phases.

\begin{figure}
\centering
\includegraphics[scale=0.34]{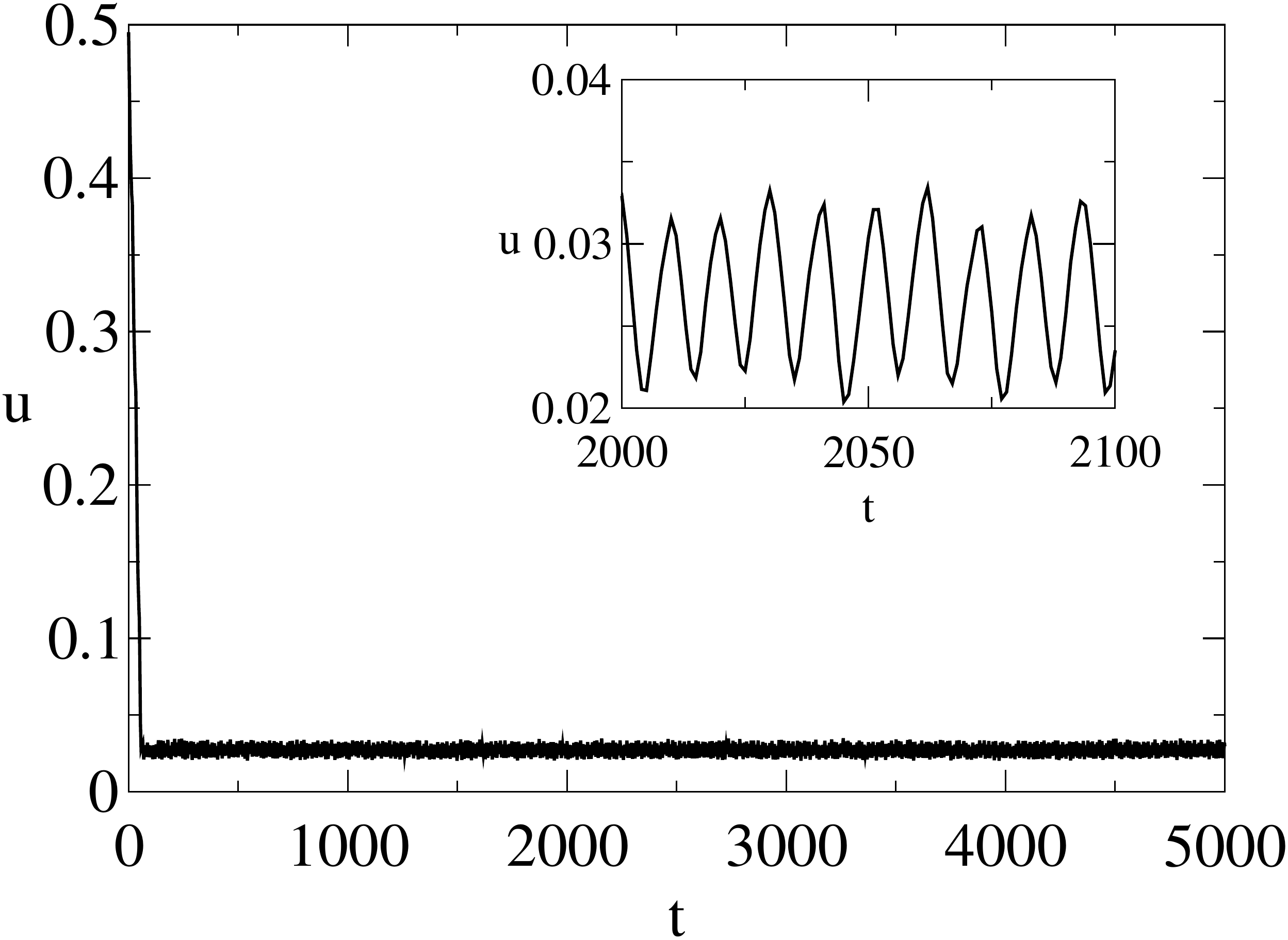}
\includegraphics[scale=0.34]{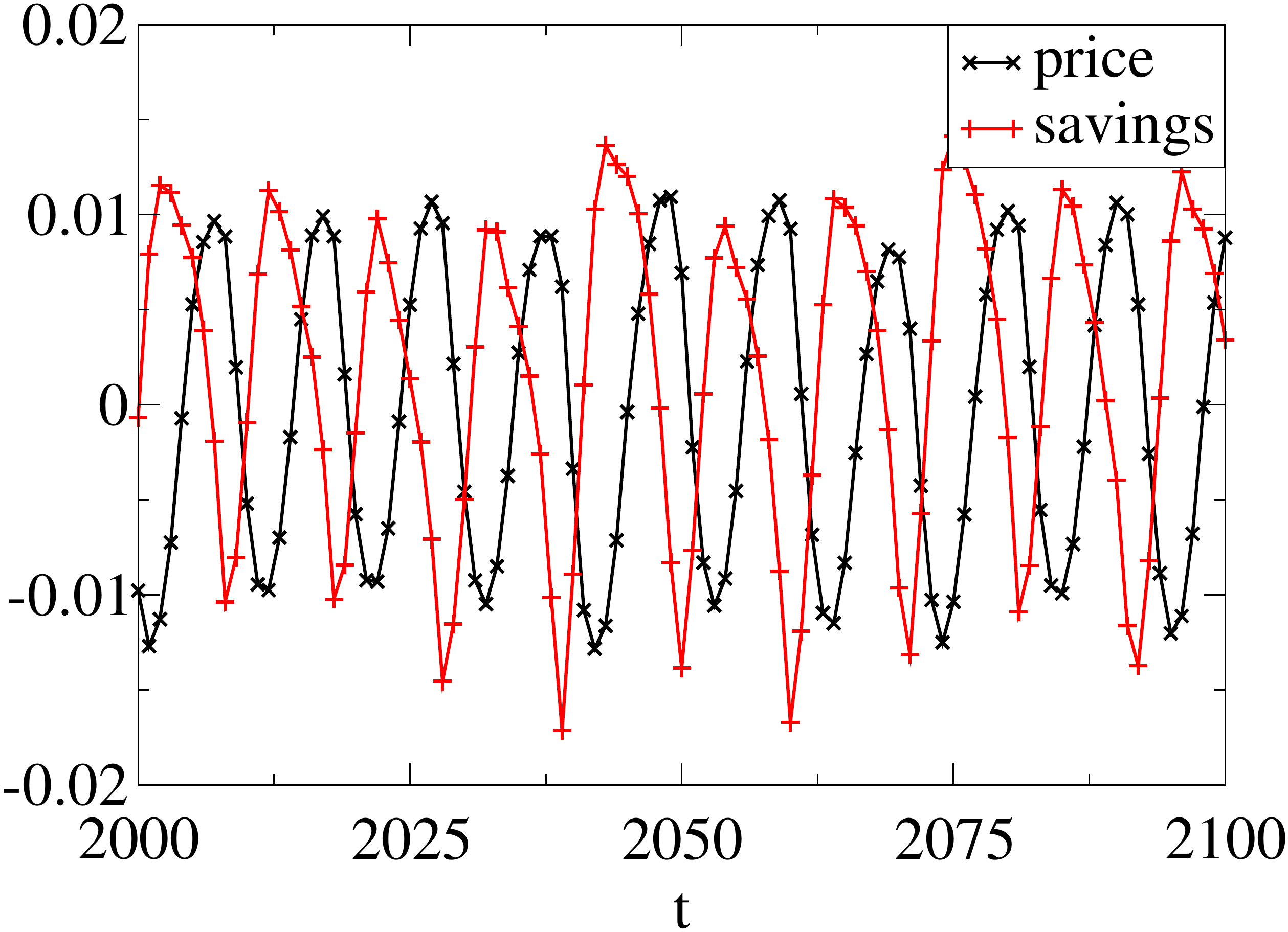}
\caption{
\emph{Left}: Typical time evolution of the unemployment $u$, starting from an 
initial condition $u = 0.5$, for the basic Mark 0 model in the Full 
Employment phase (region 4). The trajectory leading to a ``good'' state of the 
economy is obtained for $\eta_+=0.5$ and $\eta_-=0.3$. 
Note the clear endogenous ``business cycles'' that appear in that case. These 
runs are performed with $N_{\rm F}=10\,000$ firms, $c=0.5$, 
$\beta=2$, $\gamma_p=0.1$, $\varphi=0.1$, $\Theta=5$, $\delta=0.02$.\\
\emph{Right}: Oscillations for the average price and the average accumulated savings per 
household (each shifted by their average values for clarity) for the same run 
as in the 
{\it Left} figure. When prices are low, savings increase, while when prices are 
high, savings decrease. See Sec.~\ref{analytical} for a more detailed 
description.
}
\label{fig:2trajectories}
\end{figure}

We also show in Fig.~\ref{fig:2trajectories} a trajectory of $u(t)$ in the good 
phase of the economy (FE, region 4) and zoom in on 
the small fluctuations of $u(t)$ around its average value.
These fluctuations reveal a clear periodic pattern in the low unemployment 
phase; recall that we had already observed these
oscillations within Mark I+. Oscillating patterns (perhaps related to the
so-called business cycle) often appear in simplified first order differential 
models of the macro-economy; one of
the best known examples is provided by the Goodwin model \cite{Goodwin, 
Flaschel}, which is akin to a
predator-prey model where these oscillations are well known. But these 
oscillations do also
show up in other ABMs, see \cite{Dosi, Gordon}. Note that these 
fluctuations/oscillations around equilibrium {\it do  not
regress} when the number of firms get larger. We have simulated the model with 
$N_{\rm F}=10\,000$ firms and $N_{\rm F}=1\,000\,000$ firms with
nearly identical amplitude and frequencies for these fluctuations. We will 
offer some insight on the origin of these oscillations in
Sec.~\ref{analytical}. {Note that we do not expect such oscillations 
to remain so regular in the real economy, in particular because exogeneous 
shocks are absent from our model and because we assume that all firms are characterised by the
very same adaptation time scale $\gamma_p$.}

\subsection{Qualitative interpretation. Position of the phase boundaries}

An important quantity that characterizes the behavior of the model in the 
``good'' phase of the economy (i.e. for $R>R_c$) is the global leverage 
(debt-to-equity) 
ratio $k$:
\be
k(t) = \frac{\EE^-(t)}{S(t)+\EE^+(t)} = 1 - \frac{M}{S(t)+\EE^+(t)}
\ee
where, due to money conservation, $k \leq 1$. The good state of the economy is 
characterized by a large average value of $k$ reflecting the natural tendency of 
the economy towards indebtment, the level of which being controlled in Mark 0 
by the parameter $\Theta$ (the average value of $k$ increases with $\Theta$).\footnote{In 
a further version of the model, a central bank will be in charge of controlling 
$k$ through a proper monetary policy.}
Interestingly, in regions 2-RU and 4-FE $k(t)$ reaches a stationary state, 
whereas in region 3-EC its dynamics is characterized by an intermittent 
behavior 
corresponding to the appearance of endogenous crises during which indebtment is 
released through bankruptcies. 

\subsubsection{The EC phase}

This phenomenology can be qualitatively explained by the dynamics of the 
distribution of firms fragilities $\Phi_i$. For $R>R_c$, firms overemploy and 
make on average negative profits, which means that the $\Phi_i$'s are on 
average drifting towards the bankruptcy threshold $\Theta$.
\begin{itemize}
\item When $\Theta$ is small enough (i.e. in region 2-RU) the drift is 
continuously compensated by the reinitialization of bankrupted 
firms and the fragility distribution reaches a stationary state. 
\item For intermediate values of $\Theta$, however, (i.e. in region 3-EC) the 
number of bankruptcies 
per unit of time becomes intermittent. Firms fragilities now collectively drift 
towards the bankruptcy threshold; as soon as firms with higher fragilities 
reach $\Theta$, 
bankruptcies start to occur. Since for $f$ large enough, bankruptcies are 
mostly financed by households, demand starts falling which has the effect of 
increasing further the negative drift.
This feedback mechanism gives rise to an avalanche of bankruptcies after which 
most of the firms are reinitialized with positive liquidities. This mechanism 
has the effect of
{\it synchronizing} the fragilities of the firms, therefore leading to cyclical 
waves of bankruptcies, corresponding to the unemployment spikes showed in 
Fig.~\ref{fig:PDGsummary}.
The distribution of firms fragilities does not reach a stationary state in this 
case.
\item When $\Theta\gg1$ (i.e. region 4-FE), households are wealthy enough to 
absorb the bankruptcy cost without pushing the demand of goods below the 
maximum level of production reached 
by the economy.\footnote{The amount
of money circulating in the economy increases with $\Theta$ and for $R>R_c$ it 
is largely channelled to households savings since firms have on average negative
liquidities.} Hence, the economy settles down to a full employment phase with a 
constant (small) rate of bankruptcies.
\end{itemize}
The above interpretation is supported by a simple one-dimensional random walk 
model for the firms assets, with a drift that slef-consistently depends on the number of 
firms that fail. This highly simplified model accurately reproduces the above 
phenomenology, and is amenable to a full analytical solution, which will 
be published separately. 

 {The existence of the EC phase is a genuinely surprising outcome of the model, which was not
 put by hand from the outset (see our discussion in the first lines of the Introduction section above).
 Crises occur there not as a result of sweeping parameters through a phase transition (as 
 is the case, we argued, of Mark I with a time dependent interest rate) but purely  
 as a result of the own dynamics of the system.}

\subsubsection{The role of $\beta$}
\label{role_of_beta}

The dependence of the phase boundaries on the different parameters is in 
general quite intuitive. 
The dependence of aggregate variables on $\beta$, for example, is interesting: 
everything else being kept equal, we find that increasing $\beta$ (i.e. 
increasing the price selectivity of buyers), 
increases the level of unemployment (see inset of 
Fig.~\ref{fig:phaseboundaries}) and the amplitude of the fluctuations around 
the average value
(a similar effect was noted in \cite{Eurace2}). Increasing $\gamma_p$ increases 
the dispersion of prices around the average value and is thus similar to 
increasing $\beta$.

Increasing $\beta$ has also, {within the present setting of Mark 0}, some counter-intuitive effects: it increases 
both the average price compared to wages and the profits of firms, hence stabilizing the FE phase and shifting its 
boundary with EC to lower values of $\Theta$ and $R_c$ (see 
Fig.~\ref{fig:phaseboundaries}
for the amplitude of region 3-EC as a function of $\beta$.
This effect can be understood by considering the demand-production gap $G_i = D_i-Y_i$ 
as a function of the price difference $\delta p_i=p_i-\bar{p}$.
For a fixed value of $G_i$ the rules for price and production updates 
are independent of $\beta$; however, the response of the demand to a price 
change is 
stronger for higher values of $\beta$: for small values of $\delta p_i$ one finds approximately 
$G_i \approx -C(\beta)\delta p_i$, where $C(\beta)>0$ is increasing with 
$\beta$. As a consequence, the absolute value of 
the gap, $|G_i|$ for a given value of $|\delta p_i|$ is on average increasing 
with $\beta$. Therefore, the total amount of unsold goods $\sum_i \max{[Y_i-D_i,0]}$ and 
households accumulated savings $S$ also increase with 
$\beta$: {households involuntarily save more when they are more selective on prices.
An increased amount of savings is in turn responsible for the average price increase 
while the households' higher wealth expands the FE region by shifting its boundary to lower 
values of $\Theta$. 
Note, however, the effect of $\beta$ on the average price is numerically very small and depends sensitively on the precise consumption rule.
For example, if we insist that households fully spend their consumption budget $C_B(t)$ by looking for 
available products at a higher price (as in Mark I), then the above effect disappears (see Section V).
}

\subsubsection{{The role of $f$}}

\begin{figure}
\label{thetac_b}
\centering
\includegraphics[scale=0.3]{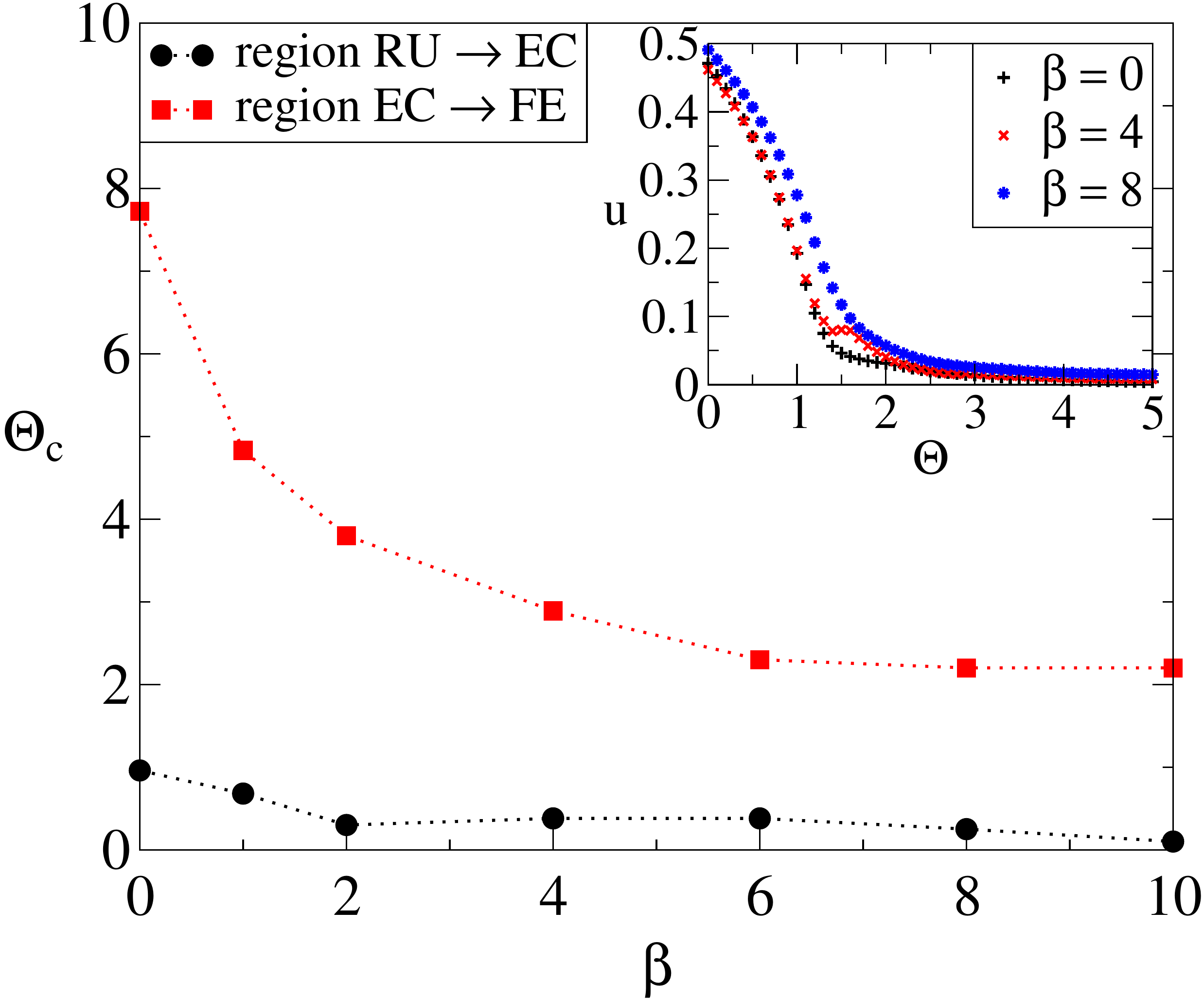}
\includegraphics[scale=0.3]{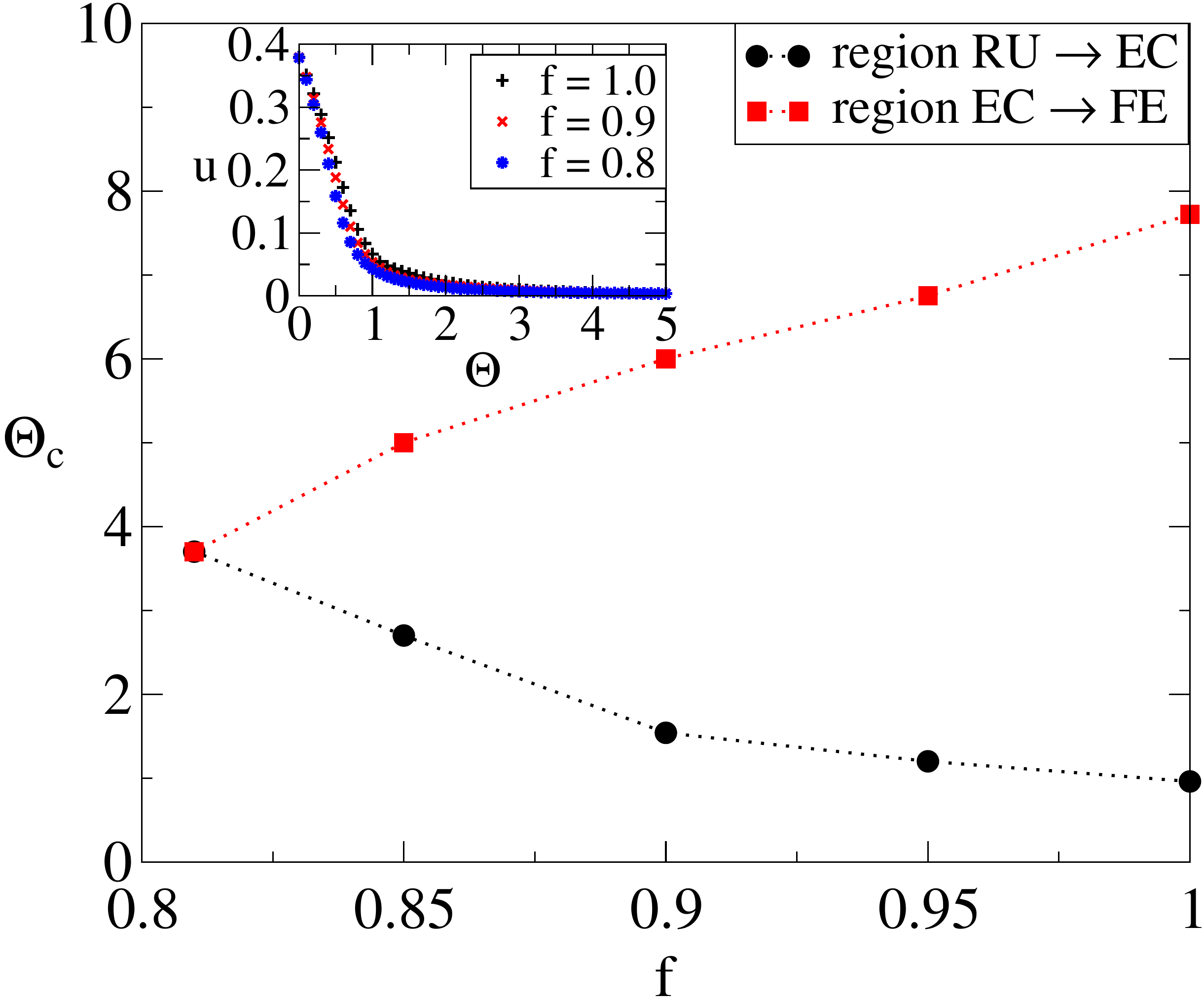}
\caption{
Phase boundary location $\Theta_c$ as a function of the price sensitivity 
$\beta$ (left) and of the debt share parameter $f$ (right). 
Circles correspond to the location of the boundary between regions 2-RU and 
3-EC while squares correspond to the boundary between regions 3-EC and 4-FE. 
In the inset of both figures we plot the average unemployment rate as a 
function of $\Theta$ for different values of $\beta$ and $f$. 
For $f\leq0.81$ region 4-EC disappears and the average value of the 
unemployment continuously goes to $0$ without the appearance of endogenous 
crises. 
Increasing $\beta$ shrinks the amplitude of region 3-EC which however remains 
finite for $\beta\gg1$. 
As a criterion for being in region 3-EC we require that the amplitude of the 
crises $\max{(u)} - \min{(u)}$ stays above $5\%$. 
Parameters are: $N=5000$, $R=3$ ($\eta_m=0.1$), $\delta=0.02$, $\f=0.1$, 
$\gamma_p=0.05$.
}
\label{fig:phaseboundaries}
\end{figure}

{A potentially more relevant discussion concerns the effect of the bankruptcies 
on the financial health of the firm sector.} Decreasing $f$ (i.e. the financial load taken up by households when 
bankruptcies occur) also stabilize, as expected, the full employment phase. Such 
a stabilization can also be achieved by distributing a fraction $\delta^+$ of the profit plus the 
total positive liquidity of the firms (instead of a fraction $\d$ of the profits only), 
which has the obvious effect of supporting the demand. In fact, we
find that as soon as $f \leq 0.81$ or $\delta^+ \geq 0.25$, the Endogenous Crisis 
phase disappears and is replaced by a continuous 
cross-over between the Residual Unemployment phase and the Full Employment 
phase (see Fig.~\ref{fig:phaseboundaries}).

\subsection{Intermediate conclusion}

The main message of the present section is that in spite of many
simplifications, and across a broad range of parameters, the phase transition
observed in Mark I as a function of the baseline interest rate is present in 
Mark 0 as well.
We find that these macroeconomic ABMs generically display two very
different phases -- high demand/low unemployment vs. low demand/high
unemployment, with a boundary between the two that is essentially controlled by
the asymmetry between the hiring and firing propensity of the firms (compare Figs. \ref{fig:M1_PT} \& \ref{fig:PDG0}). 

Moreover, in the Mark 0 model there is an additional splitting of the low 
unemployment phase
in several regions characterized by a different dynamical behaviour of the 
unemployment rate, depending on 
the level of debt firms can accumulate before being forced into bankruptcy. We 
find in particular
an intermediate debt region where endogenous crises appear, characterized by 
acute unemployment spikes. This Endogenous Crisis
phase disappears when households are spared from the financial load of 
bankruptcies and/or when capital does not accumulate within firms, but
is transferred to households through dividends. {Clearly, these different phases 
will coexist if the model parameters themselves evolve with time, which one should expect in 
a more realistic version of the model.}

{
We now turn to the first extension of Mark 0 allowing for a wage 
dynamics that lead to long term inflation or deflation,
absent in the above version of the model. Finally, several other extensions of 
the model will be briefly discussed and presented 
in a separate publication.
}

\section{Extension of Mark 0: Wage Update}
\label{section:extensions}

As emphasized above, the Mark I+ and Mark 0 models investigated up to
now both reveal a generic phase transition between a ``good'' and a ``bad'' 
state of the
economy. However, many features are clearly 
missing to make these models convincing -- setting up a full-blown, realistic 
macroeconomic
Agent-Based Model is of course a long and thorny endeavour which is precisely
what we want to avoid at this stage, focusing instead on simple mechanisms.
{
Still, it is interesting to progressively enrich these simplified models not 
only to test for
robustness of our phase diagram but also to investigate new effects that are
of economic significance. We consider here wage dynamics, which is {obviously an 
important ingredient in reality}. Wage dynamics leads to some relevant effects, in 
particular the appearance of inflation. 
Wage dynamics is indeed an item missing from both the basic Mark 0 and 
Mark I models considered above,
which assume fixed wages across time and across firms. Clearly, the
ability to modulate the wages is complementary to deciding whether to hire or 
to fire, and should play a
central role in the trajectory of the economy as well as in determining 
inflation rates. 
}

Introducing wages in Mark 0 again involves a number of arbitrary assumptions and
choices. Here, we follow (in spirit) the choices made in Mark I for price and
production update, and  propose that at each time step firm $i$  updates
its wage as:
\beq
\begin{split}
\label{update0++}
W^T_i(t+1)=W_i(t)[1+\gamma_w \varepsilon \xi^\prime_i(t)]
\quad\mbox{if}\quad
\begin{cases}
Y_i(t) &< D_i(t)\\
\PP_i(t) &> 0 
\end{cases}
 \\
W_i(t+1)=W_i(t)[1-\gamma_w u \xi^\prime_i(t)]
\quad\mbox{if}\quad
\begin{cases}
Y_i(t) &> D_i(t)\\
\PP_i(t) &< 0 
\end{cases}
\end{split}
\eeq
where $u=1-\varepsilon$ is the unemployment rate and $\gamma_w$ a certain 
parameter;
$\PP_i(t)=\min(D_i(t),Y_i(t))p_i(t) - W_i(t)Y_i(t)$ is the profit of the firm at
time $t$ and $\xi^\prime_i(t)$ an independent $U[0,1]$ random variable.
If $W^T_i(t+1)$ is such that the profit of firm $i$ at time $t$ with this
amount of wages would have been negative, $W_i(t+1)$ is chosen to be exactly at
the equilibrium point where $\PP_i=0$, hence $W_i(t+1) =  
\min(D_i(t),Y_i(t))p_i(t) / Y_i(t)$;
otherwise $W_i(t+1) = W^T_i(t+1)$. 

The above rules are intuitive: if a firm makes a profit and it has a large 
demand for
its good, it will increase the pay of its workers. 
{
The pay rise is expected to be larger if unemployment is low (i.e. if 
$\varepsilon$ is large) because pressure on
salaries is high. Conversely, if the firm makes a loss and has a low demand for 
its good, it will reduce the wages. 
This reduction is larger when unemployment is high because pressure on salaries 
is low. In all other
cases, wages are not updated.
}

When a firm is revived from bankruptcy (with probability $\varphi$ per unit
time), its wage level is set to the production weighted average wage of all 
firms in
activity.

The parameters $\gamma_{p,w}$ allow us to simulate different price/wage update 
timescales. In the following 
we set $\gamma_p=0.05$ and $\gamma_w= z \gamma_p$ with $z \in [0,1]$. 
The case $z=0$ clearly corresponds to removing completely the wage update rule, 
such that the basic version of 
Mark 0 is recovered. The extended version of Mark 0 that we consider below is therefore 
characterized by {\it single additional parameter} 
$\gamma_w= z \gamma_p$, describing the frequency of wage updates.

\subsection{Results: variable wages and the appearance of inflation}
\label{wages}

In our \emph{money conserving} toy economies\footnote{Recall that the physical 
money is conserved in the model, but virtual money creation is still possible
through firms' indebtement.}, 
a stationary inflation rate different from zero is possible as long as the 
ratio 
$\overline{p}(t)/\overline{W}(t)$ fluctuates around a steady value. In absence 
of wage update, we have a fixed $W \equiv 1$ and 
inflation is therefore impossible. The main effect induced by wage dynamics is therefore the possibility of inflation.

Using the wage update rules defined below, we found that the average inflation 
rate 
depends on parameters such as the households propensity to consume 
$c$ and the price/wage adjustment parameters $\gamma_{p,w}$. Most 
interestingly, we observe a strong dependence of 
the inflation rate upon the bankruptcy threshold $\Theta$, with large 
$\Theta$'s triggering high inflation and low $\Theta$'s corresponding to zero 
inflation.
For intermediate $\Theta$'s, periods of inflation and deflation may alternate 
and the model displays interesting instabilities.

\begin{figure}
\centering
\includegraphics[scale=0.35]{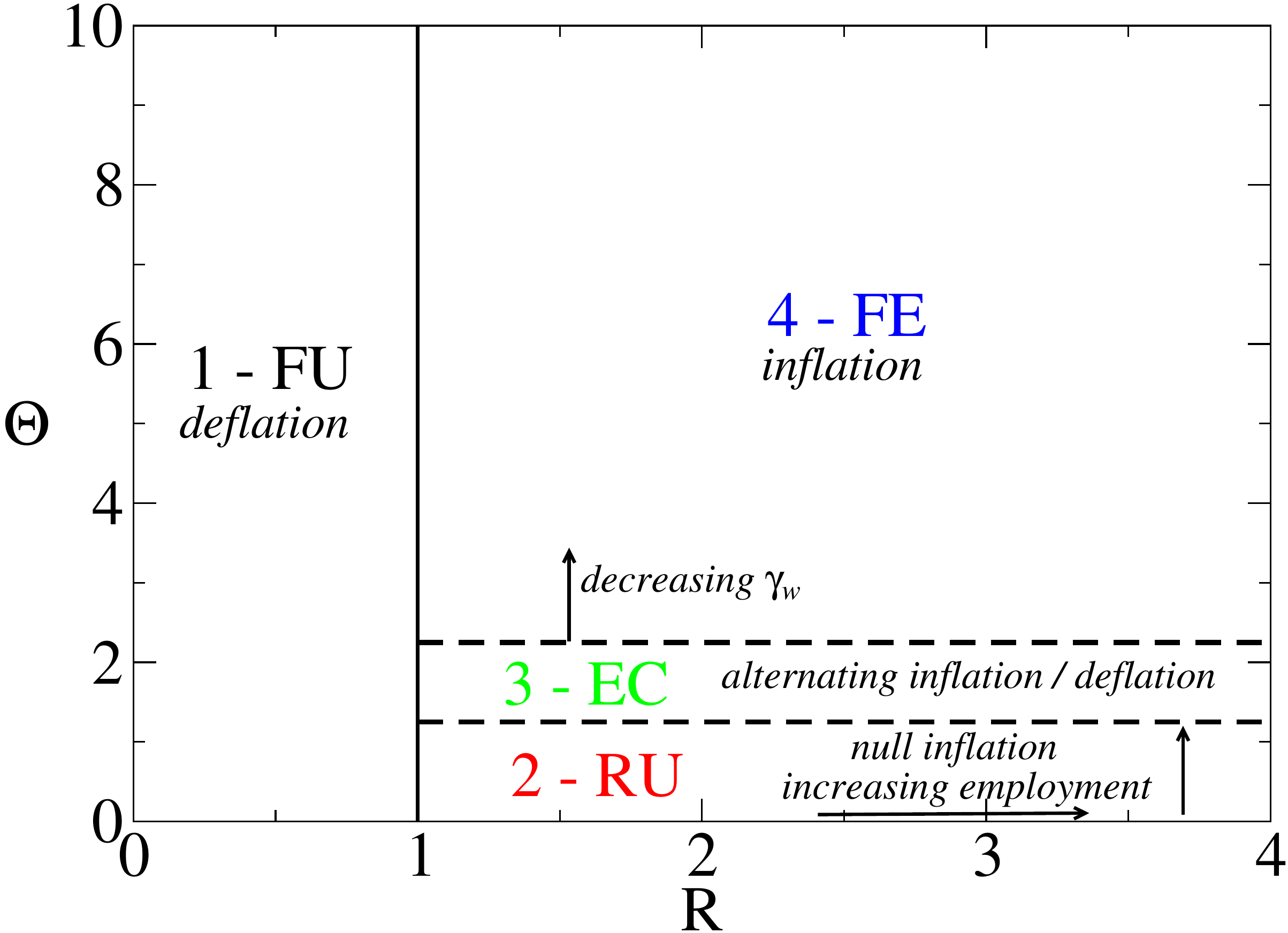}
\includegraphics[scale=0.35]{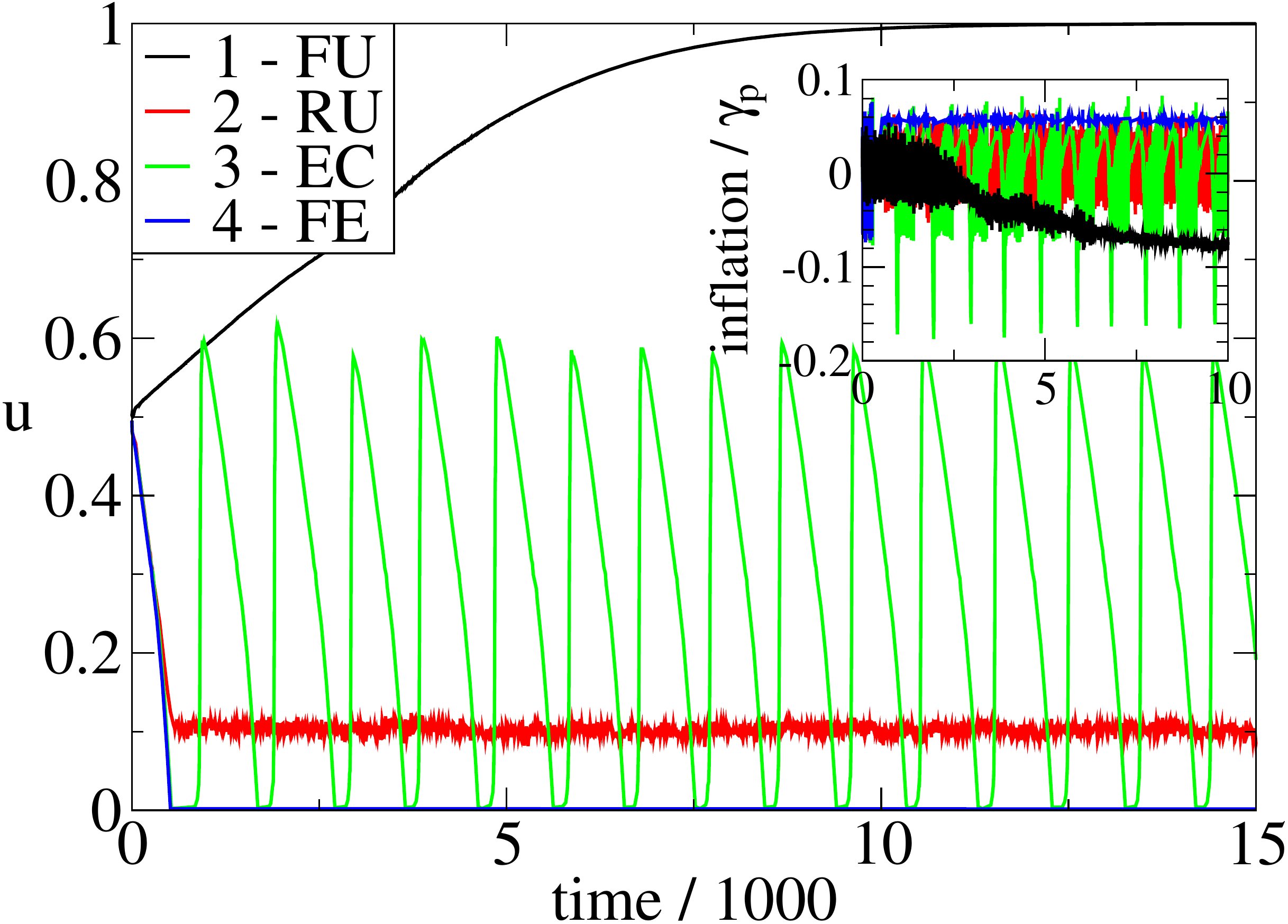}
\caption{
({\it Left})
Phase diagram in the $R-\Theta$ plane of the extended Mark 0 model with wage 
update with $\gamma_w=\gamma_p$.
The parameters of the basic model are the same as in Fig.~\ref{fig:PDGsummary}: 
$N_{\rm F}=5000$, $c=0.5$, $\g_p=0.05$, $\d=0.02$, $\f=0.1$, $f=1$, $\beta=0$.
As for the basic model, there are four distinct phases separated by critical 
lines.
Wage update brings in inflation in the Full Employment phase, and deflation in 
the full unemployment phase. Endogenous crises are characterized by alternating 
cycles
of inflation and deflation.
Interestingly, we find that the location of phase boundaries is in this 
case almost untouched by changes in both $\beta$ and $f$; the other parameters 
are irrelevant. 
Note that the wage update strongly stabilizes the full employment phase.
({\it Right})
A typical trajectory of $u(t)$ for each of the phases. In the inset, the price 
dynamics is shown, displaying inflation and deflation.
}
\label{fig:PD_wages}
\end{figure}

{
We now analyze the influence of wage adjustments 
on the phase transitions discussed in the previous sections.
}
The phase diagram for $z=1$ ($\gamma_w=\gamma_p$)  is reported in Fig.~\ref{fig:PD_wages}.
The phenomenology that we find is again very similar 
to the simple Mark 0 without wage update, except for inflation. The most 
interesting effect is the appearance of inflation in the ``good'' phases
of the economy, and deflation in the ``bad'' phases, as shown in 
Fig.~\ref{fig:PD_wages}. 
When $\Theta \gg 1$, we find again a first order critical boundary at $R=R_c$ 
that separates 
a high unemployment phase (with deflation) from a low unemployment phase (with 
inflation).
For $R > R_c$ we see again two additional phases: ``EC'', with 
endogenous crises and, correspondingly,
alternating periods of inflation and deflation but stable prices on the long 
run, and ``RU'', for small $\Theta$, where there is no 
inflation but a substantial residual unemployment rate (see 
Fig.~\ref{fig:PD_wages}).

The appearance of endogenous crises is consistent with what discussed in 
section~\ref{numerical_results0} and is related to situations in which the 
debt-to-savings
ratio $k(t)$ grows faster than prices. From this point of view, increasing 
$\gamma_w$ allows firms to better adapt wages (and thus prices) and to absorb 
the indebtment through inflation; region EC indeed shrinks when 
increasing $\gamma_w$.
{
Note that relating the parameters $\gamma_{p,w}$ to the flexibility of the
labor and goods markets is not straightforward. 
In this sense, it would be instead useful to study the effects of 
\emph{asymmetric}
upward/downward wage and price flexibilities (for example by defining
different $\gamma^\pm_{p,w}$) in order to understand whether improving
the flexibility of the labor and goods markets destabilizes the 
economy~\cite{eggersston,greenwald,napoletano}.
If, however, one considers the ratio $z=\gamma_w/\gamma_p$ as an indicator of 
wages flexibility (relative to prices flexibility), our results suggests that higher 
labor flexibility has a stabilizing effect.
}

We also observe that the oscillatory pattern found for the basic model persists 
as long as $\gamma_w\ll \gamma_p$, i.e. when wage updates are much 
less frequent that price updates. The power spectrum of the model for $R>R_c$ 
and $\Theta\gg 1$ is still characterized by 
the appearance of a peak (roughly corresponding to a period of $7$ time steps). 
Interestingly, 
we find that upon increasing the ratio $z$, the peak in the 
frequency spectrum disappears 
but not in a monotonous fashion; intermediate values of $z$ give rise to even 
more pronounced oscillations than for $z=0$, before these 
oscillations disappear for $z > z_c \approx 0.25$. 

In conclusion, the comparison between Figs.~\ref{fig:PD_wages} and 
\ref{fig:PDGsummary} demonstrates the robustness of our phase diagram 
against changes; introducing wages is a rather drastic modification since it 
allows inflation to set in, but still does not affect 
the phase transition at $R_c$, nor the overall topology of the phase diagram, 
which confirms its relevance. 
Interestingly, inflation is present in the good phases of the economy and 
deflation in the bad phases.
Our analytical understanding of these effects is however still poor; we 
feel it
would be important to bolster the above numerical results by solving simpler 
``toy models'' as we do for the basic Mark 0 (see section \ref{analytical} and 
Appendix~\ref{app:D1}).

\subsection{Other extensions and policy experiments}

\begin{figure}
\centering
\includegraphics[scale=0.35]{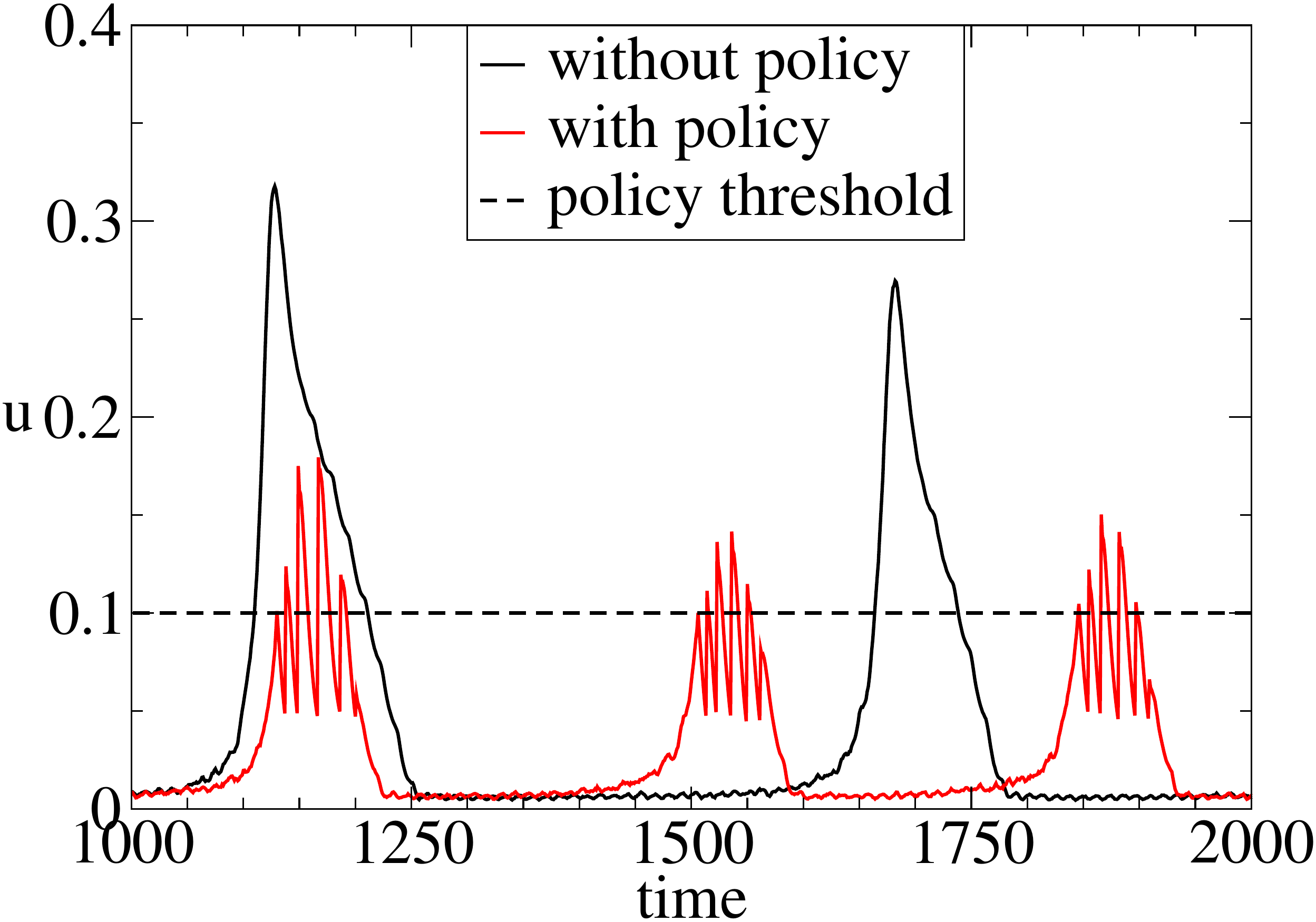}
\caption{Here we show an example of a ``toy'' policy experiment in the Mark 0 
model without wage update (i.e $\g_w=0$). 
We first run a simulation (the ``without policy'' line) with a constant value 
$\Theta=2$ lying in the 3-EC region of Fig. \ref{fig:PDG0}.
We then run the same simulation with a prototype central bank which increases 
$\Theta$ to $10$  as long as $u$ exceeds the threshold of $10\%$.
Note that unemployment is partially contained, but the crisis frequency 
concomitantly increases. 
The other parameters values are: $N=10\,000$, $R=2$, $\gamma_p=0.05$, $\f=0.1$, 
$\delta=0.02$, $f=1$.
}
\label{fig:policy}
\end{figure}

{
One of the final goal of our study is to have a prototype framework where
one is able to run meaningful policy experiments. In order to do that the model described so far is lacking a number of 
important ingredients, namely a central bank exogenously setting interest rates level and 
the amount of money in circulation. This is the project on which we are currently pursuing, with 
encouraging results \cite{ustocome}. Still, even within the simplistic framework of Mark 0, one can envisage an 
interesting prototype policy experiment
which consists in allowing the ``central bank'' to temporarily increase the 
bankruptcy threshold $\Theta$ in times of high unemployment. 
Fig.~\ref{fig:policy}
shows an example of this: the economy is, in its normal functioning mode, in 
the EC (region 3) of the phase diagram, with $R=2$ and (say) $\Theta=2$. 
This leads in general to a rather low unemployment rate $u$ but, as repeatedly 
emphasized above, this is interrupted by acute endogenous crises. The central 
bank then decides that whenever $u$ exceeds $10 \%$, its monetary policy 
becomes accommodating, and amounts to raising $\Theta$ from its normal value 
$2$ 
to -- say -- 
$\Theta=10$. As shown in  
Fig.~\ref{fig:policy}, this allows the bank to partially contain the 
unemployment bursts. However, quite interestingly, it also increases the crisis 
frequency, 
as if it did not allow the economy to fully release the accumulated stress. We 
expect that this phenomenology will survive in a more realistic framework.
}

We have also explored other potentially interesting extensions of Mark 0, for 
example adding trust or confidence, 
that may appear and disappear on time scales much shorter than the evolution 
time scale of any ``true'' economic factor, and 
can lead to market instabilities and crises (see e.g. 
\cite{Marsili,trust,JPB}). There are again many ways to model the
potentially destabilizing feedback of confidence. One of the most important
channel is the loss of confidence induced by raising unemployment, that 
increases the saving propensity of households and
reduces the demand. The simplest way to encode this in Mark 0 is to let the
``$c$'' parameter, that determines the fraction of wages and accumulated savings that is 
devoted to consumption, be an increasing
function of the employment rate $\varepsilon=1-u$. We indeed find that the 
confidence feedback loop can again induce  purely endogenous swings 
of economic activity. Similarly, a strong dependence of $c$ on the recent 
inflation can induce instabilities \cite{ustocome}.

\section{Analytical description} 
\label{analytical}

We attempt here to describe analytically some aspects of the dynamics of Mark 0 
in its simplest version, namely without bankruptcies ($\Theta=\infty$ for which 
$\f$ and $f$ become irrelevant),
with $\b \geq 0$, fixed wages $W=1$, and no dividends ($\d=0$). For simplicity, 
we also fix $\mu=1$ and $c=1/2$.
The only relevant parameters are therefore $\beta$, $\g_p$ and $\h_\pm$.
The equations of motion of this very minimal model are:
\beq\label{update0minimal}
\begin{split}
    \text{If } Y_i(t) < D_i(t)  &\hskip10pt \Rightarrow \hskip10pt 
    \begin{cases}
&     Y_i(t+1)=Y_i(t)+ \min\{ \eta_+ [ D_i(t)-Y_i(t)], 1 - \overline{Y}(t) \} \\
&  \text{If } p_i(t) < \overline{p}(t) \hskip10pt \Rightarrow \hskip10pt   
p_i(t+1) = p_i(t) (1 + \g_p \xi_i(t) ) \\
&  \text{If } p_i(t) \geq \overline{p}(t) \hskip10pt \Rightarrow \hskip10pt   
p_i(t+1) = p_i(t)  \\
\end{cases}
\\
  \text{If }   Y_i(t) > D_i(t)  &\hskip10pt  \Rightarrow \hskip10pt
  \begin{cases}
&    Y_i(t+1)=\max\{ Y_i(t) - \eta_- [Y_i(t)-D_i(t)] , 0 \} \\
&  \text{If } p_i(t) > \overline{p}(t) \hskip10pt \Rightarrow \hskip10pt   
p_i(t+1) = p_i(t) (1 - \g_p \xi_i(t) ) \\
&  \text{If } p_i(t) \leq \overline{p}(t) \hskip10pt \Rightarrow \hskip10pt   
p_i(t+1) = p_i(t) \\
    \end{cases} 
\\
  D_i(t) &= \frac{c}{p_i(t)} \frac{e^{-\beta(p_i(t) - \overline{p}(t))}}{Z(t)}  
 [ \max\{ s(t) , 0 \} + \overline{Y}(t) ],\qquad Z(t) := \sum_i e^{-\beta(p_i(t) - \overline{p}(t))} \\
  \EE_i(t+1) &= \EE_i(t) - Y_i(t) + p_i(t) \min\{ Y_i(t), D_i(t) \} \\
  s(t) &= M_0 - \overline{\EE}(t) \ .
\end{split}
\eeq
Here $M_0$ is the total money in circulation, whose precise value is irrelevant 
for this discussion, and $s = S/N_{\rm F}$ are the savings per agent. 
Overlines denote an average over firms, 
which is flat for $\overline{Y} = N_{\rm F}^{-1} \sum_i Y_i(t)$ and 
$\overline{\EE} =N_{\rm F}^{-1} \sum_i \EE_i(t)$ while it is 
weighted by production for $\overline{p}$, see Eq.~\eqref{pbardef}.
Note that the basic variables here are $\{ p_i, Y_i, \EE_i \}$, all the other 
quantities are deduced from these ones.

In the high employment phase, the model admits a stationary state with 
$\overline{Y}_{st} \sim 1$.
To show this, let us focus on the case where $\g_p$ is very small, but not 
exactly zero, otherwise of course the price dynamics is frozen.
The stationary state is attained after a transient of duration $\sim 1/\g_p$, 
and in the stationary state fluctuations between firms are very
small in such a way that $p_i \sim \overline{p}$ and $Y_i \sim \overline{Y}$.
A stationary state of Eq.~\eqref{update0minimal} has $Y_i(t+1) = Y_i(t)$ and 
therefore $Y_i = D_i$, hence $Y_i = D_i = \overline{Y}$. Furthermore,
from $\EE_i(t+1) = \EE_i(t)$ we deduce that $p_i = \overline{p} = 1$. Finally, 
we have $D_i = \overline{Y} = (s + \overline{Y}) / 2$, which gives
$s = \overline{Y}$ and $\EE = M_0 - \overline{Y}$. One obtains therefore a 
continuum of stationary states, because the production $\overline{Y}$
or equivalently the employment are not determined by requiring stationarity.
However, we will show in the following that these equilibria become unstable as 
soon as fluctuations are taken into account ($\g_p>0$). We will see
that fluctuations can induce either an exponentially fast decrease of 
$\overline{Y}$ towards zero (corresponding to the full unemployment phase), 
or an exponentially fast grow of $\overline{Y}$, which is therefore only cutoff 
by the requirement $\overline{Y} \leq \mu=1$, corresponding
to full employment. It is therefore natural to choose the stationary state with 
$\overline{Y}=1$ as a reference, and study the effect of fluctuations around
this state.
We therefore consider the stationary state with $Y_i = D_i = p_i = 1$, and 
$\EE_i = M_0 -1$ in
such a way that $s = \overline{Y} = 1$.

For analytical purposes, we will focus below on the limit in which 
$\eta_\pm, \g_p \to 0$, in such a way that we can expand around the high employment stationary state 
and obtain results for the phase transition point. Numerically, the fact that $\g_p$ and/or $\eta_\pm$ are small does not change the qualitative 
behavior of the model.

In order to consider small fluctuations around the high employment stationary 
state, we define the following variables:
\beq\label{exp_def}
\begin{split}
Y_i(t) &= 1 - \g_p \e_i(t) \ , \\
p_i(t) &=  e^{ \g_p \l_i(t)} \ , \\     
\EE_i(t) &= M_0 -1  - \g_p \a_i(t) \ .
\end{split}\eeq
Note that in the following, overlines over  $\e, \a, \l$ always denote flat 
averages over firms. Using $p_i(t) =  1+ \g_p \l_i(t) + (\g_p \l_i(t))^2/2$ for $\g_p \to 0$, one finds that, to order $\g_p^2$:
\beq
\frac{e^{-\beta p_i(t)}}{Z(t)} \approx 1  - \beta \g_p (\l_i(t) - 
\overline{\l}(t)) - \frac12 \beta \g_p^2 \bigg[ 
(1-\beta)(\l_i^2(t)-\overline{\l^2}(t))+2\beta \overline{\l}(t)(\l_i(t) - 
\overline{\l}(t))\bigg].
\eeq
Since we will find later that $\overline{\l}(t)$ is itself of order $\g_p$, we 
will drop the part of the last term in the above expression, which is 
$O(\g_p^3)$, i.e.
\beq
\frac{e^{-\beta p_i(t)}}{Z(t)} \approx 1  - \beta \g_p (\l_i(t) - 
\overline{\l}(t)) - \frac12 \beta(1-\beta) \g_p^2\l_i^2(t).
\eeq

\subsection{Stability of the high employment phase}

To study the stability of the high employment phase, we make two further 
simplifications.
\begin{enumerate}
\item We neglect the fluctuations of the savings and fix $\a_i \equiv 0$.
This is justified if the employment rate varies slowly over the time scale
$\tau_c=-1/\ln(1-c)$ that characterizes the dynamics of the savings. But, as
we shall show below, the dynamics of employment becomes 
very slow in the vicinity of the phase transition, hence this assumption seems 
justified.
\item {In order to proceed analytically, we also consider the limit $\eta_+, \eta_- \to 0$ with $\eta_+ = R \eta_-$ and $R$ of order one.
In fact, we will even assume below that $\eta_\pm \ll \g_p$. Then, it is not difficult to see that $\e_i$ is a random variable 
of order $\eta_+$ and is therefore also very small. We define $\e_i = \eta_+ z_i$.
We believe that these approximations are actually quite accurate, as is confirmed by the comparison of 
the theoretical transition line with numerical data (see Fig.~\ref{fig:PDG0}). }
\end{enumerate}

With these further simplifications (i.e. setting $\a_i=0$ and assuming that $\g_p$, $\h_\pm$ are small and $z_i = O(1)$), 
we have, together with Eq.~\eqref{exp_def}:
\beq\begin{split}
D_i(t) &= 1 - \g_p (\l_i(t) + \beta (\l_i(t) - \overline{\l}(t))) + \frac12 
\g_p^2  \l_i^2 (1 + \beta  + \beta^2) + O(\g_p \h_+) + O(\g_p^3) \ , \\
\overline{p}(t) &= 1 + \g_p \overline{\l}(t) + \frac12 \g_p^2 
\overline{\l^2}(t) + O(\g_p \h_+) + O(\g_p^3) \ , \\
\end{split}\eeq
If, as announced above we further assume that  $\h_+ \ll \g_p$, it is justified to keep the terms of order $\g_p^2$ while neglecting the terms 
of order $\g_p \h_+$ and $\g_p^3$. 

We also note that if $|\lambda_i |$ is bounded by a quantity of order 1 (which we will 
find below), then the condition $\l_i(t) + \beta (\l_i(t) - \overline{\l}(t))) 
- \frac12 \g_p \l_i(t)^2 >0 $ is equivalent to $\l_i(t) + \beta (\l_i(t) - 
\overline{\l}(t))) >0$. 
Then the first two lines of Eq.~\eqref{update0minimal} become
\beq
\label{z_lam_ev}
\begin{split}
    \text{If } \l_i(t) < \frac{\beta}{1+\beta} \overline{\l}(t)  &\hskip10pt 
\Rightarrow \hskip10pt 
    \begin{cases}
&     z_i(t+1)= z_i(t) -  \min\{
-  \l_i(t) - \beta (\l_i(t) - \overline{\l}(t)) + \frac12 \g_p \hat{\beta}   
\l_i^2,  \overline{z}(t) \} \\
&   \l_i(t+1) = \l_i(t) +  \xi_i(t) - \frac12 \g_p \xi_i(t)^2 \\
\end{cases}
\\
  \text{If } \l_i(t) > \frac{\beta}{1+\beta} \overline{\l}(t)   &\hskip10pt  
\Rightarrow \hskip10pt
  \begin{cases}
&   z_i(t+1) = z_i(t) + \frac{1}R [ \l_i(t) + \beta (\l_i(t) - 
\overline{\l}(t)) - \frac12 \g_p  \hat{\beta}  \l_i^2 ]     \\
&   \l_i(t+1) = \l_i(t) -  \xi_i(t) - \frac12 \g_p \xi_i(t)^2  \\
    \end{cases}
\end{split}
\eeq
where $\hat{\beta}=1 + \beta  + \beta^2$.
Therefore, in this limit, the evolution of $\l$ decouples from that of $\e$ (or 
$z$). Note that from Eq.~\ref{z_lam_ev} one has $-1-\frac{\beta}{1+\beta} 
\overline{\l}(t)-\gamma_p/2<\lambda<1-\frac{\beta}{1+\beta} 
\overline{\l}(t)-\gamma_p/2$ (which we rewrite in the form $\lambda_{\min}<\lambda<\lambda_{\max}$).

From the simplified evolution equation  Eq.~\ref{z_lam_ev} we now obtain the evolution of the probability 
distribution $P_t(\l)$ and for $\overline{z}(t)$, which are by definition:
\beq \label{zetaeq} 
P_{t+1}(\l) =\int_0^1 d\xi\, \int_{\frac{\beta}{1+\beta} 
\overline{\l}(t)}^{\lambda_{\max}}  d\l' P_t(\l')
\delta\left(\l-\l'+\xi+\frac{\gamma_p}{2} \xi^2\right)
         + \int_0^1 d\xi\, \int_{\lambda_{\min}}^{\frac{\beta}{1+\beta} 
\overline{\l}(t)}  d\l' P_t(\l')
\delta\left(\l-\l'-\xi+\frac{\gamma_p}{2} \xi^2\right) \ ,
\eeq 
and
\beq\label{eq:evoZ}
\overline{z}(t+1) = \overline{z}(t) + \frac{1}R \int_{\frac{\beta}{1+\beta} 
\overline{\l}(t)}^{\lambda_{\max}} d\l  \, P_t(\l)\, \left(   \l + \beta (\l- \overline{\l}) 
- \frac12 \g_p \hat{\beta}   \l^2    \right) - 
\int_{\lambda_{\min}}^{\frac{\beta}{1+\beta} \overline{\l}(t)} d\l \, P_t(\l) \, 
\min\left\{- \l - \beta (\l - \overline{\l})  +   \frac12 \g_p \hat{\beta}  
\l^2  , \overline{z}(t) \right\} \ .
\eeq
Since the dynamics of $\l$ is decoupled from the one of $\overline{z}$, we can 
assume that $P_t(\l)$ reaches a stationary
state. In Appendix~\ref{app:D1} we show that, at first order in $\gamma_p$, we 
have for the stationary distribution
\beq\label{eq:Pst}
P_{\rm st}(\l) = \left( 1 - |\l + \frac{\beta \gamma_p}{4}|- \frac{\g_p}2 
\text{sgn}(\l) \l^2 \right) \ .
\eeq
% \th\left( 1 - |\l + \frac{\beta \gamma_p}{4}| - 
% \frac{\g_p}2 \text{sgn}(\l) \l^2 \right) \
Hence, using the condition $P_{\rm st}(\l)\geq0$ we find that $| \l | \leq 1 + O(\g_p)$, 
consistently with the assumptions we made above. 
From this result, we also find that $\overline{\l} = -\gamma_p (1+\beta)/4$. 

{
Note that the average
price here decreases as $\beta$ increases, confirming the discussion in Section \ref{role_of_beta}. 
Indeed, we have neglected the fluctuations of $S$ in the above calculation, while, as explained 
in Section \ref{role_of_beta}, it is the increase of involuntary savings (and thus of $S$) with $\beta$ that leads to
the average price increase. As a further confirmation, one can easily simulate Eqs.~(\ref{z_lam_ev})
with $\a_i\neq0$, and find $d\overline{\l}/d\beta>0$ in this case.
}

Let us now discuss the dynamics of $\overline{z}(t)$. If at some time $t$ we 
have $\overline{z}(t)=0$, then it is clear from Eq.~\eqref{eq:evoZ} that 
$\overline{z}(t)$ will grow with time, and it will continue to do so unless the 
last term in the equation becomes sufficiently large. 
Recalling that $|\l| \leq 1 + O(\g_p)$, it is clear that even if 
$\overline{z}(t)$ becomes very large, the last term in Eq.~\eqref{eq:evoZ}
can be at most given by $\int_{-\io}^0 d\l \, P_{\rm st}(\l) \, \left( - \l -  
\beta (\l - \overline{\l}) +   \frac12 \g_p \l^2 \right)$. Therefore, if
\beq\label{eq:critic_cond}
\frac{1}R \int_{-\g_p \beta/4}^{\lambda_{\max}} d\l  \, P_{\rm st}(\l)\, \left( \l + \beta 
(\l - \overline{\l}) - \frac12 \g_p  \hat{\beta} \l^2 \right) > 
\int_{\lambda_{\min}}^{-\g_p \beta/4} 
d\l \, P_{\rm st}(\l) \, \left( - \l -  \beta (\l - \overline{\l}) +   \frac12 
\g_p \hat{\beta}  \l^2  \right) \ , 
\eeq
then $\overline{z}(t)$ will continue to grow with time, and $\overline{Y}(t) = 
1 - \g_p \h_+ \overline{z}(t)$ will become very small and the economy will 
collapse.
Conversely, if the condition in Eq.~\eqref{eq:critic_cond} is not satisfied, 
then Eq.~\eqref{eq:evoZ} admits a stationary solution with 
$0< \overline{z}_{\rm st} < \io$.
Using Eq.~\eqref{eq:Pst}, 
the critical boundary line finally reads:\footnote{It would be interesting to 
compute the first non trivial corrections in $\eta$ to the transition line.
We leave this for a later study.}
\be 
R_c = \frac{\eta_+}{\eta_-} 
= \frac{
\int_{-\g_p \beta/4}^{\lambda_{\max}} d\l  \, P_{\rm st}(\l)\, \left( \l +  \beta (\l - 
\overline{\l}) - \frac12 \g_p  \hat{\beta} \l^2 \right) 
}{
\int_{\lambda_{\min}}^{-\g_p \beta/4} d\l \, P_{\rm st}(\l) \, \left( - \l -  \beta (\l - 
\overline{\l}) +   \frac12 \g_p \hat{\beta}  \l^2  \right)
}
\approx 1 - \gamma_p \frac{(2+\beta)^2}{2(1+\beta)} + ...
\label{critic_line}
\ee 
where we used the condition $\gamma_p<4/(\beta+2)$ which is always verified for $0<\beta<+\infty$
and $\gamma_p\to0$. As one can see in Fig.~\ref{fig:PDG0} 
this results is in good agreement with numerical result for $\beta=2$. 
We therefore find, interestingly, that increasing $\beta$ {\it decreases} the
value of $R_c$, i.e. stabilizes the high employment phase, as indeed discussed above, see Fig. 4
and the discussion around it.

In other words, the $\e_i$ perform a {\it biased} random walk, in presence of a 
noise whose average is
given by the last two terms in Eq.~\eqref{eq:evoZ}
(of course, when $\overline{\e}$ is too small the minimum in the last term is 
important, because it is there to prevent $\overline{\e}$ from becoming 
negative).
The system will evolve towards full employment, with $\overline{\e}= O(\eta_+)$ 
and $\overline{Y} = 1 - O(\g_p \eta_+)$, whenever the average noise is 
negative, 
and to full collapse, with $\overline{\e} \to \io$ and $\overline{Y}=0$, 
otherwise. 
The critical line is given by the equality condition, such that the average 
noise vanishes. 
Therefore, 
right at the critical point, the unemployment rate makes an unbiased random walk
in time, meaning that 
its temporal fluctuations are large and slow. This justifies the ``adiabatic''
approximation\footnote{This refers, in physics,  to a situation where a system is driven by an infinitely slow process. One can then 
consider the system to be always close to equilibrium during the process.}
made above, that lead us to neglect the
dynamics of the savings. 

\subsection{Oscillations in the high employment phase}

As discussed above, in the FE phase macroeconomic variables display an 
oscillatory dynamics, see Fig.~\ref{fig:2trajectories}.
Intuitively, the mechanism behind these oscillations is the following. 
When prices are low, demand is higher than production and firms increase the 
prices.
But at the same time, households cannot consume what they demand, so they 
involuntarily save: savings increase when prices are low. 
These savings keep the demand high for a few rounds even while
prices are increasing, therefore prices keep increasing above their equilibrium 
value. When prices are too high, households need to use their savings
to consume, and therefore savings start to fall. Increase of prices and 
decrease of savings determine a contraction of the demand.
At some point demand falls below production and prices start to decrease again, 
with savings decreasing at the same time.
When prices are low enough, demand becomes again higher then production and the 
cycle is restarted.
An example is shown in Fig.~\ref{fig:2trajectories}.

Based on this argument, it is clear that to study these oscillations, 
we need to take into account the dynamics of the savings so we cannot assume 
$\a_i=0$ as in the previous section.
However, here it is enough to consider the first order terms in $\g_p$.
In terms of the basic variables in Eq.~\eqref{exp_def}, the other variables 
that appear in Eq.~\eqref{update0minimal} are easily written as follows:
\beq\begin{split}
\overline{Y}(t) &= 1 - \g_p \overline{\e}(t) \ , \\
s(t) &= 1 + \g_p \overline{\a}(t) \ , \\
\overline{p}(t) &= 1 + \g_p \overline{\l}(t)  \ , \\
\end{split}\eeq
where for $\e, \a, \l$, overlines denote flat averages over firms,
and
\beq
D_i(t) = 1 + \g_p \left( \frac12 \overline{\a}(t) - \frac12 \overline{\e}(t) - 
\l_i(t) \right) \ .
\eeq
Inserting this in Eq.~\eqref{update0minimal}, we arrive to the following 
equations that hold at the lowest order in $\gamma_p$:
\beq\label{updatelin}
\begin{split}
    \text{If } -\e_i(t) < \frac12 \overline{\a}(t) - \frac12 \overline{\e}(t) - 
\l_i(t)  &\hskip10pt \Rightarrow \hskip10pt 
    \begin{cases}
&     \e_i(t+1)= \e_i(t) -  \min\{ \eta_+ [  \frac12 \overline{\a}(t) - \frac12 
\overline{\e}(t) - \l_i(t) + \e_i(t)    ],  \overline{\e}(t) \} \\
&  \text{If } \l_i(t) < \overline{\l}(t) \hskip10pt \Rightarrow \hskip10pt   
\l_i(t+1) = \l_i(t) + \xi_i(t)   \\
&   \a_i(t+1) = \a_i(t)  - \l_i(t) \\
\end{cases}
\\
  \text{If }   -\e_i(t) > \frac12 \overline{\a}(t) - \frac12 \overline{\e}(t) - 
\l_i(t)   &\hskip10pt  \Rightarrow \hskip10pt
  \begin{cases}
&    \e_i(t+1) = \e_i(t) - \eta_- [   \frac12 \overline{\a}(t) - \frac12 
\overline{\e}(t) - \l_i(t) + \e_i(t)  ]     \\
&  \text{If } \l_i(t) > \overline{\l}(t) \hskip10pt \Rightarrow \hskip10pt   
\l_i(t+1) = \l_i(t) - \xi_i(t)    \\
&\a_i(t+1) = \a_i(t) - \e_i(t)   -   \frac12 \overline{\a}(t) + \frac12 
\overline{\e}(t) \\
    \end{cases} 
\end{split}
\eeq
From an analytical point of view, the above model is still to complex to make 
progress. 
We make therefore a further simplification, by assuming that 
\beq\label{eq:appr}
\e_i = C \l_i \ ,
\eeq
where $C$ is a certain numerical constant. 
In terms of the original Mark 0 variables, this approximation is equivalent to, 
roughly speaking,
\be
p_i - {\overline p} \propto (Y_i - D_i) - ({\overline Y} - {\overline D}).
\ee
i.e. fluctuations of the prices are proportional to supply-demand gaps. 
Although numerical simulations
only show a weak correlation, the approximation~\eqref{eq:appr} allows us to 
obtain a more tractable model
that retains the basic phenomenology of the oscillatory cycles and reads:
\beq\label{updatedelta}
\begin{split}
    \text{If } 2(1-C) \l_i(t) <  \overline{\a}(t)-C \overline{\l}(t)   
&\hskip10pt \Rightarrow \hskip10pt 
    \begin{cases}
&  \text{If } \l_i(t) < \overline{\l}(t) \hskip10pt \Rightarrow \hskip10pt   
\l_i(t+1) = \l_i(t) + \xi_i(t)  \\
&   \a_i(t+1) = \a_i(t)  - \l_i(t) \\
\end{cases}
\\
  \text{If } 2(1-C)  \l_i(t) > \overline{\a}(t)-C \overline{\l}(t)     
&\hskip10pt  \Rightarrow \hskip10pt
  \begin{cases}
&  \text{If } \l_i(t) > \overline{\l}(t) \hskip10pt \Rightarrow \hskip10pt   
\l_i(t+1) = \l_i(t) - \xi_i(t)  \\
&\a_i(t+1) = \a_i(t) - C \l_i(t)  -   \frac12 \overline{\a}(t) + \frac12 C 
\overline{\l}(t)    \\
    \end{cases} 
\end{split}
\eeq
This very minimal model, when simulated numerically, indeed gives persistent 
oscillations, independent on $N$, when $C > C^* \approx 0.45$, 
and can also
be partially investigated analytically, see Appendix~\ref{app:D1}. 

\subsection{The ``representative firm'' approximation}

To conclude this section, we observe that there is a further simplification 
that allows one to retain some
of the phenomenology of Mark 0. 
It consists in describing the firm sector by a unique
``representative firm'', $N_{\rm F}=1$,
with production $\overline{Y}(t)$, price $p(t)$ and demand 
$\overline{Y}(t)/p(t)$. The
dynamics of the production and price are given by the same rule as above, but
now the dynamics of the price completely decouples: 
\bea 
p(t) < 1  &\Rightarrow& 
\begin{cases}
\overline{Y}(t+1) = \overline{Y}(t)(1 + \eta_+ (\frac{1}{p(t)} - 1))\\
p(t+1)=p(t) (1 + \gamma \xi(t))
\end{cases}\\
p(t) > 1  &\Rightarrow& 
\begin{cases}
\overline{Y}(t+1) = \overline{Y}(t)(1 + \eta_- (\frac{1}{p(t)} - 1))\\
p(t+1)=p(t) (1 - \gamma \xi(t)). 
\end{cases}
\eea 
Of course, this simple model
misses several important effects: most notably those associated to $\Theta$, 
hence the transition between
the Full Employment, Endogenous Crises, and Residual Unemployment phases in 
Fig.~\ref{fig:PDGsummary}.
In particular, endogenous crises are never present in this case, because of the 
absence of a bankrupt/revival mechanism,
and also the oscillatory pattern in the Full Employment
phase disappears, because in this model savings are not considered.
Still,
this model is able to capture the transition between the Full Unemployment and 
Full Employment regions as a function of $R$
(see Fig.~\ref{fig:PDGsummary}), as
confirmed by the analytical solution, which is in fact identical to the one of the 
model with $N_{\rm F}>1$ when $\beta=0$ and $\eta$ is small. 
And since the model is so simple, one can hope that some of the extensions 
discussed in
the previous section can be at least partly understood analytically 
within this ``representative firm'' framework (we will give a few explicit. 
This would be an important step to
put the rich phenomenology that we observe on a firmer basis.

\section{Summary, Conclusion}

The aim of our work (which is part of the CRISIS project and still ongoing) 
was to explore the possible types
of {\it phenomena} that simple macroeconomic Agent-Based Models can reproduce, 
and to 
map out the corresponding {\it phase diagram} of these models, as 
Figs.~\ref{fig:PD_wages} and \ref{fig:PDGsummary} exemplify.
The precise motivation 
for our study was to understand in detail the nature of the macro-economic
fluctuations observed in the ``Mark I'' model devised by D. Delli Gatti and
collaborators \cite{MarkIref,MarkIbook}. One of our central findings is the generic 
existence, in
Mark I (and variations around that model) of a first order, discontinuous phase
transition between 
a ``good economy'' where unemployment is low, and a ``bad economy'' where
unemployment is high. By studying a simpler hybrid model (Mark 0), where the
household sector is 
described by aggregate variables and not at the level of agents\footnote{For a 
recent study exploring the idea of hybrid models, see \cite{Assenza}.}, we have 
argued
that this transition is induced by an {\it asymmetry between the rate of hiring 
and
the rate of firing} of 
the firms. This asymmetry can have many causes at the micro-level, for example different
hiring and firing costs. In Mark I,
for example, it reflects the reluctance of firms to take loans when the interest
rate is too high. 
As the interest rate increases, the unemployment level remains small until a
tipping point beyond which the economy suddenly collapses. If the parameters are
such that the system 
is close to this transition, any small fluctuations (for example in the level of
interest rates) is amplified as the system jumps between the two equilibria. It 
is 
actually possible that the central bank policy (absent in our current model), when 
attempting to stabilize the economy,  in fact
bring the system close to this transition. Indeed, too low an interest rate leads 
to overheating and inflation, and too high an interest rate leads to large 
unemployment.
The task of the central bank is therefore to control the system in the vicinity 
of 
an instability and could therefore be a natural realization of the enticing
`self-organized criticality' 
scenario recently proposed in \cite{Felix} (see also \cite{Bak}).

Mark 0 is  simple
enough to be partly amenable to analytic treatments, that allow us to compute
approximately the location of the transition line as a function of the
hiring/firing propensity of firms, and characterize the oscillations and the 
crises 
that are observed. Mark 0 can furthermore be extended in
several natural directions. One is to allow this
hiring/firing propensity to depend on the financial fragility of firms -- hiring
more when firms 
are financially healthy and firing more when they are close to bankruptcy. 
We find that in this case, the above transition survives but
becomes {\it second order}. 
As the transition is approached, unemployment fluctuations become larger and
larger, and the corresponding correlation time becomes infinite, leading to 
very low
frequency fluctuations. There again, we are able to give
some analytical arguments to locate the transition line \cite{ustocome}. Other stabilizing 
mechanisms,
such as the bankruptcy of indebted firms and their replacement by healthy firms 
(financed by 
the accumulated savings of households), lead to a similar phenomenology.

The role of the bankruptcy threshold $\Theta$, which is the only 
``proto-monetary'' effect in Mark 0, turns
out to be crucial in the model, and leads to the phase diagram shown in 
Fig.~\ref{fig:PDGsummary}. We generically
find not one but three ``good'' phases of the economy: one is where full 
employment prevails (FE), 
the second one is where some residual unemployment is present (RU), and the 
third, intermediate one is
prone to {\it acute endogenous crises} (EC), during
which the unemployment rate shoots up before the economy recovers. Note again 
that there are no exogeneous productivity shocks in Mark 0; interestingly, crises (when they occur) are
of endogeneous origin and not as a result of sweeping parameters through a phase transition (as 
 is the case, we argued, of Mark I with a time dependent interest rate) but purely 
 as a result of the own dynamics of the system. The existence 
of endogenous crises driven by feedback loops in such simple settings is 
quite interesting from a 
general standpoint (see also \cite{Dosi,Eurace2}), and reinforces the idea that 
many economic and financial crises may
{\it not} require large exogenous shocks (see 
\cite{Minsky,MG,MarkIbook,Sornette,Anand,Battiston,JPBerd,JPB,Julius} for related 
discussions).
We have shown that the endogenous crises can be defanged (in the model) if the household 
sector does not carry the full burden of firms
bankruptcies and/or when the profits of firms is efficiently re-injected in the 
economy (see e.g. Fig. \ref{fig:phaseboundaries}).

We have then allowed firms to vary the wages
of their employees according to some plausible rules of thumb (wages in Mark I
and Mark 0 are fixed); this leads to inflation or deflation but leaves the 
above picture essentially 
unchanged (see Fig.~\ref{fig:PD_wages}). Several other extensions are of obvious interest, 
and we plan to study them in the near future in the
same stylized way as above. The most obvious ones is to understand how a 
central bank 
that prints money and sets exogenously the interest rate can control the 
unemployment rate and the inflation rate in the vicinity of an unstable point, 
as we mentioned just above \cite{ustocome}. Other interesting topics are: modeling research and 
innovation, 
allowing firms to produce different types of goods, and introducing a financial 
sector and a housing market \cite{Farmer}.  

Beyond the generic phase diagram discussed in the whole paper, we found another 
notable, robust feature: the low unemployment phase 
of all the ABMs we considered are characterized by endogenous 
oscillations that do not vanish as the system size becomes large, with a period 
corresponding, in real time, to $\sim 5-10$ years~\cite{Dosi}.
It is tempting to interpret these oscillations as real and corresponding to the 
``business cycle'', as they arise from a very plausible 
loop between prices, demand and savings. These oscillations actually also 
appear in highly simplified models, 
where both the household and the firm sectors are represented by aggregate 
variables \cite{Goodwin,Macromodels-us}, as well as in network models \cite{Julius}.

Building upon this last remark, a very important question, it seems to us, is how 
much can be understood of the phenomenology of ABMs using ``mean-field'' 
approaches \cite{guilmi,aoki}, i.e. dynamical 
equations for aggregate variables of the type considered, for example, in 
\cite{Goodwin,Flaschel}? A preliminary analysis reveals that the dynamical 
equations 
corresponding to Mark 0 or Mark I
already lead to an amazingly complex phase-diagram \cite{Macromodels-us}. Are 
these mean-field descriptions quantitatively accurate?  When do we really need 
agents and 
when is an aggregate description sufficient?
The answer to this question is quite important, since it would allow one to 
devise faithful ``hybrid'' ABMs, where whole sectors of the economy would be 
effectively described in terms of these 
aggregate variables, only keeping agents where they are most needed. 

Another nagging question concerns the calibration of macroeconomic ABMs. It 
seems to us that before attempting any kind of quantitative calibration, 
exploring and making a catalogue of the different 
possible qualitative ``phases'' of the model is mandatory. Is the model 
qualitatively plausible or is the dynamics clearly unrealistic? 
In what ``phase'' is the true economy likely to be? On this 
point, one of the surprise of the present study is the appearance of very long 
time scales. For example, even in the case of perfectly stable economy 
with wage update rule (\ref{update0++}) and all $\gamma$ parameters equal to 
$10 \%$ (a rather large value), the equilibrium state of the economy 
(starting from an arbitrary initial condition) is only reached after $\approx 
200$ time steps. 
If one thinks that the elementary time scale in these models is of the order of 
three months, 
this means that the physical equilibration time of the economy is 20-50 years, 
or even much longer, see e.g. Figs.~\ref{fig:PDGsummary} \& \ref{fig:PD_wages}. 
But there is no reason to believe that on these long periods all 
the micro-rules and their associated parameters are stable in time. Therefore, 
studying
the {\it stationary state} of macroeconomic ABMs might be completely irrelevant 
to understand the real economy. The economy could be in a 
perpetual transient state (aka ``non ergodic''), unless one is able to 
endogenise the time evolution of all the relevant parameters governing its 
evolution 
(see the conclusion of \cite{Roventini} for a related discussion).

If this is the case, is there any use in studying ABMs at all? We strongly 
believe that ABMs would still be genuinely helpful. 
ABMs allow us to experiment and scan the universe of possible outcomes -- 
not missing important scenarios is already very good macroeconomics. Human 
imagination turns out to be very limited, and that is the reason we like models 
and equations, that help 
us guessing what {\it can} happen, especially in the presence of collective 
effects that are often very counterintuitive. In this respect, ABMs provide 
extremely valuable tools 
for generating scenarios, that can be used to test the effect of policy 
decisions (see e.g. the pleas by Buchanan \cite{Buchanan}, and Farmer \& Foley 
\cite{Foley}). 
In order to become more quantitative, we think that ABMs will have to be 
calibrated at the 
micro-level, using detailed behavioural experiments and field studies to fine 
tune all the micro-rules needed to build the economy from bottom up (see 
\cite{Hommes,Farmer}
for work in this direction.) Calibrating on historical data the emergent 
macro-dynamics of ABMs will most probably fail, because of the dimensionality 
curse 
and of the Lucas critique  (i.e. the feedback between the trajectory of the 
economy and policy decisions that dynamically change the parameters of the 
model).

\section*{Acknowledgements} This work was partially financed by the CRISIS 
project. We want to thank all the members of CRISIS for 
most useful discussions, in particular during the CRISIS meetings. The input 
and comments of P. Aliferis, T. Assenza, J. Batista, E. Beinhocker, M. Buchanan, D. Challet, 
D. Delli Gatti,
D. Farmer, J. Grazzini, C. Hommes, F. Lillo, G. Tedeschi, S. Battiston, A. 
Kirman, A. Mandel, M. Marsili and A. Roventini are warmly acknowledged. 
We are especially grateful to F. Altarelli and G. Cencetti for a careful 
critical reading of the manuscript and for their suggestions. The comments and
criticisms of the referees were also useful to improve the manuscript. 
Finally, J.P.B. wants to thank J.C. Trichet for his warm support for this type 
of research when he was president of the ECB. 

\clearpage

\appendix

\section{Pseudo-code for Mark I+}
\label{app:Mark1+}

We describe here the pseudo-code of our version of Mark I, which we call Mark 
I+. To keep the length reasonable, a few irrelevant details will be omitted,
but the information given here is enough to reproduce the results presented in 
the paper.
In particular, we describe here only the part of the code that is needed to 
generate the dynamical evolution of the model,
and we omit the part that is needed to generate the data output. The source 
code is available on the site of the CRISIS project (www.crisis-economics.eu).

\subsection{Notations}

We describe the algorithm in an object-oriented fashion, where the different 
agents are described as representatives of a few classes. 
We use an object oriented syntax. This syntax should be very intuitive and easy 
to follow.
However, it is useful to clarify a few notational conventions:
\begin{itemize}
\item
The declaration of a variable $a$ (for example of integer type {\tt int}) will 
be written in C syntax as {\tt int} $a$. 
\item
If $a$ is an object of some class, then $a.f(x)$ means that we are calling the 
method (function) $f(x)$ of object $a$ with argument $x$. 
\item
For simplicity, in the {\bf for} loops we will use the C syntax where, for 
example,
{\bf for}($t \leftarrow 1; t \leq T;t\leftarrow t+1$) means that $t$ is set to 
one before the loop starts, $t$ is increased by 1 at the end of each iteration,
and the loop continues if the condition $t\leq T$ is true.
\item
Instead of arrays we will use {\it vectors} of objects, and we will follow the 
notation of C++\footnote{See for example 
\url{http://www.cplusplus.com/reference/vector/vector/}}. 
For example, {\tt vector$<$int$>$} will denote an ordered set (array) of 
integers. Moreover:
\begin{itemize}
\item A vector of size $N$ will be declared as {\tt vector$<$int$>$} $A$($N$), 
and by default the declaration {\tt vector$<$int$>$} $A$ means that the set $A$ 
is initiated as empty.
\item $|A|$ will denote the size of the set $A$
\item The notation $A$[$i$] will denote the $i$-th element of the set (with 
$i=0, \cdots, |A|-1$). 
\item The notation $A \leftarrow a$ will denote the operation of adding the 
element $a$ to
the set $A$, therefore increasing its size by one (this correspond to the 
``push\_back'' operation in C++).
\item The notation $a \leftarrow_R A$ will denote the extraction of a random 
element from $A$, which is set equal to $a$. Note that the element is not 
removed from $A$ so the size
of $A$ remains constant. The notation $A' \leftarrow_{R,M} A$ denotes 
extraction at random of $M$ different elements from the set $A$, that constitute 
the set $A'$.
\item The notation $A \not\rightarrow a$ will denote the removal of element $a$ 
from set $A$, therefore reducing the size of $A$ by one.
\item
To denote an ordered iteration over an ordered set $A$, we will use a loop {\bf 
for}($a \in A$).
\item We define a function average($a.f$(), $a \in A$) that returns the average 
of the $|A|$ numbers $a[i].f()$, $i = 0,\cdots,|A|-1$
\end{itemize}
\item For logical operators, we use the C++ 
convention\footnote{\url{
http://en.wikipedia.org/wiki/Operators_in_C_and_C++#Logical_operators}} which 
should be quite transparent.
\end{itemize}

\subsection{Classes}

The classes are:
\begin{itemize}
\item {\it Firm}
\item{\it Household}
\item {\it Bank}
\end{itemize}
The role of the bank is very limited at this level and this class is mainly 
included for the purpose of future extension.
One household has a special role as it is the ``owner'' of the firms, which 
will pay dividends to this one household.
The main loop is described in Algorithm~\ref{Mark1+} below. The implementation 
of the firm class is in Algorithm~\ref{firm},
the household class in Algorithm~\ref{household} and the bank class in 
Algorithm~\ref{bank}.

%%%%%%%%%%%%%%%%%%%%%%%%%%%%%%%%%%%%%%MAIN%%%%%%%%%%%%%%%%%%%%%%%%%%%%%%%%%%
%%%%%%%%%%%%%%%%%%%%%%%%%%%%%%%%%%%%%%MAIN%%%%%%%%%%%%%%%%%%%%%%%%%%%%%%%%%%
%%%%%%%%%%%%%%%%%%%%%%%%%%%%%%%%%%%%%%MAIN%%%%%%%%%%%%%%%%%%%%%%%%%%%%%%%%%%
%%%%%%%%%%%%%%%%%%%%%%%%%%%%%%%%%%%%%%MAIN%%%%%%%%%%%%%%%%%%%%%%%%%%%%%%%%%%

\begin{algorithm}[h]
\caption{Main loop of Mark I+}         
\label{Mark1+}                          
\begin{algorithmic} 
\Require$N_{\rm F}$ Number of firms; $N_{\rm H}$ Number of households; $\r_0$ 
baseline interest rate; $T$ total evolution time;
\State
\State 
\Comment{ {\bf\color{red} Initialisation}}
\State {\tt vector$<$household$>$} $H$($N_{\rm H}$)
\State {\tt household} $O$
\Comment{ The owner of all firms}
\State {\tt vector$<$firm$>$} $F$($N_{\rm F}$)
\State {\tt bank} $B$
\State
\State $\overline p \leftarrow 1$
\Comment{Average price}
\For {$f \in F$}
\State $f$.set\_owner($O$)
\EndFor
\State $B$.set$\_\r$($\r_0$)
\State
\Comment{ {\bf\color{red} Main loop}}
\For {($t \leftarrow 1; t \leq T;t\leftarrow t+1$)}  %% main loop starts
\State
\State {\tt vector$<$firm$>$} $E$,$D$
\For {$f \in F$}
\Comment{ {\bf \color{blue} Firm decide new strategy on prices and production} }
\State $f$.set\_new\_strategy($\overline p$)
\State $f$.get\_loans($B$)
\State $f$.compute\_interests()
\State $ f$.define\_labor\_demand()
\If { $f$.n\_vacancies() $>$ 0 }
$D \leftarrow f$
\Comment{Firms in $D$ demand workforce}
\ElsIf { $f$.n\_vacancies() $<$ 0 }
$E \leftarrow f$
\Comment{Firms in $E$ have an excess of workforce}
\EndIf
\EndFor
\State
\For{$f \in E$}
\Comment{ {\bf \color{blue} Job market and production} }
\While{$f$.n\_vacancies() $<$ 0 }
$f$.fire\_random\_worker()
\Comment{Firms with excess workforce fire random workers}
\EndWhile
\EndFor
\State {\tt vector$<$household$>$} $U$
\For { $h\in H$ }
\If { !$h$.working() } $U \leftarrow h$
\Comment{$U$ is the set of unemployed households}
\EndIf
\EndFor
\While { $|U|>0$ \&\& $|D|>0$ }
\Comment{Random match of unemployed households and demanding firms}
\State $h \leftarrow_R U$
\State $f \leftarrow_R D$
\State $f$.hire($h$)
\State $U \not\rightarrow h$
\If {$f$.n\_vacancies()==0} $D \not\rightarrow f$
\EndIf
\EndWhile

%pagebreak
      \algstore{myalg}
  \end{algorithmic}
\end{algorithm}

\clearpage

\begin{algorithm}
  \caption{ Main loop of Mark I+ (continued)}
  \begin{algorithmic}
      \algrestore{myalg}
%endpagebreak
      
%
\For {$f\in F$}
\Comment{Firms produce and pay workers}
\State  $f$.produce()
\State   $f$.pay\_workers()
\If  {$f$.age()$<$100} f.markup\_rule()
\Comment{Young firms apply a markup rule to avoid bankrupt}
\EndIf
\EndFor
\State
\Comment{{\bf \color{blue} Goods market}}
\State $H \leftarrow $random\_permutation($H$)
\For { $h \in H$ }
$h$.consume($F$)
\Comment{Consume in random order}
\EndFor      
\State
\Comment{{\bf \color{blue} Accounting and bankrupts}}
\State bad\_debts $\leftarrow 0$
\State {\tt vector$<$firm$>$} $L$
\For { $f\in F$ }
\State  $f$.accounting($B$)
\If { $f$.liquidity() $<$ 0 }
\Comment{Firms with negative liquidity go bankrupt} 
\State bad\_debts $\leftarrow$ bad\_debts + $f$.liquidity()
\Comment{Note: bad\_debts is negative!}
\State $\overline{L} \leftarrow f$
\Comment{$\overline{L}$ is the sent of bankrupt firms}
\Else \hskip5pt$L \leftarrow f$
\Comment{$L$ is the set of healthy firms}
\EndIf
\EndFor
\If { $|L|==0$   } {\tt break}
\Comment{If all firms are bankrupt, exit the program}
\EndIf
\State $\overline{p}_b \leftarrow $ average($f[i]$.price(), $f\in L$)
\State $\overline{Y}^T \leftarrow $ average($f[i]$.target\_production(), $f\in 
L$)
\State $\overline{Y} \leftarrow $ average ($f[i]$.production(), $f\in L$)
\For  { $f\in \overline{L}$ } 
$f$.reinit($\overline{p}_b,\overline{Y}^T,\overline{Y}$)
\Comment{Bankrupt firms are reinitialized with the average parameters of 
healthy firms}
\EndFor
\State total\_liquidity $\leftarrow \sum_{i=0}^{N_{\rm F}} f[i]$.equity() $+ 
\sum_{i=0}^{N_{\rm H}} h[i]$.wealth()
\For { $f\in F$ } $f$.get\_money(bad\_debts * $f[i]$.equity() / 
total\_liquidity)
\EndFor
\For { $h \in H$ } $h$.get\_money(bad\_debts * $h[i]$.wealth() / 
total\_liquidity)
\EndFor
\Comment{Bad debt is spread over firms and households proportionally to their 
wealth}
\State $\overline p \leftarrow \frac{\sum_{i=0}^{|F|-1} f[i].\text{price}() \,  
f[i].\text{sales}() }{\sum_{i=0}^{|F|-1}   f[i].\text{sales}()} $
\Comment{Update average price}
\EndFor    %% this is the main loop
\State
\end{algorithmic}
\end{algorithm}

%%%%%%%%%%%%%%%%%%%%%%%%%%%%%%%%%%%%%%%%%%%%FIRM%%%%%%%%%%%%%%%%%%%%%%%%%%%%%%%%
%%%%%
%%%%%%%%%%%%%%%%%%%%%%%%%%%%%%%%%%%%%%%%%%%%FIRM%%%%%%%%%%%%%%%%%%%%%%%%%%%%%%%%
%%%%%
%%%%%%%%%%%%%%%%%%%%%%%%%%%%%%%%%%%%%%%%%%%%FIRM%%%%%%%%%%%%%%%%%%%%%%%%%%%%%%%%
%%%%%
%%%%%%%%%%%%%%%%%%%%%%%%%%%%%%%%%%%%%%%%%%%%FIRM%%%%%%%%%%%%%%%%%%%%%%%%%%%%%%%%
%%%%%

\begin{algorithm}
\caption{The class {\tt firm}}         
\label{firm}                          
\renewcommand{\algorithmicrequire}{\textbf{Dynamic variables}}
\renewcommand{\algorithmicensure}{\textbf{Parameters}}
\begin{algorithmic}
\Ensure: $W=1$, $\a=1$, $\g_p=0.1$, $\g_y=0.1$, $\mu=0$, $\d=0.2$, $\t=0.05$
\Require: {\tt vector$<$household$>$} $E$; {\tt household} $O; p, Y, Y^T, D, 
\LL, v, \DD^T, t$
\Require {\bf (auxiliary)}: $L_d, \r, \II$ 
\State 
\Comment{{\bf \color{blue} Initialization methods}}
\Function{init}{}
\State $E \leftarrow $empty
\Comment{The set $E$ is the list of employees and is initialized as empty}
\State $p \leftarrow 1$
\Comment{Price}
\State $Y \leftarrow 1$
\Comment{Production}
\State $Y^T \leftarrow 1$
\Comment{Target production}
\State $D \leftarrow 1$
\Comment{Demand}
\State $\LL \leftarrow 50$
\Comment{Liquidity}
\State $v \leftarrow 0$
\Comment{Number of vacancies}
\State $\DD^T \leftarrow 0$
\Comment{Total debt}
\State $t \leftarrow 0$
\Comment{Internal clock}
\EndFunction
\State 
\Function{set\_owner}{{\tt household $\wt O$}}
\State $O \leftarrow \wt O$
\EndFunction
\State 
\Function{reinit}{$\wt p, \wt Y^T, \wt Y$}
\State $p\leftarrow \wt p$
\State $Y\leftarrow \wt Y$
\State $Y^T \leftarrow \wt Y^T$
\State $D \leftarrow 0$
\State $\LL \leftarrow \min\{ O$.wealth(), $Y/\a$\}
\Comment{The owner injects money to restart the bankrupt firm}
\State $O$.get\_money(-$\LL$)
\State $v \leftarrow 0$
\State $\DD^T \leftarrow 0$
\State $t \leftarrow 0$
\State $v \leftarrow 0$
\For {$h\in E$} 
\State fire($h$)
\EndFor
\EndFunction
\State 
\Comment{{\bf \color{blue} Output methods}}
\Function{price}{}
\State \Return $p$
\EndFunction
\State 
\Function{production}{}
\State \Return $Y$
\EndFunction
\State 
\Function{stock}{}
\State \Return $Y-D$
\EndFunction
\State 
\Function{sales}{}
\State \Return $D$
\EndFunction
\State 
\Function{target\_production}{}
\State \Return $Y^T$
\EndFunction
\State 
\Function{equity}{}
\State \Return $\LL - \DD^T$
\EndFunction
\State 
\Function{liquidity}{}
\State \Return $\LL$
\EndFunction
\State 
\Function{n\_vacancies}{}
\State \Return $v$
\EndFunction
%

%pagebreak
      \algstore{myalgfirm}
  \end{algorithmic}
\end{algorithm}

\clearpage

\begin{algorithm}
  \caption{ The class {\tt firm} (continued)}
  \begin{algorithmic}
      \algrestore{myalgfirm}
%endpagebreak

\State 
\Function{age}{}
\State \Return $t$
\EndFunction
\State 
\Comment{{\bf \color{blue} Accounting methods}}
\Function{get\_money}{$\wt m$}
\State $\LL \leftarrow \LL + \wt m$
\EndFunction
\State 
\Function{get\_loans}{{\tt bank} $\wt B$}
\State $\LL_n \leftarrow W L_d - \LL$
\Comment Financial need
\If {$\LL_n > 0$}
\State $\ell \leftarrow (\DD^T + \LL_n)/(\LL + 0.001)$
\Comment{This is the leverage}
\State $\r_{\rm offer} \leftarrow \wt B$.compute\_offer\_rate($\ell$)
\Comment{New offered interest rate}
\State $\DD^c \leftarrow \LL_n \, F(\r_{\rm offer})$
\Comment{The function $F(\r)$ can be whatever decreasing function of $\r$, see 
Eq.~\eqref{creditcontraction}}
\If {$\DD^c > 0$}
\State $\r \leftarrow \r_{\rm offer}$
\Comment{If new credit is contracted, the interest rate is updated}
\State $\DD^T \leftarrow \DD^T + \DD^c$
\Comment{Total debt is increased by current debt $\DD^c$}
\State $\LL \leftarrow \LL + \DD^c$
\State $\wt B$.get\_money(-$\DD^c$)
\EndIf
\EndIf
\EndFunction
\State 
\Function{compute\_interests}{}
\State $\II \leftarrow \r \DD^T$
\Comment{Interests to be paid in this round}
\EndFunction
\State 
\Function{pay\_workers}{}
\For {$h \in E$}
\State $h$.get\_money($W$)
\EndFor
\State $\LL \leftarrow \LL - W |E|$
\EndFunction
\State 
\Function{markup\_rule}{}
\If { $Y>0$ }
\State $p_{\rm markup} \leftarrow (1+\mu) (W |E| + \II )/Y$
\State $p \leftarrow \max\{ p , p_{\rm markup} \}$
\EndIf
\EndFunction
\State 
\Function{accounting}{{\tt bank} $\wt B$}
\State $\LL \leftarrow \LL  - \II -\t \DD^T$
\Comment{Firm pays interests and repays a fraction $\t$ of its debt}
\State $\wt B$.get\_money($\II + \t \DD^T$)
\State $\DD^T \leftarrow (1-\t) \DD^T $
\State $\PP \leftarrow p D - W |E| - \II$
\Comment{Profit}
\If {$\PP > 0$}
\State $O$.get\_money($\d \PP$)
\Comment{Firm pays dividends to the owner}
\State $\LL \leftarrow \LL - \d\PP$
\EndIf
\EndFunction
\State 
\Function{sell}{$\wt q$}
\State $D \leftarrow D + \wt q$
\State $\LL \leftarrow \LL + p \wt q$
\EndFunction

%pagebreak
      \algstore{myalgfirm}
  \end{algorithmic}
\end{algorithm}

\clearpage

\begin{algorithm}
  \caption{ The class {\tt firm} (continued)}
  \begin{algorithmic}
      \algrestore{myalgfirm}
%endpagebreak

%
\State
\State 
\Comment{{\bf \color{blue} Production and job market methods}}
\Function{set\_new\_strategy}{$\wt p$}
\State $t \leftarrow t+1$
\If { $Y=D$ \&\&  $p \geq \wt p$ } $Y^T \leftarrow Y ( 1 + \g_y \,${\tt random})
\Comment{This is Eq.~\eqref{update} in the main text}
\ElsIf{ $Y=D$ \&\&  $p < \wt p$} $p \leftarrow p ( 1 + \g_p \,${\tt random})
\ElsIf{ $Y > D$ \&\&  $p \geq \wt p$ } $p \leftarrow p ( 1 - \g_p \,${\tt 
random})
\ElsIf{ $Y > D$ \&\&  $p < \wt p$ } $Y^T \leftarrow Y ( 1 - \g_y \,${\tt 
random})
\EndIf
\State $Y^T \leftarrow \max\{ Y^T, \a \}$
\State $L_d \leftarrow $ceil$(Y^T/\a)$
\EndFunction
\State 
\Function{define\_labor\_demand}{}
\State $L_d \leftarrow \min\{ L_d, \text{floor}(\LL/W) \}$
\State $L_d \leftarrow \max\{ L_d, 0 \}$
\State $v \leftarrow L_d - |E|$
\EndFunction
\State 
\Function{produce}{}
\State $Y = \min\{ Y^T, \a |E| \}$
\State $D=0$
\Comment{The demand is reset to zero at each production cycle}
\EndFunction
\State 
\Function{hire}{{\tt household $h$}}
\State $E \leftarrow h$
\State $h$.get\_job($W$)
\State $v \leftarrow v-1$
\EndFunction
\State 
\Function{fire}{{\tt household $h$}}
\State $h$.lose\_job()
\State $E \not\rightarrow h$
\State $v \leftarrow v+1$
\EndFunction
\State 
\Function{fire\_random\_worker}{}
\If { $|E| > 0$ }
\State $h \leftarrow_R E$
\State fire($h$)
\EndIf
\EndFunction

\end{algorithmic}
\end{algorithm}

%%%%%%%%%%%%%%%%%%%%%%%%%%%%%%%%%%%%HOUSEHOLD%%%%%%%%%%%%%%%%%%%%%%%%%%%%%%%%%%%
%%%%
%%%%%%%%%%%%%%%%%%%%%%%%%%%%%%%%%%%%HOUSEHOLD%%%%%%%%%%%%%%%%%%%%%%%%%%%%%%%%%%%
%%%%
%%%%%%%%%%%%%%%%%%%%%%%%%%%%%%%%%%%%HOUSEHOLD%%%%%%%%%%%%%%%%%%%%%%%%%%%%%%%%%%%
%%%%
%%%%%%%%%%%%%%%%%%%%%%%%%%%%%%%%%%%%HOUSEHOLD%%%%%%%%%%%%%%%%%%%%%%%%%%%%%%%%%%%
%%%%

\begin{algorithm}[h]
\caption{The class {\tt household}}         
\label{household}                          
\renewcommand{\algorithmicrequire}{\textbf{Dynamic variables}}
\renewcommand{\algorithmicensure}{\textbf{Parameters}}
\begin{algorithmic}
\Ensure: $M=3$, $c=0.8$
\Require: $S$, $W$
%\Require {\bf (auxiliary)}: budget
%
\State 
\Comment{{\bf \color{blue} Initialization methods}}
\Function{init}{}
\State $S \leftarrow 0$
\Comment{Savings}
\State $W \leftarrow 0$
\Comment{Salary}
\EndFunction
\State 
\Comment{{\bf \color{blue} Accounting methods}}
\Function{get\_money}{$\wt m$}
\State $S \leftarrow S + \wt m$
\EndFunction
\State 
\Comment{ {\bf \color{blue} Output methods}}
\Function{wealth}{}
\State \Return $S$ 
\EndFunction
\State 
\Function{working}{}
\If { $W>0$}
 \Return True 
\Else \hskip3pt
 \Return False
 \EndIf
\EndFunction
\State 
\Comment{ {\bf \color{blue} Job and goods market methods}}
\Function{get\_job}{$\wt W$}
\State $W \leftarrow \wt W$
\EndFunction
\State 
\Function{lose\_job}{}
\State $W \leftarrow 0$
\EndFunction
\State 
\Function{consume}{ {\tt vector$<$firm$>$ $\wt F$} }
\State budget $\leftarrow c S$
\If {budget $>0$ }
\State $F_c \leftarrow_{R,M} \wt F$
\Comment{Extract $M$ random firms from $\wt F$ and put them in the set $F_c$}
\State $F_c \leftarrow$ sort($f\in F_c$, $f$.price())
\Comment{Order the set $F_c$ according to firms' prices}
\State spent $\leftarrow 0$
\For {($i \leftarrow 0;  i < |F_c| \ \&\& \ \text{spent} < \text{budget}   ; 
i\leftarrow i+1$) }
\State $s \leftarrow f[i]$.stock() 
\Comment{ $s$ is the stock available from this firm}
\If { $s> 0$ }
\State  $q \leftarrow$ (budget - spent)$/f[i]$.price()
\Comment{Maximum possible consumption from this firm}
\If {  $s > q$ }
\State $f[i]$.sell($q$)
\State spent $\leftarrow$ budget
\Comment{The household has finished the budget, the loop ends}
\Else
\State $f[i]$.sell($s$)
\State spent $\leftarrow$ spent $+ \, s \, f[i].$price()
\EndIf
\EndIf
\EndFor
\EndIf
\State $S \leftarrow S - $spent
\EndFunction

\end{algorithmic}
\end{algorithm}

%%%%%%%%%%%%%%%%%%%%%%%%%%%%%%%%%%%%BANK%%%%%%%%%%%%%%%%%%%%%%%%%%%%%%%%%%%%%%%
%%%%%%%%%%%%%%%%%%%%%%%%%%%%%%%%%%%%BANK%%%%%%%%%%%%%%%%%%%%%%%%%%%%%%%%%%%%%%%
%%%%%%%%%%%%%%%%%%%%%%%%%%%%%%%%%%%%BANK%%%%%%%%%%%%%%%%%%%%%%%%%%%%%%%%%%%%%%%
%%%%%%%%%%%%%%%%%%%%%%%%%%%%%%%%%%%%BANK%%%%%%%%%%%%%%%%%%%%%%%%%%%%%%%%%%%%%%%

\begin{algorithm}[h]
\caption{The class {\tt bank}} 
\label{bank}                          
\renewcommand{\algorithmicrequire}{\textbf{Dynamic variables}}
\renewcommand{\algorithmicensure}{\textbf{Parameters}}
\begin{algorithmic}
\Require: $E$, $\r_b$
%\Require {\bf (auxiliary)}: budget

\State 
\Function{init}{}
\State $E \leftarrow 0$
\Comment{Bank liquidity}
\State $\r_b \leftarrow 0$
\Comment{Baseline interest rate}
\EndFunction
\State 
\Function{set\_$\r$}{$\wt \r$}
\State $\r_b \leftarrow \wt\r$
\EndFunction
\State 
\Function{compute\_offer\_rate}{$\wt\ell$}
\State \Return $\r_b  \, G(\wt\ell) $
\Comment{We chose $G(\ell) = 1 + \log(1+\ell)$} 
\EndFunction
\State 
\Function{get\_money}{$\wt m$}
\State $ E \leftarrow E + \wt m$
\EndFunction

\end{algorithmic}
\end{algorithm}

\clearpage

\section{Pseudo-code of Mark 0}
\label{app:Mark0}

We present here the pseudo-code for the Mark 0 model discussed in 
Sec.~\ref{sec:Mark0_desc} and in Sec.~\ref{section:extensions}. 
The source code is available on the site of the CRISIS project 
(www.crisis-economics.eu).

\begin{algorithm}[h]
\caption{The basic Mark 0}         
\label{alg:Mark0}                          
\begin{algorithmic} 
\Require$N_{\rm F}$ Number of firms; 
$\mu,c,\beta,\gamma_p,\eta^0_+,\eta^0_-,\delta,\Theta,\varphi,f$; $T$ total 
evolution time; 
\State
\State
\Comment{ {\bf\color{red} Initialization}}
\For {($i \leftarrow 0; i< N_{\rm F} ;i\leftarrow i+1$)}
\State $W[i] \leftarrow 1$
\Comment{Salaries are always fixed to one}
\State $p[i] \leftarrow 1+0.2 (\text{\tt random} - 0.5) $
\State $Y[i] \leftarrow \mu [1 +0.2 (\text{\tt random} - 0.5)] /2$
\Comment{Initial employment is $0.5$}
\State $\EE[i] \leftarrow W[i] Y[i] \, 2 \, \text{\tt random}$
\State $a[i] \leftarrow 1$
\EndFor
\State
\State $S\leftarrow N_{\rm F} - \sum_i \EE[i]$
\State
\Comment{ {\bf\color{red} Main loop}}
\For {($t \leftarrow 1; t \leq T;t\leftarrow t+1$)}  %% main loop starts
%\State
\State $u \leftarrow 1-\frac1{\mu N_{\rm F}} \sum_i Y[i]$
%\State
\State $\varepsilon\leftarrow 1-u$
%\State
\State $\overline p \leftarrow \frac{\sum_i p[i] Y[i]}{\sum_i Y[i]}$ 
%\State
\State $\overline w \leftarrow \frac{\sum_i W[i] Y[i]}{\sum_i Y[i]}$ 
%\State
\State $\tilde{u}[i] \leftarrow \frac{\exp(\b W[i]/\overline w)}{\sum_i 
a[i]\exp(\b W[i]/\overline w)}N_{\rm F}u $
\State
\Comment{{\bf \color{blue} Firms update prices, productions and wages}}
\For {($i \leftarrow 0; i< N_{\rm F} ;i\leftarrow i+1$)}
\State
\If { $a[i]==1$ }
\State
\If { $Y[i] < D[i]$ }
%\State
\Comment{Wage update}
\If {$\PP[i]>0$}
\State $W[i]\leftarrow W[i][1+\g_w\varepsilon$ {\tt random}$]$
\State $W[i]\leftarrow \min{\{W[i],P[i]\min{[D[i],Y[i]]}/Y[i] \}}$
\EndIf
%\State
\Comment{This is Eq.~\eqref{update0} in the main text}
\State $Y[i] \leftarrow Y[i] + \min\{ \h_+  (D[i] - Y[i] ), \mu \tilde{u}[i] \}$
%\State
\If{ $p[i] < \overline p$} $p[i] \leftarrow p[i] ( 1 + \g_p \,${\tt random})
\EndIf
\State
\ElsIf{ $Y[i] > D[i]$ }
%\State
\If {$\PP[i]<0$}
\State $W[i]\leftarrow W[i][1-\g_wu$ {\tt random}$]$
\EndIf
%\State
\State $Y[i] \leftarrow \max\{ 0, Y[i] - \h_- ( D[i] - Y[i] ) \}$
%\State
\If{ $p[i] < \overline p$ } $p[i] \leftarrow p[i] ( 1 - \g_p \,${\tt random})
\EndIf
\EndIf
\EndIf
\EndFor
\State
\State $u \leftarrow 1-\frac1{\mu N_{\rm F}} \sum_i Y[i]$
\Comment{Update $u$ and $\overline p$}
\State $\overline p \leftarrow \frac{\sum_i p[i] Y[i]}{\sum_i Y[i]}$ 
%
%
%
%pagebreak
      \algstore{Mark0}
  \end{algorithmic}
\end{algorithm}

\clearpage

\begin{algorithm}
  \caption{ The basic Mark0 (continued)}
  \begin{algorithmic}
      \algrestore{Mark0}
%endpagebreak
%
%
%
\State
\Comment{{\bf \color{blue} Households decide the demand}}
\State $C \leftarrow c (\max\{S,0\} + \sum_i W[i] Y[i])$
\State
\For {($i \leftarrow 0; i< N_{\rm F} ;i\leftarrow i+1$)}
\State $D[i] \leftarrow \frac{C a[i] \exp(-\b p[i]/\overline p)}{p[i] \sum_i 
a[i]\exp(-\b p[i]/\overline p)} $
\Comment{Inactive firms have no demand}
\EndFor
\State
\Comment{{\bf \color{blue} Accounting}}
\For {($i \leftarrow 0; i< N_{\rm F} ;i\leftarrow i+1$)}
\If { $a[i]==1$ }
\State $\PP[i] \leftarrow p[i] \min\{ Y[i], D[i] \} - W[i] Y[i]$
\State $S \leftarrow S - \PP[i]$
\State $\EE[i] \leftarrow \EE[i] + \PP[i]$
\If { $\PP[i] > 0$ \&\& $\EE[i]>0$ }
\Comment{Pay dividends}
\State $S \leftarrow S + \d \, \PP[i]$
\State $\EE[i] \leftarrow \EE[i] - \d \, \PP[i]$
\EndIf
\If { $\EE[i]>\Theta W[i]Y[i]$ }
\Comment{Set of healthy firms}
\State $\mathcal{H}\leftarrow i$
\EndIf
\EndIf
\EndFor
\State
\Comment{{\bf \color{blue} Defaults}}
\State deficit $= 0$
\For {($i \leftarrow 0; i< N_{\rm F} ;i\leftarrow i+1$)}
\If { $a[i]==1$ \&\& $\EE[i] < -\Th Y[i] W[i]$ }
\State $j \leftarrow_R \mathcal{H}$
\If { {\tt random }$< 1-f$ \&\& $\EE[j]>-\EE[i]$}
\Comment{Bailed out}
\State $\EE[j]\leftarrow\EE[j]+\EE[i]$
\State $\EE[i]\leftarrow 0$
\State $p[i] \leftarrow p[j]$
\State $W[i] \leftarrow W[j]$
\Else
\Comment{Bankrupted}
\State deficit $\leftarrow$ deficit $-\EE[i] $
\State $a[i] \leftarrow 0$
\State $Y[i] \leftarrow 0$
\State $\EE[i] \leftarrow 0$
\EndIf
\EndIf
\EndFor
\State
\Comment{{\bf \color{blue} Revivals}}
\State $\EE^+\leftarrow0$
\For {($i \leftarrow 0; i< N_{\rm F} ;i\leftarrow i+1$)}
\If { $a[i]==0$ \&\& {\tt random }$< \varphi$ }
\Comment{Reactivate firm}
\State $a[i] \leftarrow 1$
\State $p[i] \leftarrow \overline p$
\State $Y[i] \leftarrow \mu u \, \text{\tt random}$
\State $\EE[i] \leftarrow W[i] Y[i]$
\State deficit $\leftarrow$ deficit $+\EE[i] $
\EndIf
\If { $a[i]==1$ \&\& $\EE[i]> 0$ }
\Comment{Firms total savings}
\State $\EE^+ \leftarrow \EE^+ + \EE[i]$
\EndIf
\EndFor
\State
\Comment{{\bf \color{blue} Debt}}
\If {deficit $>S$}
\Comment{Households cannot be indebted}
\State deficit $\leftarrow $ deficit $-S$
\State $S\leftarrow 0$
\For {($i \leftarrow 0; i< N_{\rm F} ;i\leftarrow i+1$)}
\If { $a[i]==1$ \&\& $\EE[i]>0$ }
\State $\EE[i]\leftarrow\EE[i]-$ $\EE[i]/\EE^+$ deficit
\EndIf
\EndFor
\Else
\State $S\leftarrow S-$ deficit
\EndIf
\State
\EndFor  %%main loop is closed

\end{algorithmic}
\end{algorithm}

\clearpage

\section{Perturbative solution of the schematic model} 

\subsection{Stationary distribution at first order in $\gamma_p$}
\label{app:D1}

From Eq.~\eqref{zetaeq}, the stationary distribution $P_{\rm st}(\l)$ satisfies
\beq \label{zetaeq_app} 
P_{\rm st}(\l) =\int_0^1 d\xi\, \int_0^\infty  d\l' P_{\rm st}(\l')
\delta\left(\l-\l'+\xi+\frac{\gamma_p}{2} \xi^2\right)
         + \int_0^1 d\xi\, \int_{-\infty}^0  d\l' P_{\rm st}(\l')
\delta\left(\l-\l'-\xi+\frac{\gamma_p}{2} \xi^2\right) \ .
\eeq 
We recall that the dynamics of $\l_i(t)$ is
\be
\begin{cases}
\l_i(t)<0 & \l_i(t+1) = \l_i(t) + \xi_i(t) - \frac12 \gamma_p \xi_i(t)^2 \\
\l_i(t)>0 & \l_i(t+1) = \l_i(t) - \xi_i(t) - \frac12 \gamma_p \xi_i(t)^2 
\end{cases}
\ee
with $\xi_i(t)$ a random variable in $[0,1]$. From these equations 
it is clear that positive and negative $\l_i$ are pushed towards the origin, and
after some transient time one necessarily has
\be\label{support}
-1 -\frac12\gamma_p \leq \l_i(t) \leq 1 - \frac12\gamma_p \ .
\ee

We first set $\gamma_p=0$ and therefore restrict $\l \in [-1,1]$ in 
Eq.~\eqref{zetaeq_app}.
We obtain
\be 
P_{\rm st}(\l) = \int_{\max(0,-\l)}^{1-\max(\l,0)} {\rm d}\xi P_{\rm 
st}(\l+\xi) +
\int_{\max(0,\l)}^{1-\max(-\l,0)} {\rm d}\xi P_{\rm st}(\l-\xi)  = {\cal L}_0 
P_{\rm st} \ , 
\ee 
where we call ${\cal L}_0$ the linear operator that appears on the right hand
side. 
When one computes the action of ${\cal L}_0$ on the functions $g_n(\l)={\rm
sign}(\l) \l^n$ and $h_n(\l)= \l^n$, one finds, for $\l > 0$: 
\be 
(n+1) {\cal L}_0 g_n(\l)= 
\begin{cases}
 1 -  (-1)^n - \sum_{k=0}^{n-1} D_n^k (-1)^k \l^{n-k}   & \l > 0 \\
 1 -  (-1)^n  + \sum_{k=0}^{n-1} D_n^k \l^{n-k}   & \l < 0 
 \end{cases}
\ee 
with $C_n^k$ the binomial coefficients and
$D_n^k = \sum_{j=k}^n C_j^k$, and
\be 
(n+1) {\cal L}_0 h_n(\l)= 
\begin{cases}
1-\l^{n+1} - (\l - 1)^{n+1}  = 1 +
(-1)^n  - 2 \l^{n+1} - \sum_{k=1}^n C_{n+1}^k (-1)^k \l^{n+1-k} & \l>0 \ , \\
 (1+\l)^{n+1} + \l^{n+1} + (-1)^{n} 
=  1 + (-1)^n  + 2 \l^{n+1} + \sum_{k=1}^n C_{n+1}^k \l^{n+1-k} & \l<0 \ . 
\end{cases}
\ee 
Let us focus on small values of $n$, useful in the following: 
\begin{eqnarray}
n=0 &\to&  {\cal L}_0 g_0=0 \ , \qquad \qquad {\cal L}_0 h_0=2(h_0-g_1)  \ ,   
\nonumber \\
n=1 &\to&  {\cal L}_0 g_1=1 - g_1 \ , \qquad {\cal L}_0 h_1=-g_2 + h_1 \ ,   \\
n=2 &\to& {\cal L}_0 g_2=h_1 - g_2 \ ,  \qquad {\cal L}_0 
h_2=\frac23(h_0-g_3)+h_2-g_1 \ .  \nonumber
\end{eqnarray}
In particular, one has: 
\be 
{\cal L}_0 (h_0-g_1) = 2(h_0-g_1) - 1 + g_1 = (h_0-g_1). 
\ee 
This shows that $P_0(\l)=h_0(\l) - g_1(\l)= 1 - |\l|$ (called the ``tent'') is 
an
eigenvector of ${\cal L}_0$ with eigenvalue $1$, i.e. this is the stationary
state for $\gamma_p=0$. 

This basic solution allows us to obtain perturbatevely the stationary solution
for small $\gamma_p$. Recalling that the support of $P_{\rm st}(\l)$ is given 
by Eq.~\eqref{support},
Eq.~\eqref{zetaeq_app} can be written as
\be 
P_{\rm st}(\l) = \int_{a_-}^{a_+} {\rm d}\xi P_{\rm st}(\l+\xi+\frac12 \gamma_p 
\xi^2) +
\int_{b_-}^{b_+} {\rm d}\xi P_{\rm st}(\l-\xi+\frac12 \gamma_p \xi^2) = {\cal 
L}_{\g_p} P_{\rm st} \ , 
\ee 
with: 
\beq
\begin{split}
&a_-=\max\left\{ 0,-\l(1+\frac{\gamma_p}{2}\l) \right\} \ , \quad
a_+=\min\left\{ 1,1-\l - \frac{\gamma_p}{2}(1+(1-\l)^2)\right\}  \ , \\
&b_-=\max\left\{ 0,\l(1+\frac{\gamma_p}{2}\l) \right\} \ , \quad
b_+=\min\left\{ 1,1+\l+\frac{\gamma_p}{2}(1 + (1+\l)^2) \right\} \ ,
\end{split}
\ee 
these integration bounds coming from the combination of the support in 
Eq.~\eqref{support}
and the integration bounds on $\l'$ in Eq.~\eqref{zetaeq_app}.
Writing $P_{\rm st}=P_0 + \gamma_p P_1$ where $P_0$ is the tent solution,
and $\LL_{\g_p} = \LL_0 + \g_p \LL_1$, we obtain at first order in $\g_p$ an
equation of the form: 
\be 
(1 - {\cal L}_0) P_1 = \LL_1 P_0 = S \ . 
\ee 
The source term $S$ can be most easily computed by computing $\LL_{\g_p} P_0$ 
and
expanding the result at first order in $\g_p$.
We find $S = \l/2 - \text{sign}(\l) \l^2 = \frac12 h_1 - g_2$. This is nice 
because one
can look for a solution involving only $n=1$ and $n=2$ that closes the 
equation: 
\be 
P_1 = \alpha h_1 + \beta g_2 
\ee 
Using: ${\cal L}_0 g_2 = h_1 - g_2 = {\cal L}_0 h_1$, one finds: 
\be 
(1 - {\cal L}_0) P_1 = \alpha h_1 + \beta g_2 - (\alpha + \beta) (h_1 - g_2) = -
\beta h_1 + (\alpha+2\beta) g_2 = S =  \frac12 h_1 - g_2 \ , 
\ee 
so the result for the coefficients is
\be 
\beta = -\frac12 \ , \qquad \alpha = 0 \ .
\ee 
The final solution for the stationary state to first order in $\gamma$ is: 
\be 
P =  1 - |\l| -\frac12 \gamma_p \,{\rm sign}(\l) \l^2. 
\ee 
Note that $P$ is still normalized, as it should and goes to zero (to first 
order in $\gamma_p$) at the
boundary of the support interval given in Eq.~\eqref{support}.

\subsection{Perturbative analysis of the oscillations}
\label{app:D2}

In order to understand oscillations, we start from Eqs. (\ref{updatedelta}) and
in the following we assume that $0< C < 1$.
We introduce $\EE(t) = \frac{ \overline{\a}(t) - C \overline{\l}(t)}{2 (1-C)}$, 
$\L(t) = \min\{ \overline{\l}(t), \EE(t) \}$ and $\Omega(t) =  \max\{ 
\overline{\l}(t),  \EE(t) \}$, 
and we write Eq.~\eqref{updatedelta} equivalently as
\beq\label{updatedelta2_app}
\begin{split}
&\a_i(t+1) = \a_i(t) - \min\{ \l_i(t),  C \l_i(t)  +   \frac12 
(\overline{\a}(t) - C \overline{\l}(t))   \} \\
&    \begin{cases}
&  \text{If } \l_i(t) < \L(t) \hskip42pt \Rightarrow \hskip10pt   \l_i(t+1) = 
\l_i(t) + \xi_i(t)  \\
&  \text{If } \Omega(t) < \l_i(t) < \L(t) \hskip10pt \Rightarrow \hskip10pt   
\l_i(t+1) = \l_i(t)  \\
&  \text{If } \l_i(t) > \Omega(t) \hskip42pt \Rightarrow \hskip10pt   \l_i(t+1) 
= \l_i(t) - \xi_i(t)  \\
\end{cases}
\end{split}
\eeq
The master equation for the distribution of $\l$ reads:
\beq\label{eqU1}
\begin{split}
P_{t+1}(\l') &= \int_{-\io}^{\L(t) } d\l  \int_0^1 d\xi \, P_t(\l) \, \d(\l'-\l 
-\xi) + \int_{\L(t)}^{\Omega(t)} d\l \, P_t(\l) \d(\l - \l') 
+ \int^{\io}_{\Omega(t)} d\l  \int_0^1 d\xi \, P_t(\l) \, \d(\l'-\l +\xi) \\
&= \int_{-\io}^{\L(t) } d\l  \int_0^1 d\xi \, P_t(\l) \, \d(\l'-\l -\xi) + \th( 
\L(t) \leq \l' \leq \Omega(t) ) P_t(\l')  
+ \int^{\io}_{\Omega(t)} d\l  \int_0^1 d\xi \, P_t(\l) \, \d(\l'-\l +\xi)
 \ .
\end{split}\eeq
while the evolution equation for $\overline{\a}$ is
\beq\label{baA}
\overline{\a}(t+1) = \overline{\a}(t) - \int_{-\io}^\io d\l \, P_t(\l) \min\{ 
\l,  C \l  +   \frac12 (\overline{\a}(t) - C \overline{\l}(t))   \} \ .
\eeq

\subsubsection{Stationary state}

Numerically, we observe that $\L(t) \sim \Omega(t)$ and their variations are 
much smaller than the width of the distributions of $\l$ and $\a$.
In the limit where
\beq\label{tuttiuguali}
\L(t) = \Omega(t) = \overline{\l}(t) = \EE(t) = \frac{ \overline{\a}(t) - C 
\overline{\l}(t)}{2 (1-C)}
\eeq
we obtain
\beq\begin{split}
P_{t+1}(\l') &= \int_{-\io}^{\L(t) } d\l  \int_0^1 d\xi \, P_t(\l) \, \d(\l'-\l 
-\xi) + \int^{\io}_{\L(t)} d\l  \int_0^1 d\xi \, P_t(\l) \, \d(\l'-\l +\xi)  \ ,
\end{split}\eeq
whose stationary solution is $P(\l) = P_0[\l - \overline{\l}^*]$ where $P_0(x) 
= (1 - |x|) \th(1-|x|)$ is the tent function, with $\overline{\l}^*$
undetermined at this stage. Plugging this result in Eq.~\eqref{baA}, we have
\beq
\overline{\a}(t+1) = \overline{\a}(t)   - \overline{\l}^* + \frac{1-C}6 \ .
\eeq
The fixed point is therefore $\overline{\l}^* = \frac{1-C}6$ and from the 
condition \eqref{tuttiuguali} we get 
$\overline{\a}^* = (2-C)\overline{\l}^* = (2-C)(1-C)/6$.

\subsubsection{Oscillation around the stationary state -- a simple 
approximation}

In order to study the small oscillations we can make a very simple 
approximation, namely
that at each time $t$ we have $P_t(\l) = P_0[\l - \bar\l(t)]$.
If we inject this approximation in Eq.~\eqref{baA} we get, provided $\bar\l(t) 
- \bar\l^*$ is not too large,
that
\beq\begin{split}
\bar\a(t+1) & = \bar\a(t)  - \int_{-\io}^\io d\l \min\{\l,C\l + (1-C) \EE(t) \} 
P_0[\l-\bar\l(t)] \\
& =
\bar\a(t)  - \bar\l(t) - \frac{1-C}6 \left[ -1 + 3 A - 3 A^2 + A^3\sgn(A) 
\right]_{A = \EE(t) - \bar\l(t)}
\end{split}\eeq
Next, injecting this approximation in Eq.~\eqref{eqU1} we get
\beq\begin{split}
\bar\l(t+1) & =  \bar\l(t) + \frac12 \int_{-\io}^{\L(t) } d\l \, P_0[\l - 
\bar\l(t)] - \frac12 \int^{\io}_{\Omega(t)} d\l\, P_0[\l - \bar\l(t)] \\
% & =  \bar\l(t) + \frac12 \int_{-\io}^{\L(t) - \bar\l(t)} dx \, T[x] - \frac12 
\int^{\io}_{\Omega(t) - \bar\l(t)} dx \, T[x] \\
 & =  \frac12 \left[
 \L(t)  + \Omega(t) 
-\frac12 [  \L(t) - \bar\l(t) ]^2 \text{sgn}[ \L(t) - \bar\l(t)]
-\frac12 [ \Omega(t) - \bar\l(t) ]^2 \text{sgn} [ \Omega(t) - \bar\l(t)]
 \right] 
\end{split}\eeq
We therefore get a closed system of two equations for $\bar\l(t)$ and 
$\bar\a(t)$.
For small $C$, this system of equations converges quickly to the fixed point. 
However,
for $C$ sufficiently close to 1 ($C\sim 0.94$) it has a limit cycle of period 
2, followed by an exponential divergence.

\subsubsection{Oscillation around the stationary state -- perturbative 
computation}

Another strategy to characterize the oscillations is to look again for a 
perturbed solution around the stationary state, with:
${\ola}(t) = \frac{1-C}{6} + {\dla}(t)$ and $\overline{\a}(t)= (2-C) 
\overline{\l} + 2(1-C){\dlaT}(t) + C {\dla}(t)$.
The distribution of  $\lambda_i$ at time $t$ is now $P_t(\l)=P_0(\l - 
\frac{1-C}{6}) + \delta P_t (\l - \frac{1-C}6)$. 
The shifted variable $\l - \frac{1-C}{6}$ will be denoted $x$.
The aim is to write an evolution for $\dP_{t+1}$ after one time step. The 
${\cal L}_0$ operator is the same as before, as well as
the set of functions $g_n$ and $h_n$. We also introduce $\Delta(x)$ as a 
$\delta$-function slightly spread out, but of unit area (its precise width
is irrelevant if it is small enough). 

To lowest order, one has:
\beq
{\cal L}_0 \Delta \approx \frac12 h_0.
\eeq
We will also need, as found above: 
\beq
{\cal L}_0 h_0 = 2(h_0 - g_1),\qquad {\cal L}_0 g_0 = 0, \qquad {\cal L}_0 g_1 
= h_0 - g_1.
\eeq
We now introduce $d=|\dlaT - \dla|$ and $s = \dlaT + \dla$. Assuming $d,s \ll 
1$, 
one finds, to first order:
\beq
\dP_{t+1} = {\cal L}_0 \dP_t - \frac12 d_t h_0 + \frac12 s_t g_0 + d_t \Delta
\eeq
Note that the integral $[\dP]=\int_{-1}^1 dx \dP(x)$ is zero and conserved in 
time, as it should be (one has $[h_0]=2$, $[g_0]=0$ and $[\Delta]=1$).

The idea now is to expand $\dP$ in terms of the $h,g$ functions. It turns out 
that only three of them are needed, plus the $\Delta$ contribution.
Indeed, assume:
\beq
\dP_t = A_t (h_0 - 2 \Delta) + B_t g_0 + C_t (g_1 - \frac12 h_0)
\eeq
Then, using the dynamical equation and the algebra above, one finds:
\beq
{\cal L}_0 \dP_t = A_t (2(h_0-g_1) - h_0) = - 2A_t (g_1 - \frac12 h_0)
\eeq
Hence, the evolution is closed on itself, with the following evolution rules:
\beq
A_{t+1} = -\frac12 d_t, \qquad B_{t+1} = \frac12 s_t, \qquad C_{t+1} = - 2A_t.
\eeq
Note that $A_t - B_t = -\frac12 (d_{t-1} + s_{t-1}) = - 
\max({\dla}_{t-1},{\dlaT}_{t-1}) = - M_{t-1}$. 

Now, by definition ${\ola}_{t} = \frac{1-C}{6} + \int_{-1}^1 dx x \dP_t$, but 
since $h_0$ and $g_1$ are even, the only contribution comes from the $g_0$
component.
Therefore:
\beq
{\dla}_t = B_t \int_{-1}^1 dx x g_0(x) = B_t.
\eeq
This leads to a first evolution equation:
\beq \label{eq:dla1}
{\dla}_{t+1} = \frac12 s_t = \frac12 ({\dla}_{t}+{\dlaT}_{t}).
\eeq
The second equation comes from the evolution of the $\a_i$'s. From:
\beq
\overline{\a}(t+1)= \overline{\a}(t) - C {\dla}(t) - \frac{1-C}{2} {\dlaT}(t) - 
(1 - C) \int_{-1}^{0} dx\, x \dP_t(x),
\eeq
we finally find:
\beq \label{eq:dla2}
{\dlaT}_{t+1} = \frac34 {\dlaT}_t - \frac14 M_{t-1} +  \frac{7}{24} d_{t-2} - 
\frac{C}{4(1-C)} ({\dla}_{t}+{\dlaT}_{t}).
\eeq

The coupled set of iterations for ${\dla}_{t}$ and ${\dlaT}_t$, 
Eqs.~\eqref{eq:dla1} and~\eqref{eq:dla2}, lead 
to damped oscillations for $C < C^{**}$ and sustained oscillations for $C > 
C^{**}$, with $C^{**} \approx 0.91$. 
The numerical value of $C^{**}$ however does not coincide with that of $C^* 
\approx 0.45$, obtained from the 
direct simulation of Eqs. (\ref{updatedelta}). This might be due to neglecting 
higher order corrections, which appear to be numerically large:
the oscillations generated by Eqs. (\ref{updatedelta}) are of amplitude $\sim 
0.1$ at the onset, suggesting a sub-critical bifurcation. For small
$C$, on the other hand, the oscillation amplitude is small and the above 
equations appear to be quantitatively correct, validating the above
calculations.

\end{document}